\documentclass[journal,11pt,onecolumn,final]{IEEEtran}
\pdfoutput=1
\usepackage{amssymb}
\usepackage{amsmath}
\usepackage{amsthm}
\usepackage{mathrsfs}
\usepackage{latexsym}
\usepackage{epsf}
\usepackage{float,amsmath}
\usepackage{amsfonts}
\usepackage{color}
\usepackage{graphicx}
\usepackage{url}
\usepackage{setspace}
\renewcommand\baselinestretch 2

\newcommand\tra{{\rm Tr}}

\newcommand\ve{{\rm vec}}
\newcommand\diag{{\rm Diag}}
\newcommand\e{{\rm E}}
\newcommand\pardef{ \stackrel{{\rm def}}{=} }

\newtheorem{remark}{Remark}
\newtheorem{result}{Result}

\begin{document}
\singlespacing
\title{\bf Iterative Sparse Asymptotic Minimum Variance Based Approaches for Array Processing}
\author{\IEEEauthorblockN{Habti Abeida\IEEEauthorrefmark{1},
		Qilin Zhang\IEEEauthorrefmark{2}, Jian Li\IEEEauthorrefmark{3} and
		Nadjim Merabtine\IEEEauthorrefmark{4}}\\
	\IEEEauthorblockA{
		\IEEEauthorrefmark{1}Department of Electrical Engineering, University of Taif, Al-Haweiah, Saudi Arabia\\
		\IEEEauthorrefmark{2}Department of Computer Science, Stevens Institue of Technology, Hoboken, NJ, USA\\
		\IEEEauthorrefmark{3}Department of Electrical and Computer Engineering, University of Florida, Gainesville, FL, USA\\
		\IEEEauthorrefmark{4}Department of Electrical Engineering, University of Taif, Al-Haweiah, Saudi Arabia}}
\maketitle
%
\begin{abstract}
This paper presents a series of user parameter-free iterative Sparse Asymptotic Minimum Variance (SAMV) approaches for array processing applications based on the asymptotically minimum variance (AMV) criterion. With the assumption of abundant snapshots in the direction-of-arrival (DOA) estimation problem, the signal powers and noise variance are jointly estimated by the proposed iterative AMV approach, which is later proved to coincide with the Maximum Likelihood (ML) estimator. We then propose a series of power-based iterative SAMV approaches, which are robust against insufficient snapshots, coherent sources and arbitrary array geometries. Moreover, to overcome the direction grid limitation on the estimation accuracy, the SAMV-Stochastic ML (SAMV-SML) approaches are derived by explicitly minimizing a closed form stochastic ML cost function with respect to one scalar parameter, eliminating the need of any additional grid refinement techniques. To assist the performance evaluation, approximate solutions to the SAMV approaches are also provided at high signal-to-noise ratio (SNR) and low SNR, respectively. Finally, numerical examples are generated to compare the performance of the proposed approaches with existing approaches. 
\end{abstract}
%

\noindent {\bf Index terms:} Array Processing, AMV estimator, Direction-Of-Arrival (DOA) estimation, Sparse parameter estimation, Covariance matrix, Iterative methods, Vectors, Arrays, Maximum likelihood estimation, Signal to noise ratio, SAMV approach.
\begin{center}
Preprint version PDF available on arXiv. Official version: {\em Abeida Habti, Qilin Zhang, Jian Li, and Nadjim Merabtine. ``Iterative sparse asymptotic minimum variance based approaches for array processing.'' IEEE Transactions on Signal Processing 61, no. 4 (2013): 933-944.}\\
Matlab implementation codes available online, \url{https://qilin-zhang.github.io/publications/}
\end{center}
%

\noindent\rule{7 in}{.005in}

\noindent

%
\section{Introduction}
\label{sec:Introduction}

Sparse signal representation has attracted a lot of attention in
recent years and it has been successfully used for solving inverse
problems in various applications such as channel equalization (e.g.,
\cite{Fevrier, Cotter, Ling, Berger}), source localization (e.g.,
\cite{Malioutov, Yardibi,zhang2012fast,zhang2011fast}) and radar imaging (e.g., \cite{Cetin01,
Chen10, Roberts, Austin11}). In its basic form, it attempts to find
the sparsest signal  ${\bf x}$ satisfying the constrain ${\bf
y}={\bf A}{\bf x}$ or ${\bf y}={\bf A}{\bf x}+{\bf e}$ where ${\bf
A}\in\mathbb{C}^{M\times K}$ is an overcomplete basis (i.e., $K>M$),
${\bf y}$ is the observation data, and ${\bf e}$ is the noise term.
Theoretically, this problem is underdetermined and has multiple
solutions. However, the additional constraint that ${\bf x}$ should
be sparse allows one to eliminate the ill-posedness (e.g.,
\cite{Donoho06, Donoho06bis}). In recent years, a number of
practical algorithms such as $\ell_1$ norm minimization (e.g.,
\cite{Tropp, Donoho03}) and focal underdetermined system solution
(FOCUSS) (e.g., \cite{Gorodnitsky, Rao}) have been proposed to
approximate the sparse solution.

Conventional subspace-based source localization algorithms such as
multiple signal classification (MUSIC) and estimation of signal
parameters via a rotational invariance technique (ESPRIT)
\cite{Schmidt, Roy} are only applicable when $M>K$, and they require
sufficient snapshots and high signal-to-noise ratio (SNR) to achieve
high spatial resolution. However, it is often unpractical to collect
a large number of snapshots, especially in fast time-varying
environment, which deteriorates the construction accuracy of the
subspaces and degrades the localization performance. In addition,
even with appropriate array calibration, subspace-based methods are
incapable of handling the source coherence due to their sensitivity
to subspace orthogonality (e.g., \cite{Schmidt,Pillai}).

%
Recently, a user parameter-free non-parametric algorithm, the
iterative adaptive approach (IAA), has been proposed in
\cite{Yardibi} and employed in various applications (e.g.,
\cite{Chen10, Roberts}). It is demonstrated in these works that the
least square fitting-based IAA algorithm provides accurate DOA and
signal power estimates, and it is insensitive to practical
impairments such as few (even one) snapshots, arbitrary array
geometries and coherent sources. However, the iterative steps are
based on the IAA covariance matrix ${\bf R}={\bf A}{\diag({\bf
p})}{\bf A}^H$, which could be singular in the noise-free scenarios
when only a few components of the power vector ${\bf p}$ are
non-zero. In addition, a regularized version of the IAA algorithm
(IAA-R) is later proposed in \cite{Roberts} for single-snapshot and
nonuniform white noise cases. Stoica {\em et al.} have recently
proposed a user parameter-free SParse Iterative Covariance-based
Estimation (SPICE) approach in \cite{Stoica10, Stoica11} based on
minimizing a covariance matrix fitting criterion. However, the SPICE
approach proposed in \cite{Stoica10} for the multiple-snapshot case
depends on the inverse of the sample covariance matrix, which exists
only if the number of snapshot $N$ is larger than $M$
\cite{Anderson}. Therefore, this approach also suffers from
insufficient snapshots when $N < M$. We note that the source
localization performance of the power-based algorithms is mostly
limited by the fineness of the direction grid \cite{Malioutov}.

In this paper, we propose a series of iterative Sparse Asymptotic
Minimum Variance (SAMV) approaches based on the asymptotically
minimum variance (AMV) approach (also called asymptotically best
consistent (ABC) estimators in \cite{Stoica85}), which is initially
proposed for DOA estimation in \cite{Delmas04a, Abeida07b}. After
presenting the sparse signal representation data model for the DOA
estimation problem in Section \ref{sec:Signal model}, we first
propose an iterative AMV approach in Section \ref{sec:Convectional
AMV approach}, which is later proven to be identical to the
stochastic Maximum Likelihood (ML) estimator. Based on this
approach, we then propose the user parameter-free iterative SAMV
approaches that can handle arbitrary number of snapshots ($N<M$ or
$N>M$), and only a few non-zero components in the power estimates
vector ${\bf p}$ in Section \ref{sec:SAMV Approaches}. In addition,
A series of SAMV-Stochastic ML (SAMV-SML) approaches are proposed in
Section \ref{sec:SAMV-ML approach} to alleviate the direction grid
limitation and enhance the performance of the power-based SAMV
approaches. In Section \ref{sec:High and low SNR approx}, we derive
approximate expressions for the SAMV powers-iteration formulas at
both high and low SNR. In Section \ref{sec:simulations}, numerical
examples are generated to compare the performances of the proposed
approaches with existing approaches. Finally, conclusions are given
in Section \ref{sec:Conclusion}.

The following notations are used throughout the paper. Matrices and
vectors are represented by bold upper case and bold lower case
characters, respectively. Vectors are by default in column
orientation, while $T$, $H$, and $*$ stand for transpose, conjugate
transpose, and conjugate, respectively. $\e(\cdot)$, $\tra(\cdot)$
and $\det(\cdot)$ are the expectation, trace and determinant
operators, respectively. $\ve (\cdot)$ is the ``vectorization''
operator that turns a matrix into a vector by stacking all columns
on top of one another, $\otimes$ denotes the Kronecker product,
${\bf I}$ is the identity matrix of appropriate dimension, and ${\bf
e}_m$ denotes the $m$th column of ${\bf I}$.
%
%
\section{Problem Formation and Data Model}
\label{sec:Signal model}
Consider an array of $M$ omnidirectional sensors receiving $K$
narrowband signals impinging from the sources located at
$\boldsymbol{\theta}\pardef(\theta_1,\ldots,\theta_K)$ where
$\theta_k$ denotes the location parameter of the $k$th signal,
$k=1,\ldots,K$. The $M \times 1$ array snapshot vectors can be
modeled as (see e.g., \cite{Yardibi, Stoica10})
\begin{equation}
\label{eq:eq1} {\bf y}(n)={\bf A}{\bf x}(n)+{\bf e}(n),
\;n=1,\ldots,N,
\end{equation}
where ${\bf A}\pardef[{\bf a}(\theta_1),\ldots,{\bf a}(\theta_K)]$
is the steering matrix with each column being a steering vector
${\bf a}_k\pardef{\bf a}(\theta_k)$, a known function of $\theta_k$.
The vector ${\bf x}(n)\pardef[{\bf x}_1(n),\ldots,{\bf x}_K(n)]^{T}$
contains the  source waveforms, and ${\bf e}(n)$ is the noise term.
Assume that $\e\left({\bf e}(n){\bf e}^H(\bar{n})\right)=
 \sigma{\bf I}_M\delta_{n,\bar{n}}$\footnote{The nonuniform white noise case is considered later in Remark
 \ref{re:remak2}.},
 where $\delta_{n,\bar{n}}$ is the Dirac delta and it equals to $1$
 only if $n=\bar{n}$ and $0$ otherwise.
 We also assume first that ${\bf e}(n)$ and ${\bf x}(n)$ are independent,
 and that  $\e\left({\bf x}(n){\bf x}^H(\bar{n})\right)={\bf P}\delta_{n,\bar{n}}$, where
 ${\bf P}\pardef \diag( {p_1,\ldots,p_K})$. Let ${\bf p}$ be a
 vector containing the unknown signal powers and noise variance,
 ${\bf p} \pardef  [p_1,\ldots,p_K, \sigma]^T$.

The covariance matrix of ${\bf y}(n)$ that conveys information
 about $\boldsymbol{\bf p} $ is given by
 %
 %
\[{\bf R}\pardef{\bf A}{\bf P}{\bf A}^{H}+\sigma{\bf I}.
\]
This covariance matrix is traditionally estimated by the sample
covariance matrix ${\bf R}_{N}\pardef {\bf Y}{\bf Y}^{H}/N$ where
${\bf Y}\pardef[{\bf y}(1),\ldots,{\bf y}(N)]$. After applying the
vectorization operator to the matrix ${\bf R}$, the obtained vector
${\bf r}(\boldsymbol{\bf p})\pardef\ve({\bf R})$ is linearly related
to the unknown parameter $\boldsymbol{\bf p}$ as
\begin{equation}
\label{eq:linearizationR} {\bf r}(\boldsymbol{\bf p})\pardef\ve({\bf
R})={\bf S}\boldsymbol{\bf p},
\end{equation}
where  ${\bf S}\pardef [{\bf S}_1,\bar{\bf a}_{K+1}]$, ${\bf S}_1
=[\bar{\bf a}_1,\ldots,\bar{\bf a}_K]$, $\bar{\bf a}_k\pardef{\bf
a}^{*}_k\otimes{\bf a}_k$, $k=1,\ldots K$, and $\bar{\bf
a}_{K+1}\pardef\ve({\bf I})$.

We note that  the Gaussian circular asymptotic covariance matrix
${\bf r}_N\pardef \ve({\bf R}_N)$ is given by \cite[Appendix
B]{Abeida05a}, \cite{Delmas04a}
\[{\bf C}_r={\bf R}^{*}\otimes{\bf R}.
\]

The number of sources, $K$, is usually unknown. The power-based
algorithms, such as the proposed SAMV approaches, use a predefined
scanning direction grid $\{\theta_k \}_{k=1}^{K}$ to cover the
entire region-of-interest ${\hbox{$\bf \Omega$}}$, and every point
in this grid is considered as a potential source whose power is to
be estimated. Consequently, $K$ is the number of points in the grid
and it is usually much larger than the actual number of sources
present, and only a few components of $\boldsymbol{\bf p}$ will be
non-zero. This is the main reason why sparse algorithms can be used
in array processing applications \cite{Stoica10, Stoica11}.

To estimate the parameter $\boldsymbol{\bf p}$ from the statistic
${\bf r}_N$, we develop a series of iterative SAMV approaches based
on the AMV approach introduced by Porat and Fridelander in
\cite{Porat85}, Stoica {\em et al.} in \cite{Stoica85} with their
asymptotically best consistent (ABC) estimator, and Delmas and
Abeida in \cite{Delmas04a, Abeida07b}.

%
\section{The asymptotically minimum variance approach}
\label{sec:Convectional AMV approach}
%
%
In this section, we develop a recursive approach to estimate the
signal powers and noise variance (i.e., $\boldsymbol{\bf p}$) based
on the AMV criterion using the statistic ${\bf r}_N$. We assume that
$\boldsymbol{\bf p}$ is identifiable from ${\bf r}(\boldsymbol{\bf
p})$. Exploiting the similarities to the works in \cite{Delmas04a,
Abeida07b}, it is straightforward to prove that the covariance
matrix ${\bf Cov}^{\rm Alg}_{\boldsymbol{p}}$ of an arbitrary
consistent estimator of $\boldsymbol{p}$ based on the second-order
statistic ${\bf r}_N$ is bounded below by the following real
symmetric positive definite matrix:
$${\bf Cov}^{\rm Alg}_{\boldsymbol{p}}\geq[{\bf
S}^H_{d}{\bf C}^{-1}_r{\bf S}_{d}]^{-1},$$
where ${\bf S}_{d}\pardef {\rm d}{\bf r}(\boldsymbol{p})/ {\rm
d}\boldsymbol{p}$. In addition, this lower bound is attained by the
covariance matrix of the asymptotic distribution of $\hat{\bf p}$
obtained by minimizing the following AMV criterion:
\begin{equation}
\hat{\boldsymbol{p}} =\arg \min_{\boldsymbol{p}} f(\boldsymbol{p}),
\nonumber
\end{equation}
where
\begin{equation}
\label{eq:MinVa} f(\boldsymbol{p})\pardef [{\bf r}_{N}-{\bf
r}(\boldsymbol{p})]^{H} {\bf C}_{r}^{-1} [{\bf r}_{N}-{\bf
r}(\boldsymbol{p})].
\end{equation}

From  \eqref{eq:MinVa} and using \eqref{eq:linearizationR}, the
estimate of $\boldsymbol{\bf p}$ is given by the following results
proved in Appendix A:
\begin{result}
\label{res: result1} The $\{\hat{p}_k\}_{k=1}^{K}$ and
$\hat{\sigma}$ that minimize \eqref{eq:MinVa} can be computed
iteratively. Assume $\hat{p}^{(i)}_k$ and $\hat{\sigma}^{(i)}$ have
been obtained in the $i$th iteration, they can be updated at the
$(i+1)$th iteration as:
\begin{eqnarray}
\label{eq:pestimate} \hat{p}^{(i+1)}_k&=&\frac{{\bf a}^{H}_k{\bf
R}^{-1{(i)}}{\bf R}_N {\bf R}^{-1{(i)}}{\bf a}_k}{ ({\bf
a}^{H}_k{\bf R}^{-1{(i)}}{\bf a}_k)^2}+\hat{p}^{(i)}_k-\frac{1}{{\bf
a}^{H}_k{\bf R}^{-1{(i)}}{\bf
a}_k}, \;\;k=1\ldots,K, \\
\label{eq:noisepower} \hat{\sigma}^{(i+1)}&=&\left(\tra({\bf
R}^{-2^{(i)}}{\bf R}_N)+\hat{\sigma}^{(i)}\tra({\bf R}^{-2^{(i)}})
 -\tra({\bf R}^{-1^{(i)}})\right)/{\tra{({\bf
R}^{-2^{(i)}})}},
\end{eqnarray}%
where the estimate of ${\bf R}$ at the $i$th iteration is given by
${\bf R}^{(i)}={\bf A}{\bf P}^{(i)}{\bf
A}^{H}+\hat{\sigma}^{(i)}{\bf I}$ with ${\bf
P}^{(i)}=\diag(\hat{p}^{(i)}_1,\ldots,\hat{p}^{(i)}_K).$
\end{result}

Assume that ${\bf x}(n)$ and ${\bf e}(n)$ are both circularly
Gaussian distributed, ${\bf y}(n)$ also has a circular Gaussian
distribution with zero-mean and covariance matrix ${\bf R}$. The
stochastic negative log-likelihood function of $\{{\bf
y}(n)\}_{n=1}^{N}$ can be expressed as (see, e.g., \cite{Stoica90,
Yardibi})
\begin{equation}
\label{eq:MLfunc} L(\boldsymbol{\bf p})=\ln(\det({\bf R}))+\tra({\bf
R}^{-1}{\bf R}_N).
\end{equation}
In lieu of the cost function \eqref{eq:MinVa} that depends linearly
on $\boldsymbol{\bf p}$ (see \eqref{eq:linearizationR}), this ML
cost-function \eqref{eq:MLfunc} depends non-linearly on the signal
powers and noise variance embedded in ${\bf R}$. Despite this
difficulty and reminiscent of \cite{Yardibi}, we prove in Appendix B
that the following result holds:
\begin{result}
\label{res: result2} The estimates given by \eqref{eq:pestimate} and
\eqref{eq:noisepower} are identical to the ML estimates.
\end{result}
Consequently, there always exists approaches that gives the same
performance as the ML estimator which is asymptotically efficient.
Returning to the Result \ref{res: result1}, first we notice that the
expression given by \eqref{eq:pestimate} remains valid regardless of
$K>M$ or $K<M$.  In the scenario where $K>M$, we observe from
numerical calculations that the $\hat{p}_k$ and $\hat{\sigma}$ given
by \eqref{eq:pestimate} and \eqref{eq:noisepower} may be negative;
therefore, the nonnegativity of the power estimates can be enforced
at each iteration by forcing the negative estimates to zero as
\cite[Eq. (30)]{Yardibi},
\begin{eqnarray}
\label{eq:positiveestimates}
 \hat{p}^{(i+1)}_k&=&\max\left(0,  \frac{{\bf a}^{H}_k{\bf R}^{-1{(i)}}{\bf R}_N {\bf R}^{-1{(i)}}{\bf
a}_k}{ ({\bf a}^{H}_k{\bf R}^{-1{(i)}}{\bf
a}_k)^2}+\hat{p}^{(i)}_k-\frac{1}{{\bf a}^{H}_k{\bf R}^{-1{(i)}}{\bf
a}_k}\right),\;\;k=1\ldots,K, \\ \nonumber
\hat{\sigma}^{(i+1)}&=&\max\left(0, \left(\tra({\bf
R}^{-2^{(i)}}{\bf R}_N)+\hat{\sigma}^{(i)}\tra({\bf R}^{-2^{(i)}})
 -\tra({\bf R}^{-1^{(i)}})\right)/{\tra{({\bf
R}^{-2^{(i)}})}}\right).
\end{eqnarray}
The above updating formulas of $\hat{p}_k$ and $\hat{\sigma}$ at
$(i+1)$th iteration require knowledge of ${\bf R}$, $\hat{p}_k$ and
$\hat{\sigma}$ at the $i$th iteration, hence this algorithm must be
implemented iteratively. The initialization of $\hat{p}_k$ can be
done with the periodogram (PER) power estimates (see, e.g.,
\cite{Stoica05b})
\begin{equation}
\label{eq:PERinti} \hat{p}^{(0)}_{k,{\rm PER}}=\frac{{\bf
a}^{H}_k{\bf R}_{N}{\bf a}_k}{\|{\bf a}_k\|^4}.
\end{equation}
The noise variance estimator $\hat{\sigma}$ can be initialized as,
for instance,
\begin{equation}
\label{eq:MLinti} \hat{\sigma} =\frac{1}{M N}\sum_{n=1}^{N}\|{\bf
y}(n)\|^2.
\end{equation}%

\begin{remark}
\label{re:remak1} In the classical scenario where there are more
sensors than sources (i.e., $K\leq M$), closed form approximate ML
estimates of a single source power and noise variance are derived in
\cite{Gershman92} and \cite{Gershman95} assuming uniform white noise
and nonuniform white noise, respectively. However, these approximate
expressions are derived at high and low SNR regimes
separately\footnote{For high and low SNR, the ML function
\eqref{eq:MLfunc} is linearized by different approximations in
\cite{Gershman92} and \cite{Gershman95}}, compared to the unified
expressions \eqref{eq:pestimate} and \eqref{eq:noisepower}
regardless of SNR or number of sources.
\end{remark}

\begin{remark}
\label{re:remak2}
Result \ref{res: result1} can be extended easily
to the nonuniform white  Gaussian noise case where the covariance
matrix is given by
\begin{equation}
\label{eq:nonunfnoisesa} \e\left({\bf e}(n){\bf
e}^H(n)\right)=\diag(\sigma_1,\ldots,\sigma_M)\pardef\sum_{m=1}^{M}\sigma_m{\bf
a}_{K+m} {\bf a}^T_{K+m},
\end{equation}
where ${\bf a}_{K+m} \pardef {\bf e}_m$, $m = 1,\ldots,M$, denote
the canonical vectors. Under these assumptions and from Result
\ref{res: result1}, the estimates of ${\bf p}$ at $(i+1)$th
iteration are given by
\begin{equation}
\label{eq:nonunfnoise} \hat{p}^{(i+1)}_k=\frac{{\bf a}^{H}_k{\bf
R}^{-1{(i)}}{\bf R}_N {\bf R}^{-1{(i)}}{\bf a}_k}{ ({\bf
a}^{H}_k{\bf R}^{-1{(i)}}{\bf a}_k)^2}+\hat{p}^{(i)}_k-\frac{1}{{\bf
a}^{H}_k{\bf R}^{-1{(i)}}{\bf a}_k}, \;k=1\ldots,K+M,
\end{equation}
where ${\bf R}^{{(i)}}={\bf A}{\bf P}^{{(i)}}{\bf A}^{H}+\sum_{m =
1}^{M}\hat{\sigma}^{{(i)}}_m {\bf a}_{K+m}  {\bf a}^T_{K+m}$, ${\bf
P}^{(i)}=\diag(\hat{p}^{(i)}_1,\ldots,\hat{p}^{(i)}_K)$ and
$\hat{\sigma}^{{(i)}}_m=\hat{p}^{{(i)}}_{K+m}$, $m=1,\ldots,M$.
\end{remark}
As mentioned before, the $\hat{p}_k$ may be negative when $K>M$,
therefore, the power estimates \label{eq:nonunfnoise} can be
iterated similar to \eqref{eq:positiveestimates} by forcing the
negative values to zero.
%
%
\section{The sparse asymptotic minimum variance Approaches}
\label{sec:SAMV Approaches}%
%
%
In this section, we propose the iterative SAMV approaches to
estimate $\boldsymbol{\bf p}$ even when $K$ exceeds the number of
sources $K$(i.e., when the steering matrix ${\bf A}$ can be viewed
as an overcomplete basis for ${\bf y}(n)$) and only a few non-zero
components are present in ${\bf p}$. This is the common case
encountered in many spectral analysis applications, where only the
estimation of ${\bf p}$ is deemed relevant (e.g., \cite{Stoica10,
Stoica11}).

As mentioned in Result \ref{res: result2}, the estimates given by
\eqref{eq:pestimate} and \eqref{eq:noisepower} may give irrational
negative values due to the presence of the non-zero terms
$p_k-1/({\bf a}^{H}_k{\bf R}^{-1}{\bf a}_k)$ and $\sigma-\tra({\bf
R}^{-1})/\tra({\bf R}^{-2})$. To resolve this difficulty, we assume
that\footnote{$p_k=1/({\bf a}^{H}_k{\bf R}^{-1}{\bf a}_k)$ is the
standard Capon power estimate \cite{Stoica05b}.} $p_k=1/({\bf
a}^{H}_k{\bf R}^{-1}{\bf a}_k)$ and $\sigma=\tra({\bf
R}^{-1})/\tra({\bf R}^{-2})$, and propose the
following SAMV approaches based on Result \ref{res: result1}:\\

\noindent {\em  SAMV-0 approach:}\\
\noindent The estimates of $p_{k}$ and ${\sigma}$  are updated at
$(i+1)$th iteration as:
\begin{eqnarray}
\label{eq:SAMV-0} \hat{p}^{(i+1)}_{k}&=&\hat{p}^{2(i)}_{k}({\bf
a}^{H}_{k} {\bf R}^{-1(i)}{\bf R}_N{\bf R}^{-1(i)}{\bf a}_{k}), \;
k=1,\ldots,K, \\\noindent
 \hat{\sigma}^{(i+1)}&=&\frac{\tra({\bf R}^{-2(i)}{\bf R}_N)}{\tra({\bf
 R}^{-2(i)})}.
\end{eqnarray}

\noindent {\em  SAMV-1 approach:}\\
\noindent The estimates of $p_{k}$ and ${\sigma}$  are updated at
$(i+1)$th iteration as:
\begin{eqnarray}
\label{eq:SAMV-1} \hat{p}^{(i+1)}_{k}&=&\frac{{\bf a}^{H}_{k} {\bf
R}^{-1(i)}{\bf R}_N{\bf R}^{-1(i)}{\bf a}_{k}}{({\bf a}^H_{k}{\bf
R}^{-1(i)}{\bf a}_{k})^2},\; k=1,\ldots,K, \\
\label{eq:noisesam}
 \hat{\sigma}^{(i+1)}&=&\frac{\tra({\bf R}^{-2(i)}{\bf R}_N)}{\tra({\bf
 R}^{-2(i)})}.
\end{eqnarray}

\noindent {\em SAMV-2 approach:}\\
%
\noindent The estimates of $p_{k}$ and ${\sigma}$  are updated at
$(i+1)$th iteration as:
\begin{eqnarray}
\label{eq:SAMV-3} \hat{p}^{(i+1)}_{k}&=&\hat{p}^{(i)}_{k}\frac{{\bf
a}^{H}_{k} {\bf R}^{-1(i)}{\bf R}_N{\bf R}^{-1(i)}{\bf a}_{k}}{{\bf
a}^H_{k}{\bf R}^{-1(i)}{\bf a}_{k}}, \;k=1,\ldots,K, \\ \nonumber
 \hat{\sigma}^{(i+1)}&=&\frac{\tra({\bf R}^{-2(i)}{\bf R}_N)}{\tra({\bf
 R}^{-2(i)})}.
\end{eqnarray}

In the case of nonuniform white Gaussian noise with covariance
matrix given in Remark \ref{re:remak2}, the SAMV noise powers
estimates can be updated alternatively as
\begin{equation}
\label{eq:nonunfnoise} \hat{\sigma}^{(i+1)}_m=\frac{{\bf e}^{H}_m
{\bf R}^{-1{(i)}}{\bf R}_N {\bf R}^{-1{(i)}}{\bf e}_m}{ ({\bf
e}^{H}_m {\bf R}^{-1{(i)}}{\bf e}_m)^2}, \;\;m=1\ldots,M,
\end{equation}
where ${\bf R}^{{(i)}}={\bf A}{\bf P}^{{(i)}}{\bf A}^{H}+\sum_{m =
1}^{M}\hat{\sigma}^{{(i)}}_m {\bf e}_m {\bf e}^T_m$, ${\bf
P}^{(i)}=\diag(\hat{p}^{(i)}_1,\ldots,\hat{p}^{(i)}_K)$ and ${\bf
e}_m$ are the canonical vectors, $m=1,\ldots,M$.

In the following Result \ref{res: result3} proved in Appendix
\ref{sec:Appendix C}, we show that the SAMV-1 signal power and noise
variance updating formulas given by \eqref{eq:SAMV-1} and
\eqref{eq:noisesam} can also be obtained by minimizing a weighted
least square (WLS) cost function.
\begin{result}
\label{res: result3} The SAMV-1 estimate is also the minimizer of
the following WLS cost function:
\begin{equation}
\hat{p}_k=\arg \min_{p_k} g(p_k),   \notag
\end{equation}
where
\begin{equation}
\label{eq:MinVb} g(p_k) \pardef \arg \min_{p_k} [{\bf
r}_N-p_k\bar{\bf a}_k]^{H}{\bf C}'^{-1}_k[{\bf r}_N-p_k\bar{\bf
a}_k].
\end{equation}
 and ${\bf C}'_k\pardef{\bf C}_r-p^2_k\bar{\bf a}_k\bar{\bf a}^{H}_k$,
$k=1,\ldots,K+1$.
\end{result}
 The implementation steps of the these SAMV approaches are summarized in Table 1.
\begin{center}\begin{tabular}{|l|c|r|}
  \hline
  \hspace{4.5cm}TABLE 1\\
\hspace{3.5cm}The  SAMV approaches\\
    \hline
  Initialization:
  \noindent
$\{{p}^{(0)}_k\}_{k=1}^{K}$ and $\hat{\sigma}^{(0)}$ using e.g., \eqref{eq:PERinti} and \eqref{eq:MLinti}. \\
 {\bf repeat}\\
 \vspace{-0.1cm}\noindent
  $\bullet$ \;Update ${\bf R}^{(i)}={\bf A}{\bf P}^{(i)}{\bf A}^H+\sigma^{(i)}{\bf I}$, \\
  $\bullet$ \;Update $\hat{p}^{(i+1)}_{k}$ using SAMV formulas \eqref{eq:SAMV-0} or \eqref{eq:SAMV-1} or \eqref{eq:SAMV-3}, \\
 $\bullet$ \;Update $\hat{\sigma}^{(i+1)}$ using \eqref{eq:noisesam}.
 \\
  \hline
\end{tabular}
\end{center}

 \begin{remark}
 \label{re:remak3}
  Since ${\bf
R}_{N}=\frac{1}{N}\sum_{n=1}^{N}{\bf y}(n){\bf y}^{H}(n)$, the
SAMV-1 source power updating formula \eqref{eq:SAMV-1} becomes
\begin{equation}
\label{eq:IAASAMV} p^{(i+1)}_{k}=\frac{1}{N{({\bf a}^H_{k}{\bf
R}^{-1(i)}{\bf a}_{k})^2}}\sum_{n=1}^{N}{|{\bf a}^{H}_{k} {\bf
R}^{-1(i)}{\bf y}(n)|^2}.
 \end{equation}
Comparing this expression with its IAA counterpart (\cite[Table
II]{Yardibi}), we can see that the difference is that the IAA power
estimate is obtained by adding up the signal magnitude estimates
$\{x_k(n)\}_{n=1}^{N}$. The matrix ${\bf R}$ in the IAA approach is
obtained as ${\bf A}{\bf P}{\bf A}^{H}$, where ${\bf P}=\diag(p_1,
\ldots, p_K)$. This ${\bf R}$ can suffer from matrix singularity
when only a few elements of $\{p_k\}_{k=1}^{K}$ are non-zero (i.e.,
the noise-free case).
\end{remark}
\begin{remark}
 \label{re:remak4} We note that the SPICE+ algorithm
derived in \cite{Stoica10} for the multiple snapshots case requires
that the matrix ${\bf R}_N$ be nonsingular, which is true with
probability 1 if $N \geq M$ \cite{Anderson}.
 This implies that this algorithm can not
be applied when $N<M$.  On contrary, this condition is not required
for the proposed SAMV approaches, which do not depend on the inverse
of ${\bf R}_N$. In addition, these SAMV approaches provide good
spatial estimates even with a few snapshots, as is shown in section
\ref{sec:simulations}.
\end{remark}%
%
\section{DOA estimation: The Sparse Asymptotic Minimum Variance-Stochastic Maximum Likelihood approaches}
\label{sec:SAMV-ML approach}
%
%
%
It has been noticed in \cite{Malioutov} that the resolution of most
power-based sparse source localization techniques is limited by the
fineness of the direction grid that covers the location parameter
space. In the sparse signal recovery model, the sparsity of the
truth is actually dependent on the distance between the adjacent
element in the overcomplete dictionary, therefore, the difficulty of
choosing the optimum overcomplete dictionary (i.e., particularly,
the DOA scanning direction grid) arises. Since the computational
complexity is proportional to the fineness of the direction grid, a
highly dense grid is not computational practical. To overcome this
resolution limitation imposed by the grid, we propose the grid-free
SAMV-SML approaches, which refine the location estimates
$\boldsymbol{\bf \theta}=(\theta_1,\ldots,\theta_K)^T$ by
iteratively minimizing a stochastic ML cost function with respect to
a single scalar parameter $\theta_k$.

Using \eqref{eq:MLfun_se}, the ML objective function can be
decomposed into $\mathcal{L}(\boldsymbol{\bf \theta}_{-k})$, the
marginal likelihood function with parameter $\theta_k$ excluded, and
$l(\theta_k)$ with terms concerning  $\theta_k$:
\begin{equation}
\label{eq:MLfuncti31} l(\theta_k)\pardef\ln\left( \frac{1}{1+p_k\alpha_{1,k}}\right)+p_k \frac{\alpha^{N}_{2,k}}{1+p_k\alpha_{1,k}},
\end{equation}
where the $\alpha_{1,k}$ and $\alpha^{N}_{2,k}$ are defined in
Appendix \ref{sec:Appendix B}. Therefore, assuming that the
parameter $\{p_k\}_{k=1}^{K}$ and $\sigma$ are estimated using the
SAMV approaches\footnote{SAMV-SML variants use different $p_k$ and
$\sigma$ estimates: AMV-SML: \eqref{eq:pestimate} and
\eqref{eq:noisepower}, SAMV1-SML: \eqref{eq:SAMV-1} and
\eqref{eq:noisesam},  SAMV2-SML: \eqref{eq:SAMV-3} and
\eqref{eq:noisesam}.}, the estimate of $\theta_k$ can be obtained by
minimizing \eqref{eq:MLfuncti31} with respect to the scalar
parameter $\theta_k$.

The classical stochastic ML estimates are obtained by minimizing the
cost function with respect to a multi-dimensional vector
$\{\theta_k\}_{k=1}^{K}$, (see e.g., \cite[Appendix B, Eq.
(B.1)]{Stoica90}). The computational complexity of the
multi-dimensional optimization is so high that the classical
stochastic ML estimation problem is usually unsolvable. On contrary,
the proposed SAMV-SML algorithms only require minimizing
\eqref{eq:MLfuncti31} with respect to a scalar $\theta_k$, which can
be efficiently implemented using derivative-free uphill search
methods such as the Nelder-Mead algorithm\footnote{The Nelder-Mead
algorithm has already been incorporated in the function
``fminsearch'' in MATLAB\textregistered.} \cite{Nelder65}.

The SAMV-SML approaches are summarized in Table 2.
\begin{center}\begin{tabular}{|l|c|r|}
  \hline
  \hspace{6.2cm}TABLE 2\\
\hspace{4.5cm}The SAMV-SML approaches\\
    \hline
  Initialization:
  \noindent
$\{{p}^{(0)}_k\}_{k=1}^{K}$, $\hat{\sigma}^{(0)}$ and
$\{\hat{\theta}^{(0)}_k\}_{k=1}^{K}$ based on
 the result of SAMV approaches, \\e.g., SAMV-3 estimates, \eqref{eq:SAMV-3} and \eqref{eq:noisesam}. \\
 {\bf repeat}\\
 \noindent
  $\bullet$ \;Compute ${\bf R}^{(i)}$ and ${\bf Q}^{(i)}_k$ given by \eqref{eq:Qkmatrix}, \\
  $\bullet$ \;Update $p_k$ using \eqref{eq:pestimate} or \eqref{eq:SAMV-1} or \eqref{eq:SAMV-3}, update
 $\sigma$ using \eqref{eq:noisepower} or \eqref{eq:noisesam}, \\
 $\bullet$ \;Minimizing  \eqref{eq:MLfuncti31} with respect to $\theta_k$ to obtain the stochastic ML estimates $\hat{\theta}_k$. \\
  \hline
\end{tabular}
\end{center}

%
%
%
%
%
%
%
\section{High and low SNR approximation}
\label{sec:High and low SNR approx}
%
%
To get more insights into the SAMV approaches, we derive the
following approximate expressions for the SAMV approaches at high
and low SNR, respectively.

\subsubsection{Zero-Order Low SNR Approximation}

Note that the inverse of the matrix ${\bf R}$ can be written as:
\begin{equation}
\label{eq:Rinv}
{\bf R}^{-1}=\left(\bar{\bf R}+\sigma{\bf I}\right)^{-1}=\bar{\bf R}^{-1}-\bar{\bf R}^{-1}\left(\frac{1}{\sigma}{\bf I} +\bar{\bf R}^{-1}\right)^{-1}\bar{\bf R}^{-1}, \;\; \mbox{where} \;\; \bar{\bf R}\pardef{\bf A}{\bf P}{\bf A}^H.
\end{equation}%

At low SNR (i.e., $\frac{p_k}{\sigma}\ll 1$), from \eqref{eq:Rinv},
we obtain ${\bf R}^{-1}\approx\frac{1}{\sigma}{\bf I}$. Thus,
\begin{eqnarray}
\label{eq:Rel1} {\bf a}^{H}_{k} {\bf R}^{-1}{\bf R}_N{\bf
R}^{-1}{\bf a}_{k}&\approx&\frac{1}{\sigma^2}({\bf a}^{H}_{k}{\bf
R}_N{\bf a}_{k}),
\\\label{eq:Rel2}{\bf a}^{H}_{k} {\bf R}^{-1} {\bf
a}_{k}&\approx&\frac{M^2}{\sigma},
\\  \label{eq:Rel3} \tra({\bf R}^{-2}{\bf R}_N)&\approx&\frac{1}{N \sigma^2}\sum_{n=1}^{N}\|{\bf
y}(n)\|^2,
\\ \label{eq:Rel4} \tra({\bf R}^{-2(i)})&\approx&\frac{M}{\sigma^2}.
\end{eqnarray}
Substituting \eqref{eq:Rel1} and \eqref{eq:Rel2} into the SAMV
updating formulas \eqref{eq:SAMV-0}-\eqref{eq:SAMV-3}, we obtain
\begin{eqnarray}
 \label{eq:lowsnr0}
\hat{p}^{(i+1)}_{k, {\rm SAMV-0}}&=&\frac{M^2}{\hat{\sigma}^{2}}\hat{p}^{2(i)}_{k, {\rm SAMV-0}} \hat{p}_{k,{\rm PER}}, \\
\label{eq:lowsnr1}
\hat{p}_{k, {\rm SAMV-1}}&=& \hat{p}_{k,{\rm PER}}, \\
\label{eq:lowsnr3} \hat{p}^{(i+1)}_{k, {\rm
SAMV-2}}&=&\frac{M}{\hat{\sigma}}\hat{p}^{(i)}_{k, {\rm SAMV-2}}
\hat{p}_{k,{\rm PER}},
\end{eqnarray}
where $\hat{p}_{k,{\rm PER}}$ is given by \eqref{eq:PERinti}. Using
\eqref{eq:Rel3} and \eqref{eq:Rel4}, the common SAMV noise updating
equation \eqref{eq:noisesam} is approximated as
$$
\hat{\sigma}=\frac{1}{M N}\sum_{n=1}^{N}\|{\bf y}(n)\|^2.
$$

From \eqref{eq:lowsnr1}, we comment that the SAMV-1 approach is
equivalent to the PER method at low SNR. In addition, we remark that
at very low SNR, the SAMV-0 and SAMV-2 power estimates given by
\eqref{eq:lowsnr0} and \eqref{eq:lowsnr3} are scaled versions of the
PER estimate $\hat{p}_{k,{\rm PER}}$, provided that they are both
initialized by PER.

\noindent
\subsubsection{Zero-Order High SNR Approximation}

At high SNR (i.e., $\frac{p_k}{\sigma}\gg 1$), from \eqref{eq:Rinv},
we obtain ${\bf R}^{-1}\approx \bar{\bf R}^{-1}$. Thus,
\begin{eqnarray}
\label{eq:Rel11} {\bf a}^{H}_{k} {\bf R}^{-1}{\bf R}_N{\bf
R}^{-1}{\bf a}_{k}&\approx&{\bf a}^{H}_{k} \bar{\bf R}^{-1}{\bf
R}_N\bar{\bf R}^{-1}{\bf a}_{k},
\\\label{eq:Rel21}{\bf a}^{H}_{k} {\bf R}^{-1} {\bf a}_{k}&\approx&{\bf a}^{H}_{k} \bar{\bf R}^{-1} {\bf
a}_{k}.
\end{eqnarray}
Substituting \eqref{eq:Rel11} and \eqref{eq:Rel21} into the SAMV
formulas \eqref{eq:SAMV-0}--\eqref{eq:SAMV-3} yields:
\begin{eqnarray}
\label{eq:Rel211} p^{(i+1)}_{k, \;{\rm SAMV-0}}&=& p^{2(i)}_{k, {\rm
SAMV-0}}({\bf a}^{H}_{k} \bar{\bf
R}^{-1(i)}{\bf R}_N\bar{\bf R}^{-1(i)}{\bf a}_{k}), \\
\label{eq:Rel212}
p^{(i+1)}_{k, \;{\rm SAMV-1}}&=&\frac{{\bf a}^{H}_{k} \bar{\bf
R}^{-1(i)}{\bf R}_N\bar{\bf R}^{-1(i)}{\bf a}_{k}}{({\bf a}^H_{k}\bar{\bf
R}^{-1(i)}{\bf a}_{k})^2}, \\
\label{eq:Rel213} p^{(i+1)}_{k, \;{\rm SAMV-2}}&=& p^{(i)}_{k, {\rm
SAMV-2}}\frac{{\bf a}^{H}_{k} \bar{\bf R}^{-1(i)}{\bf R}_N\bar{\bf
R}^{-1(i)}{\bf a}_{k}}{{\bf a}^H_{k}\bar{\bf R}^{-1(i)}{\bf a}_{k}}.
\end{eqnarray}

From \eqref{eq:Rel212}, we get
\begin{equation}
\label{eq:IAA-SAM1}
p^{(i+1)}_{k}=\frac{{\bf a}^{H}_{k} \bar{\bf
R}^{-1(i)}{\bf R}_N\bar{\bf R}^{-1(i)}{\bf a}_{k}}{({\bf a}^H_{k}\bar{\bf
R}^{-1(i)}{\bf a}_{k})^2}=\frac{1}{N}\sum_{n=1}^{N}|x^{(i)}_{k, {\rm IAA}}(n)|^2,
\end{equation}
where $x^{(i)}_{k, {\rm IAA}}(n)\pardef \frac{{\bf a}^H_k\bar{\bf
R}^{-1(i)} {\bf y}(n)}{{\bf a}^H_k\bar{\bf R}^{-1(i)} {\bf a}_k}$ is
the the signal waveform estimate at the direction $\theta_k$ and
$n$th snapshot\cite[Eq. (7)]{Yardibi}. From \eqref{eq:IAA-SAM1}, we
comment that SAMV-1 and IAA are equivalent at high SNR. The only
difference is that the IAA powers estimates are obtained by summing
up the signal magnitude estimates $\{x_{k, {\rm
IAA}}(n)\}_{n=1}^{N}$.

\section{Simulation Results}
\label{sec:simulations}
\subsection{Source Localization}

This subsection focuses on evaluating the performances of the
proposed SAMV and SAMV-SML algorithms using an $M = 12$ element
uniform linear array (ULA) with half-wavelength inter-element
spacing, since the application of the proposed algorithms to
arbitrary arrays is straightforward. For all the considered
power-based approaches, the scanning direction grid $\{\theta_k
\}_{k=1}^{K}$ is chosen to uniformly cover the entire
region-of-interest ${\hbox{$\bf \Omega$}} = [0^\circ \; 180^\circ )$
with the step size of $0.2^\circ$.  The various SNR values are
achieved by adjusting the noise variance $\sigma$, and the SNR is
defined as:
\begin{align}
\textrm{SNR} \triangleq 10 \log_{10} \left(
\frac{p_{\textrm{avg}}}{\sigma} \right) \; [\textrm{dB}],
\label{SNRdefine}
\end{align}
where $p_{\textrm{avg}}$ denotes the average power of all sources.
For $K$ sources,
$
p_{\textrm{avg}} \triangleq \frac{1}{K} \sum_{k=1}^{K} p_k.
$

\begin{figure}[tbp]
\centering 
\begin{tabular}{cc}
{\includegraphics[width=2.0in,height=1.5in]{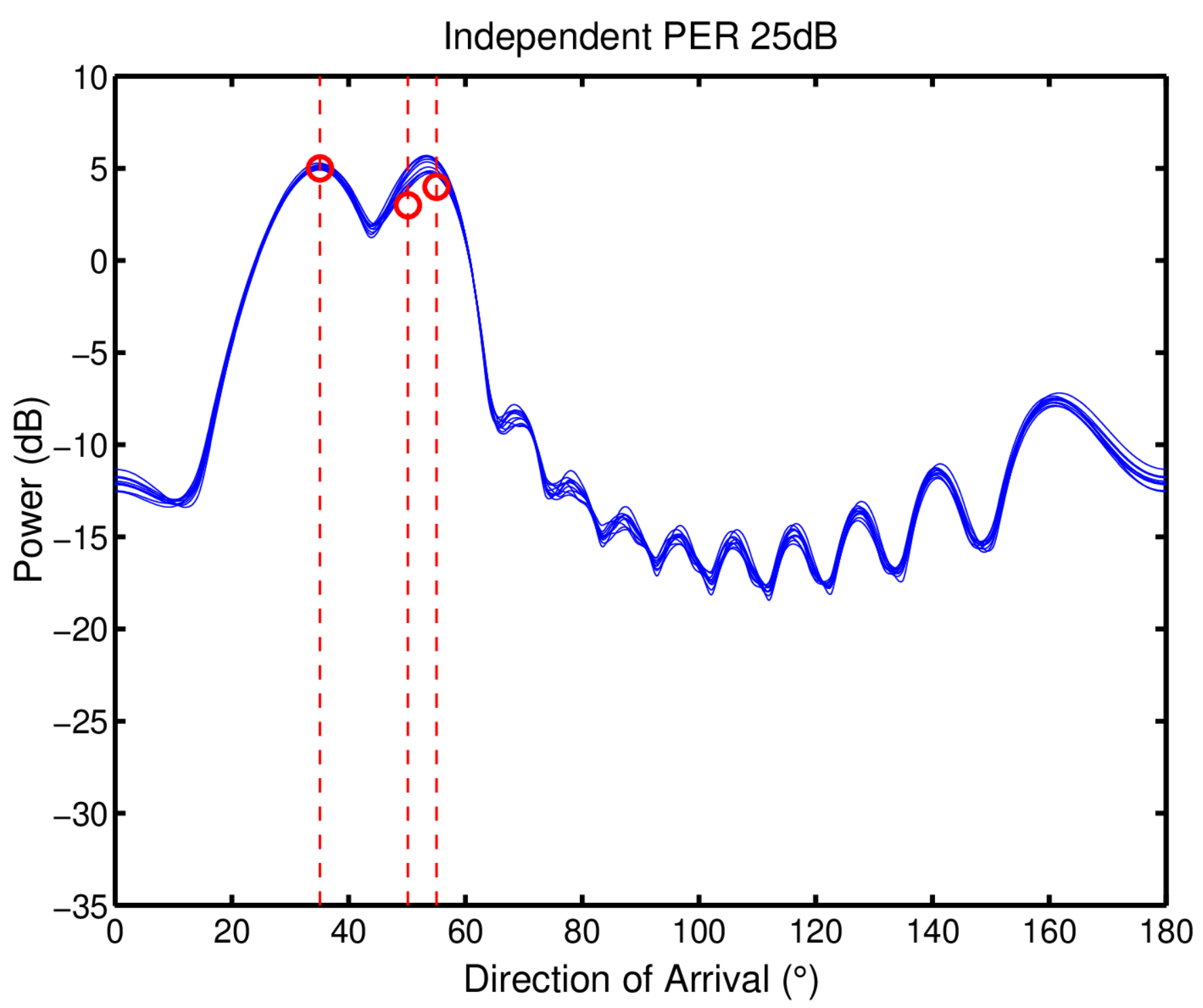}}
&
{\includegraphics[width=2.0in,height=1.5in]{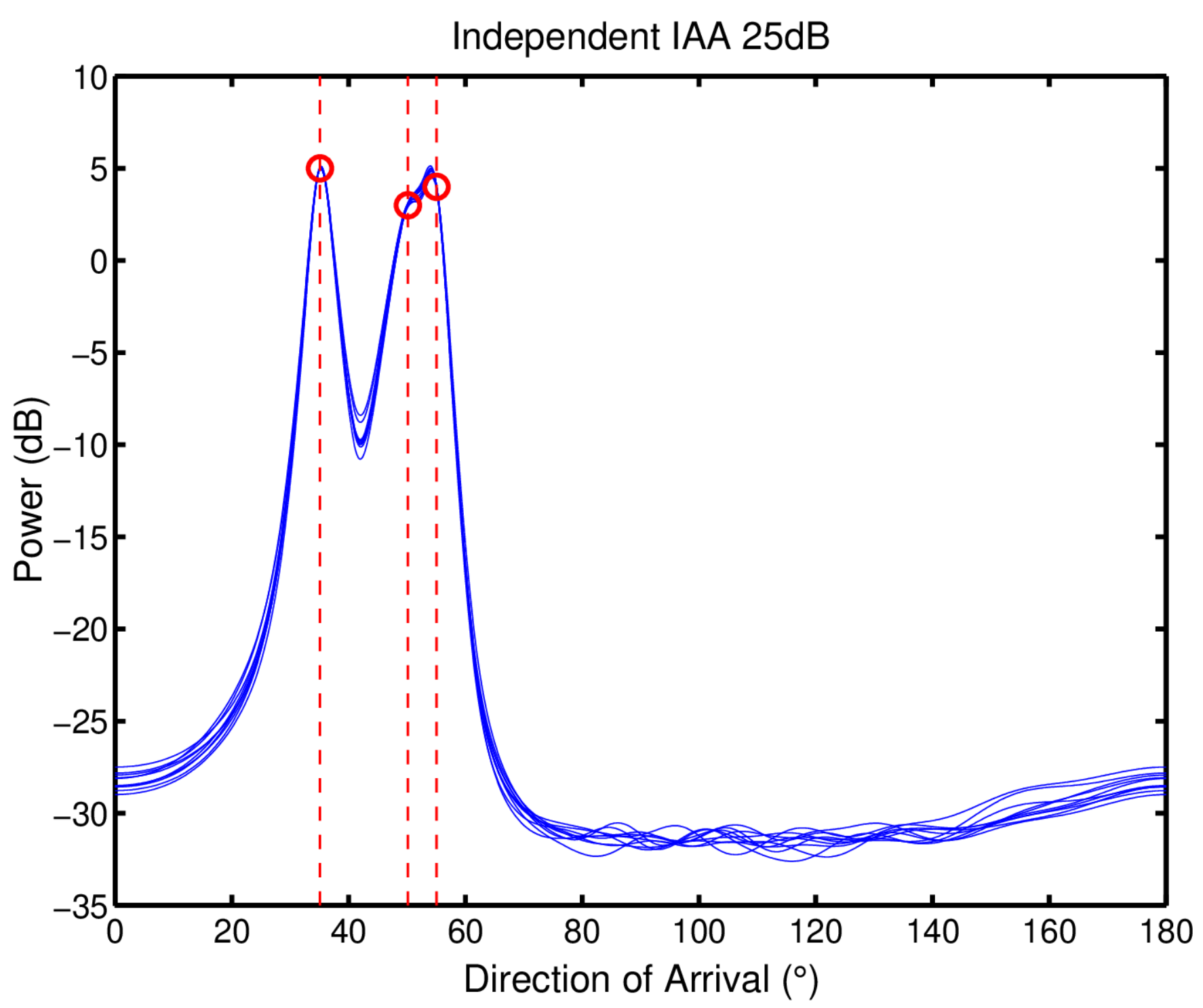}}\\
(a) & (b) \\
{\includegraphics[width=2.0in,height=1.5in]{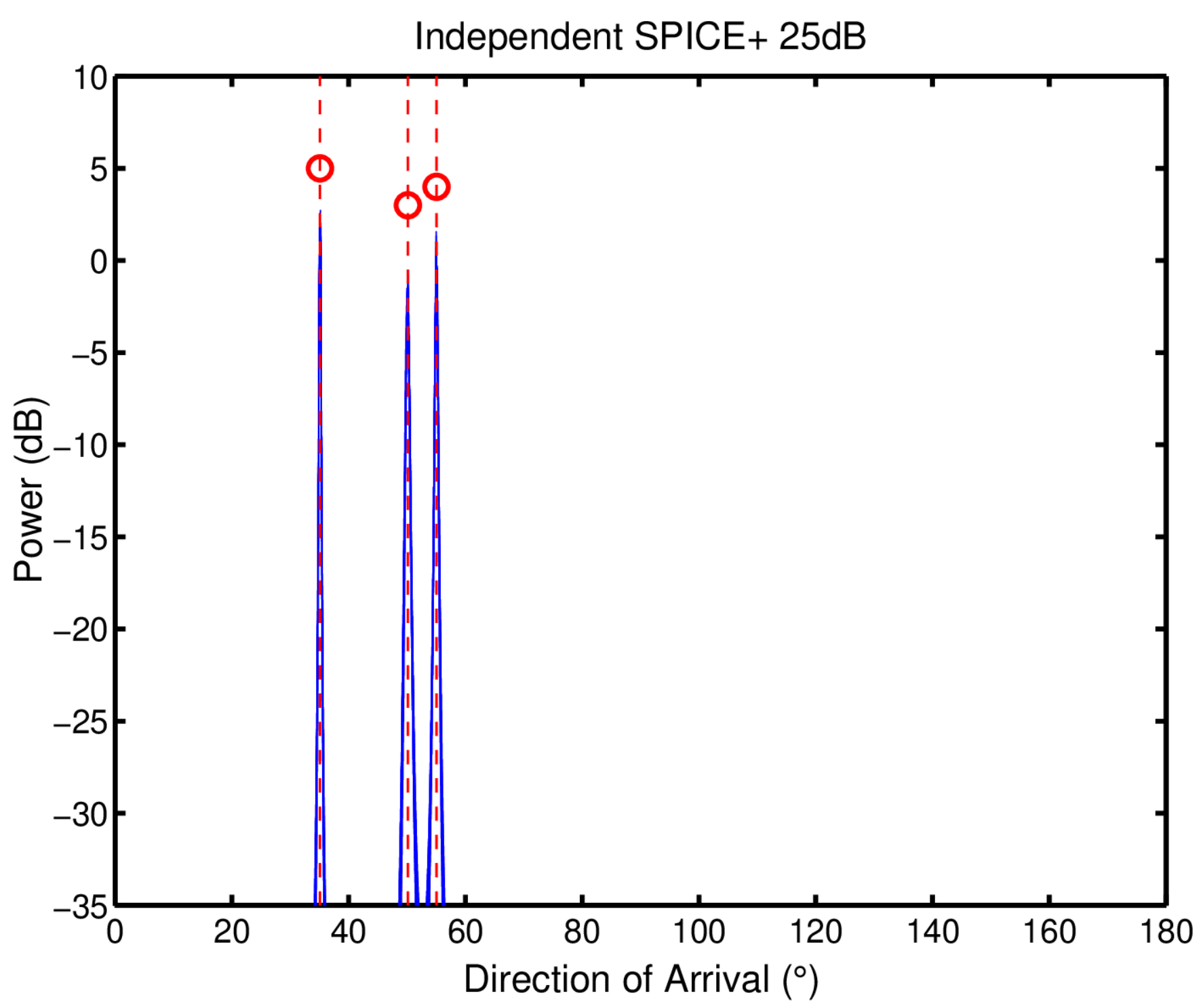}}
&
{\includegraphics[width=2.0in,height=1.5in]{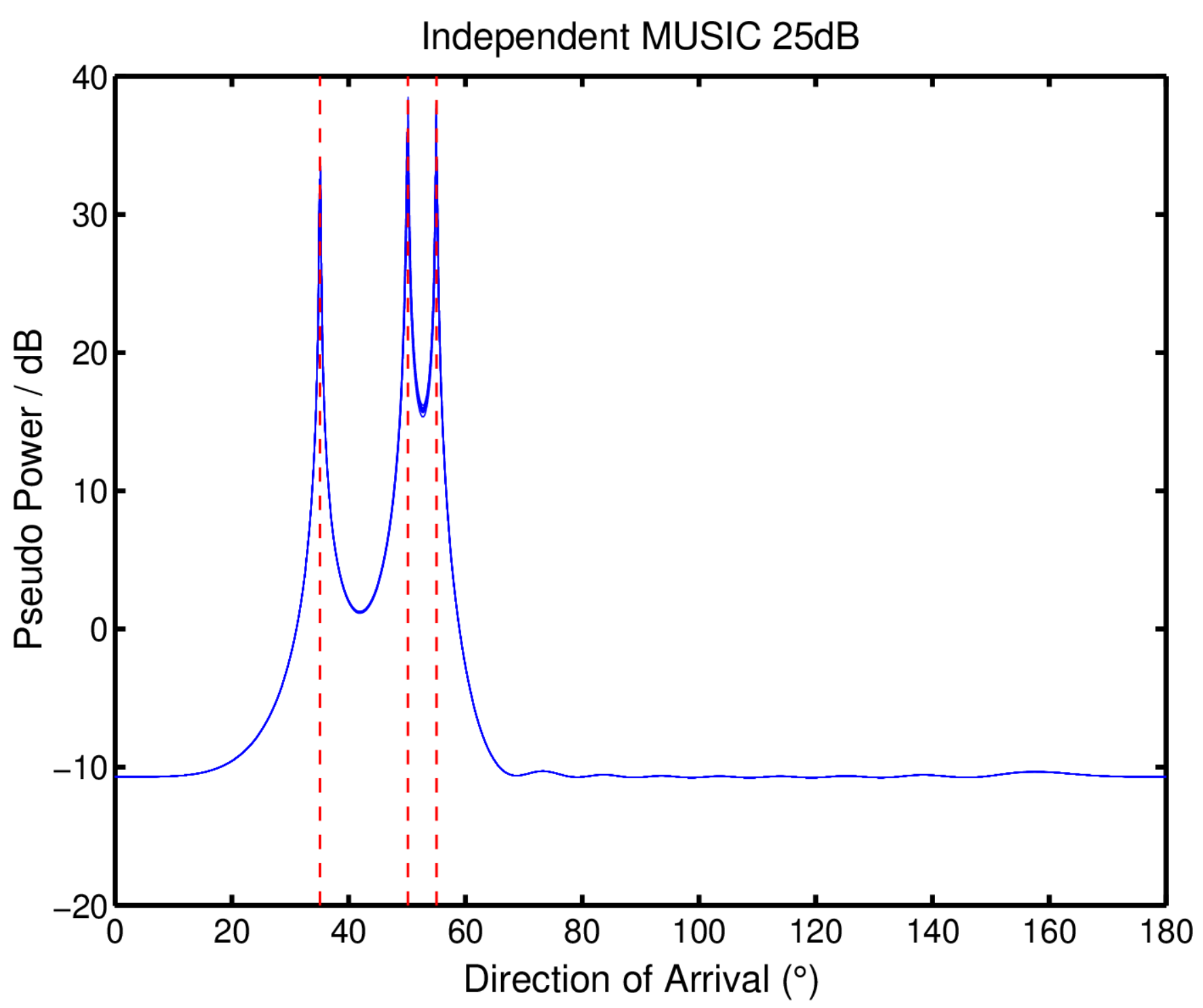}}\\
(c) & (d) \\
\end{tabular}
\begin{tabular}{ccc}
{\includegraphics[width=2.0in,height=1.5in]{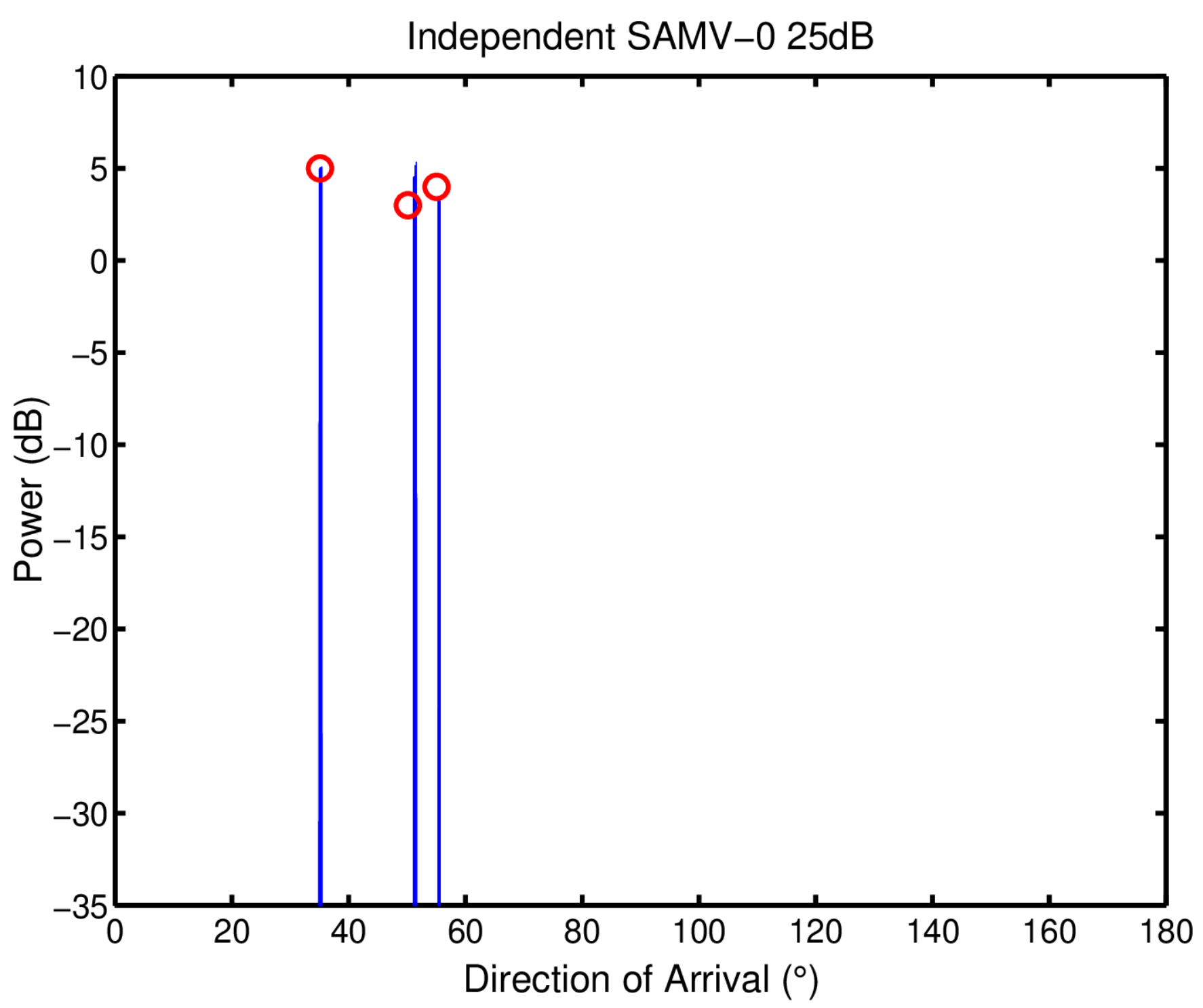}}
&
{\includegraphics[width=2.0in,height=1.5in]{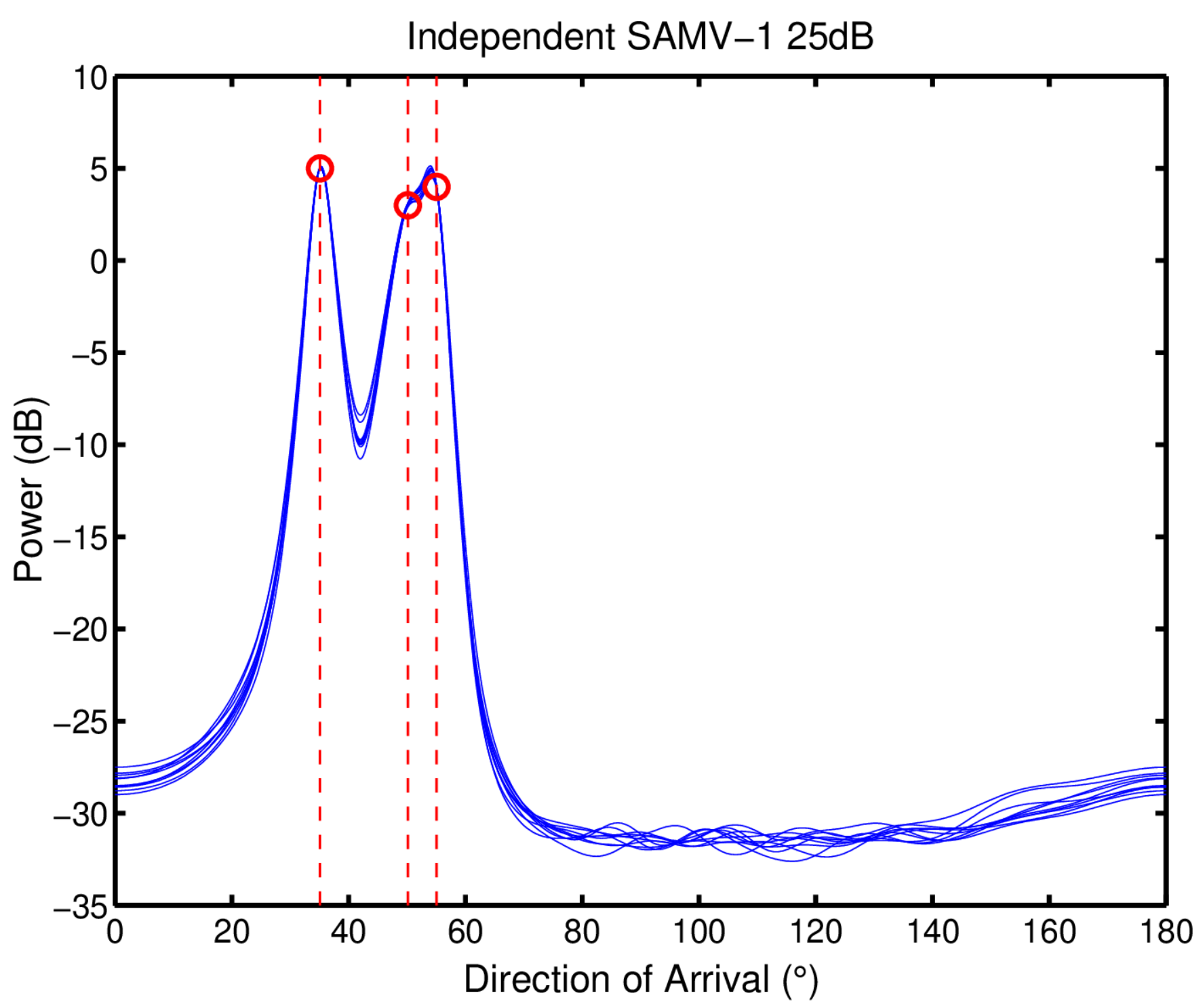}}
&
{\includegraphics[width=2.0in,height=1.5in]{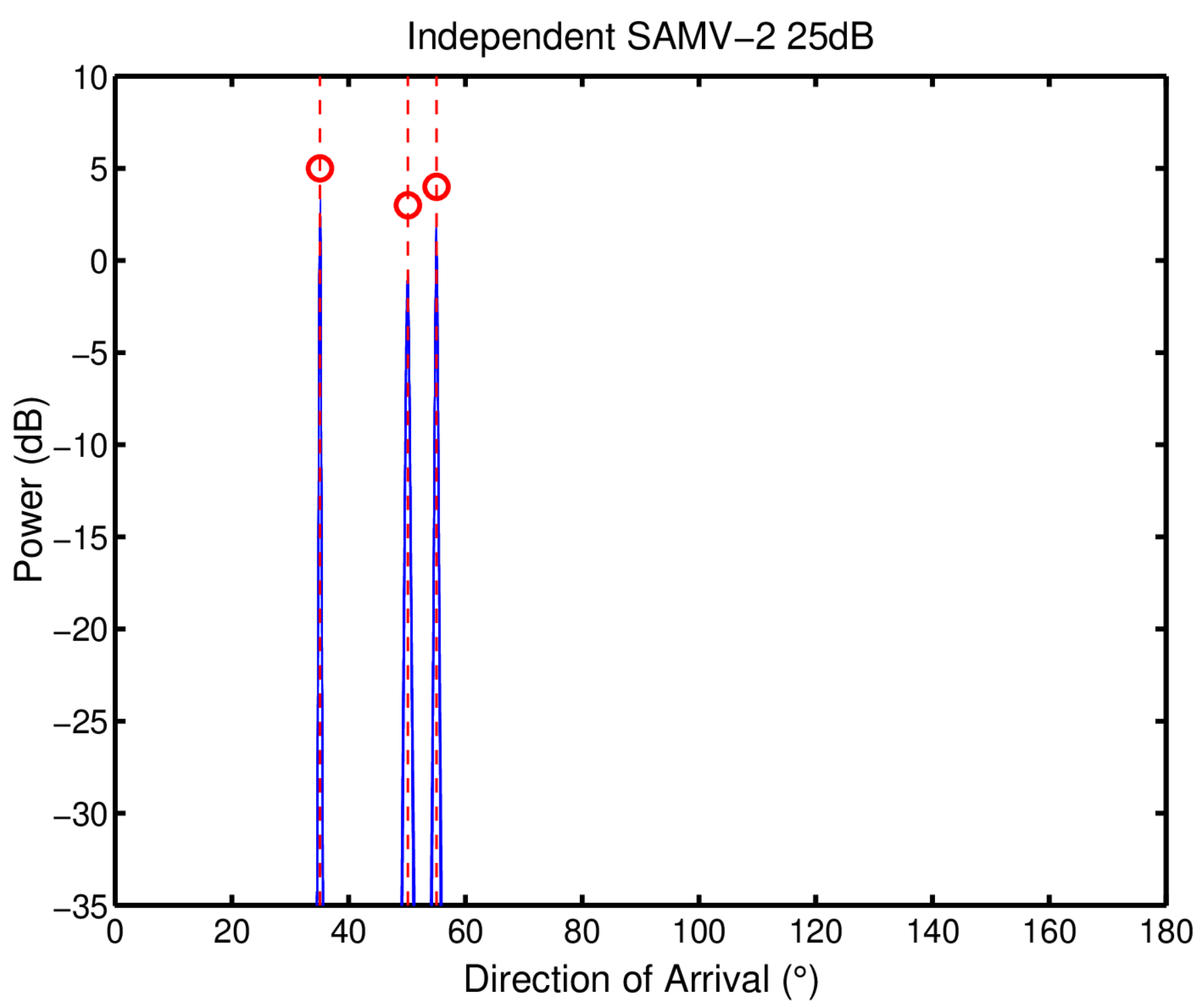}}\\
(e) & (f) & (g) \\
{\includegraphics[width=2.0in,height=1.5in]{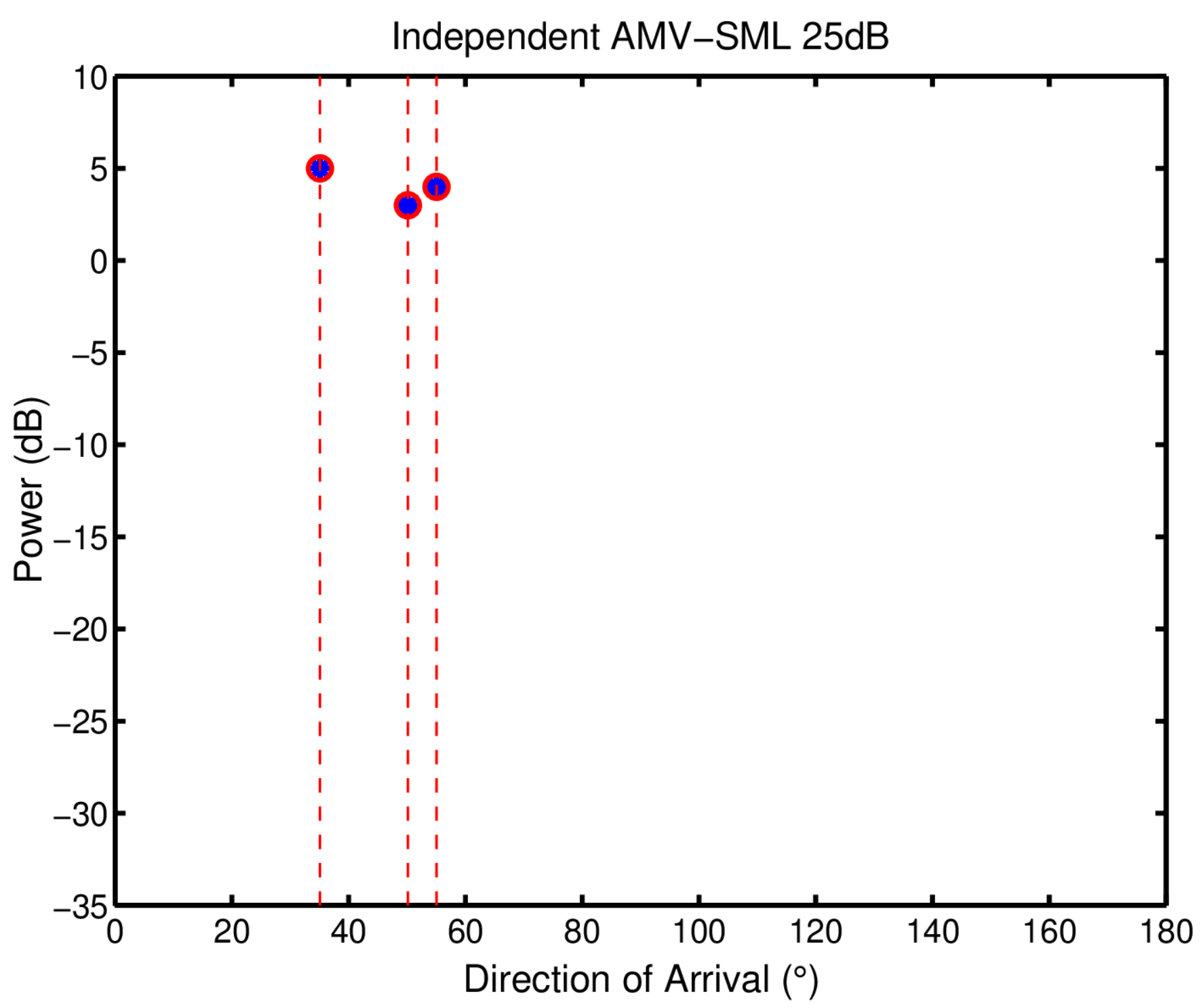}}
&
{\includegraphics[width=2.0in,height=1.5in]{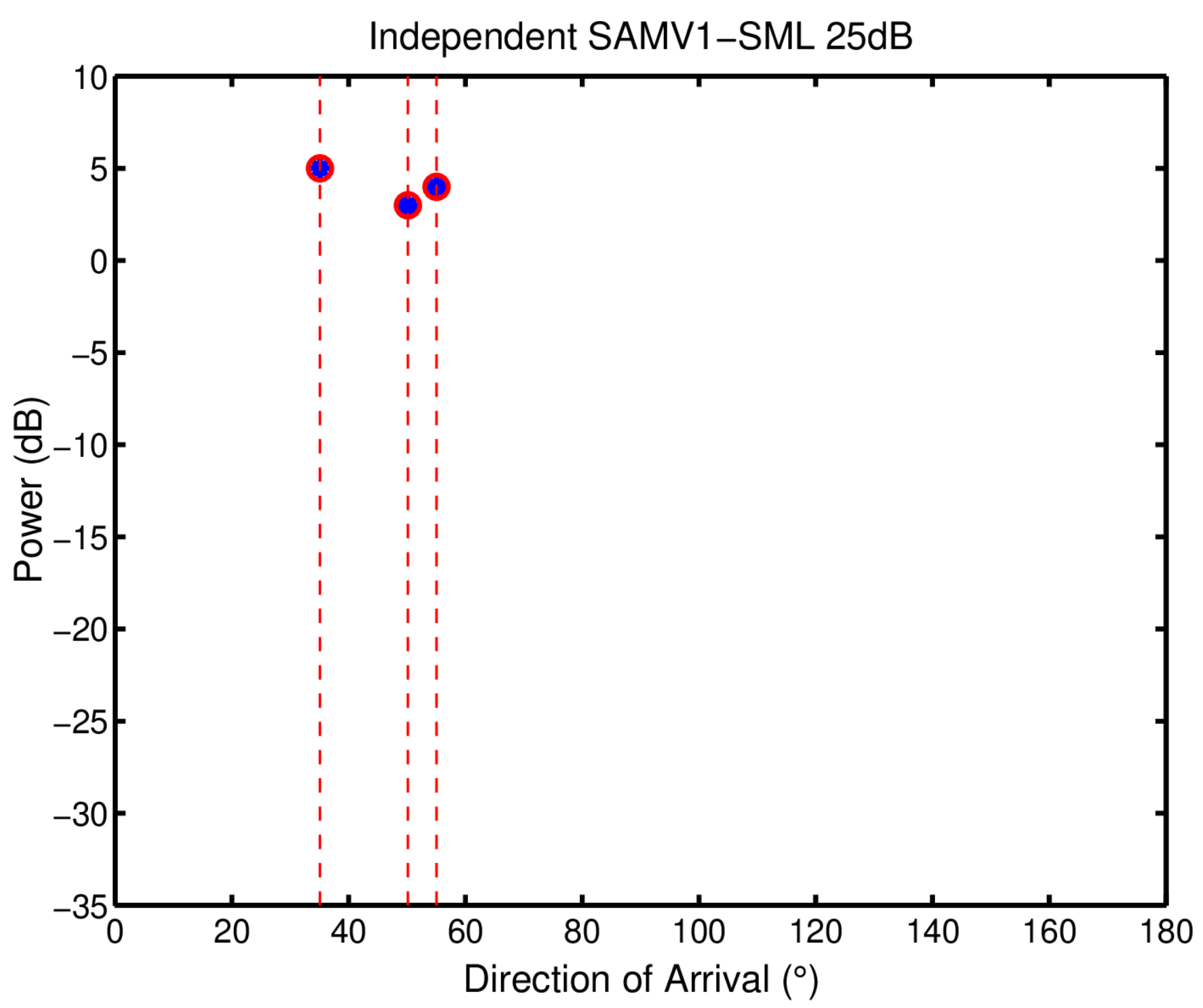}}
&
{\includegraphics[width=2.0in,height=1.5in]{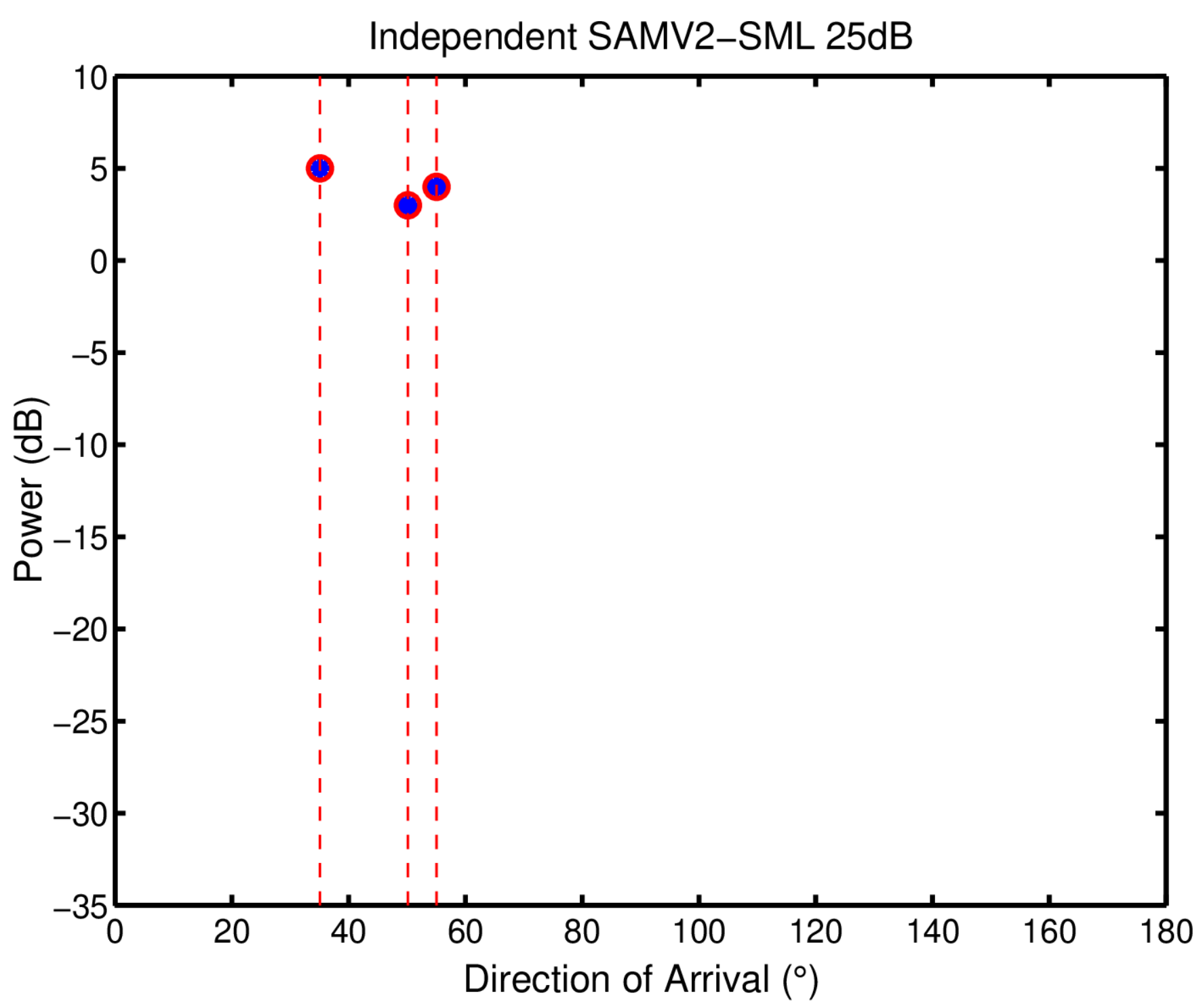}}\\
(h) & (i) & (j) \\
\end{tabular}
\centering \caption{Source localization with a ULA of $M=12$ sensors
and $N = 120$ snapshots, SNR = $25$ dB: Three uncorrelated sources
at $35.11^\circ$, $50.15^\circ$ and $55.05^\circ$, respectively, as
represented by the red circles and vertical dashed lines in each
plot. $10$ Monte Carlo trials are shown in each plot. Spatial
estimates are shown with (a) Periodogram (PER), (b) IAA, (c) SPICE+,
(d) MUSIC, (e) SAMV-0, (f) SAMV-1, (g) SAMV-2, (h) AMV-SML, (i)
SAMV1-SML and (j) SAMV2-SML.} \label{DOAspatial_indp}
\end{figure}

\begin{figure}[tbp]
\centering 
\begin{tabular}{cc}
{\includegraphics[width=2.0in,height=1.5in]{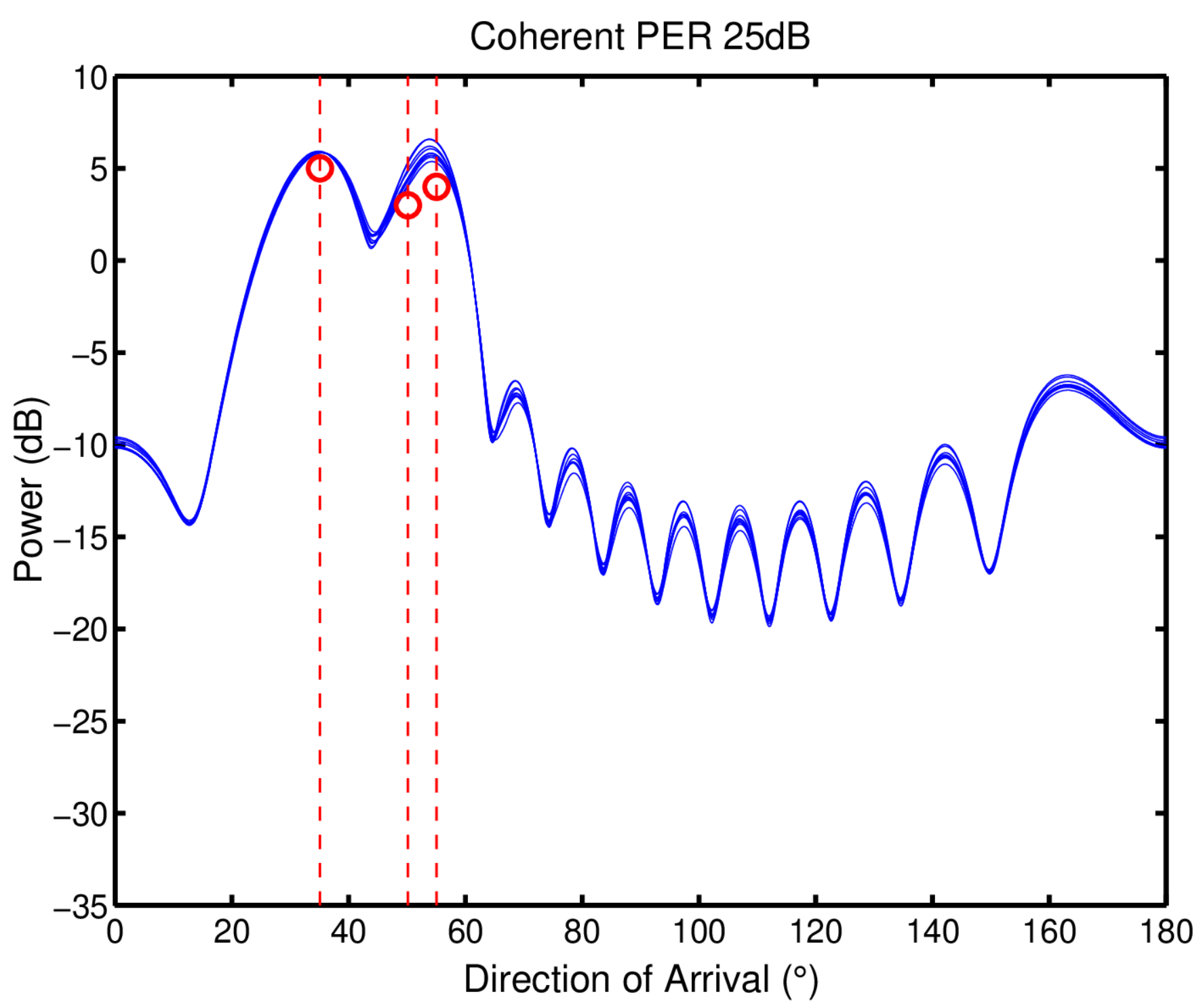}}
&
{\includegraphics[width=2.0in,height=1.5in]{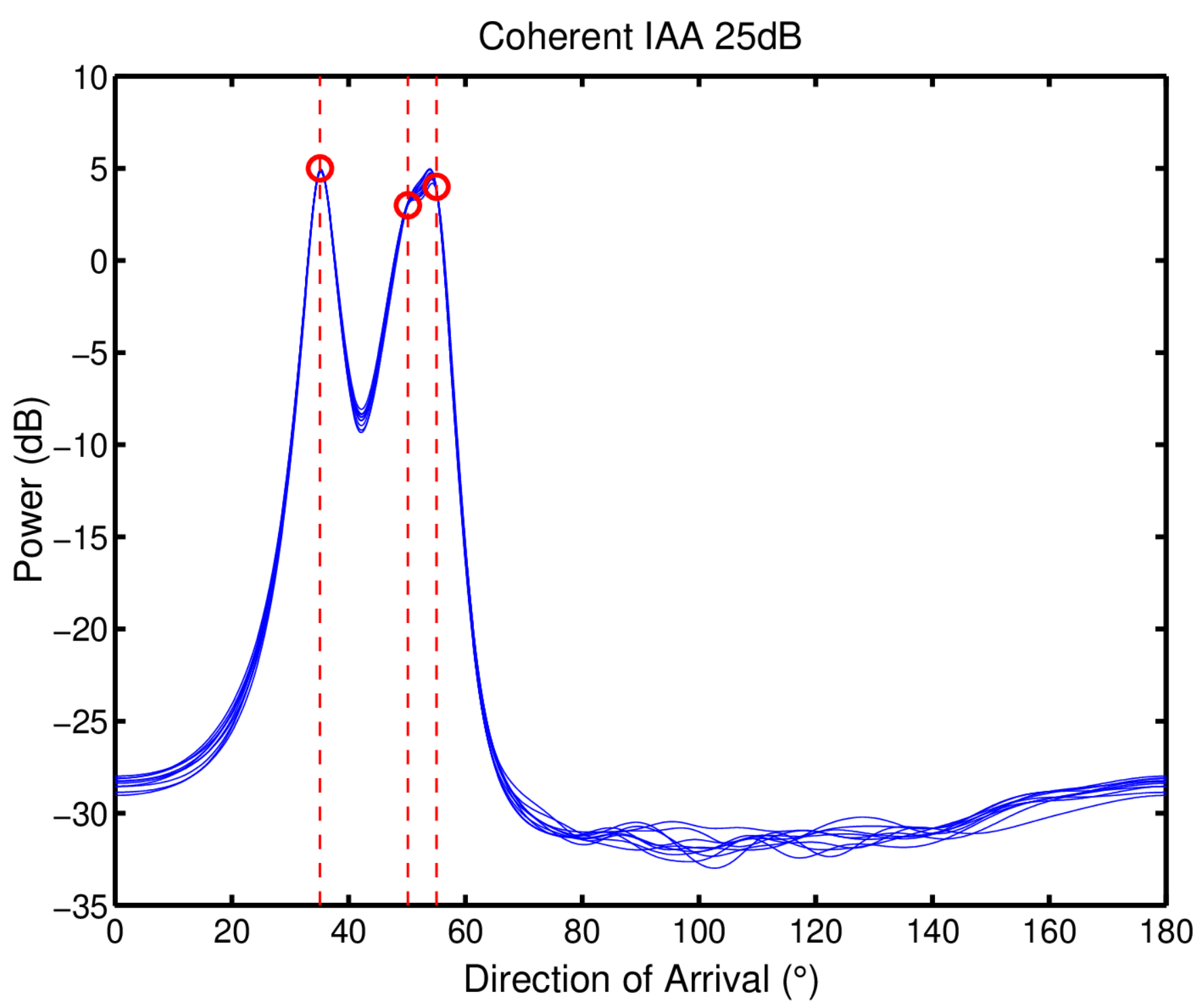}}\\
(a) & (b) \\
{\includegraphics[width=2.0in,height=1.5in]{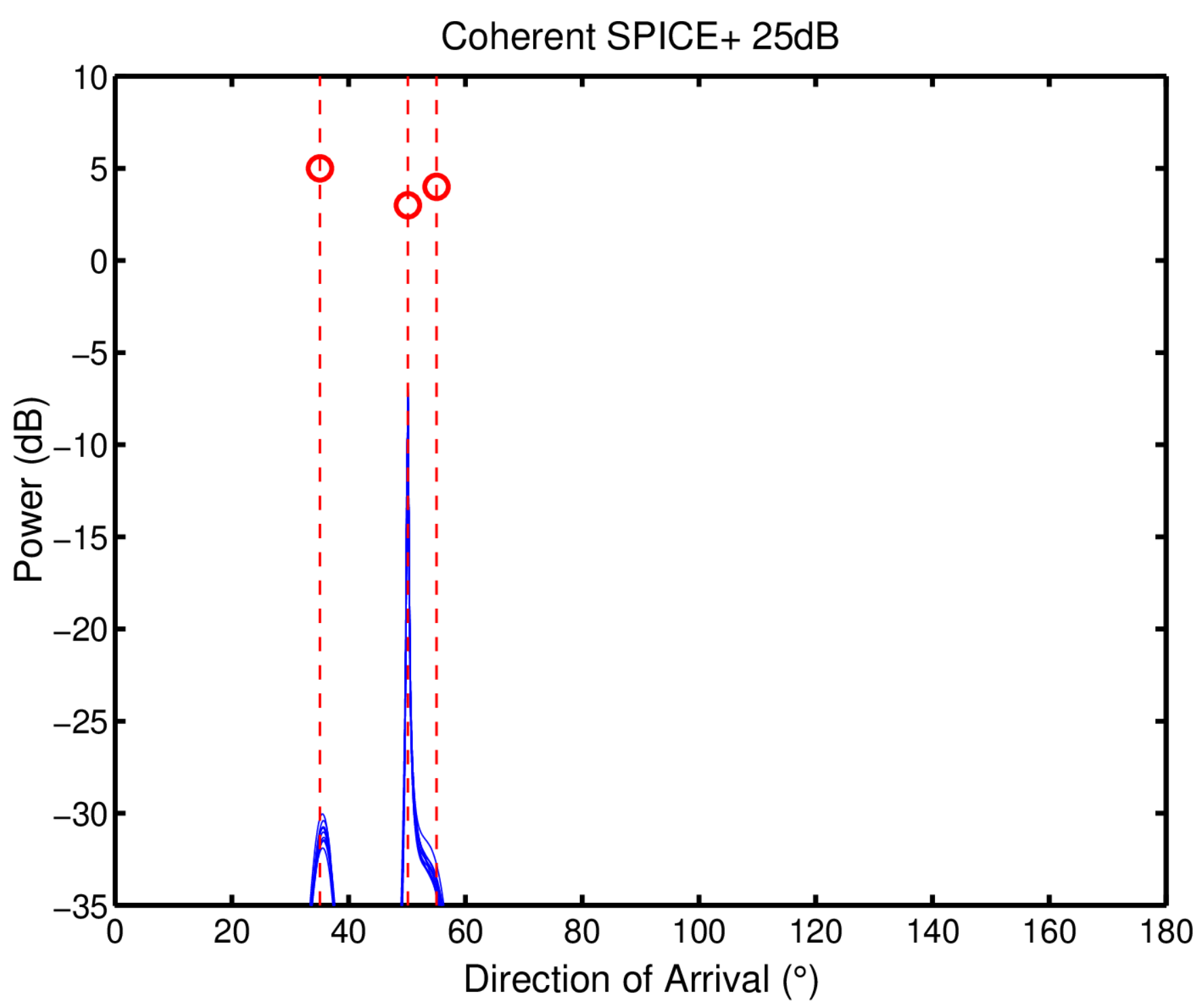}}
&
{\includegraphics[width=2.0in,height=1.5in]{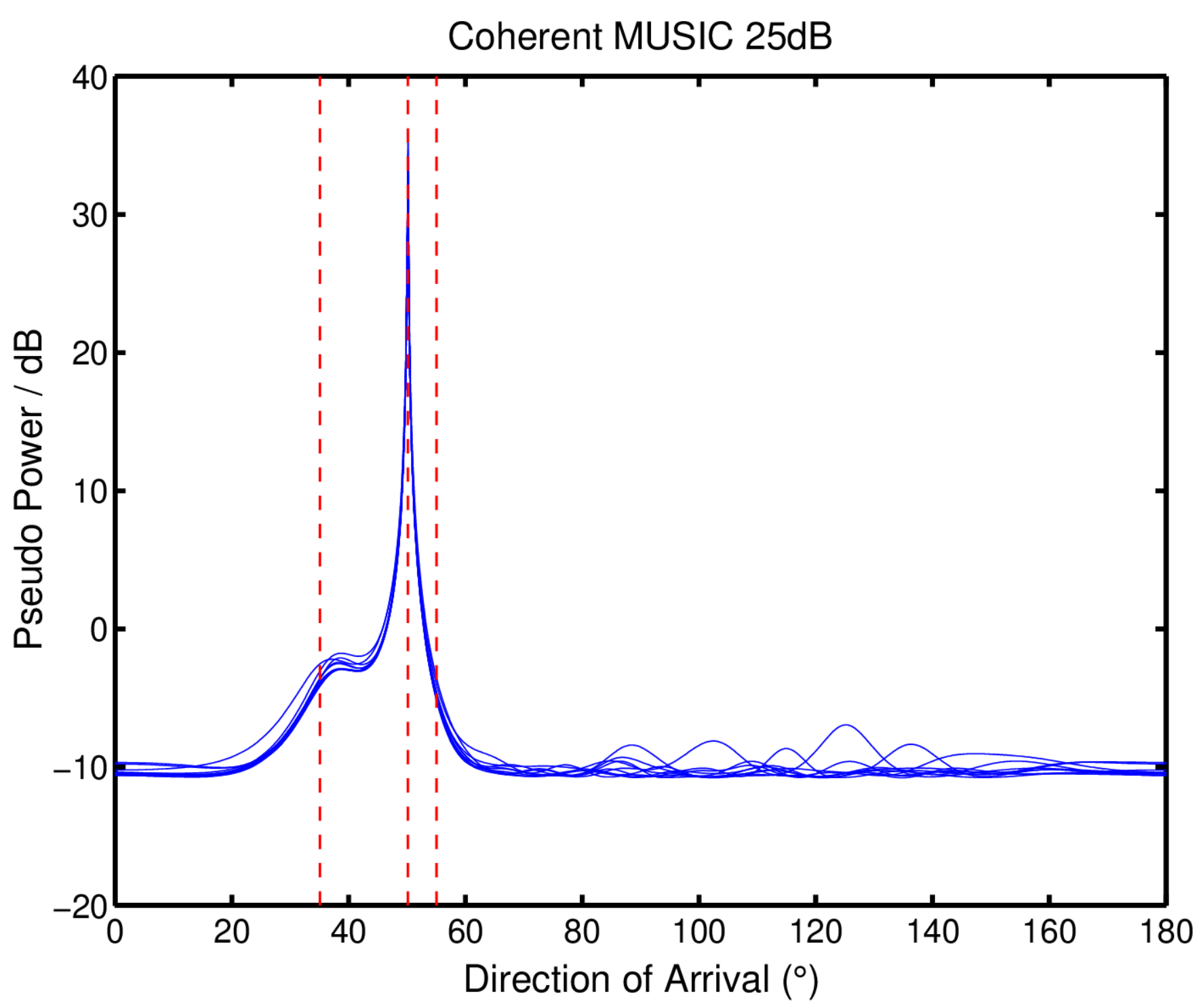}}\\
(c) & (d) \\
\end{tabular}
\begin{tabular}{ccc}
{\includegraphics[width=2.0in,height=1.5in]{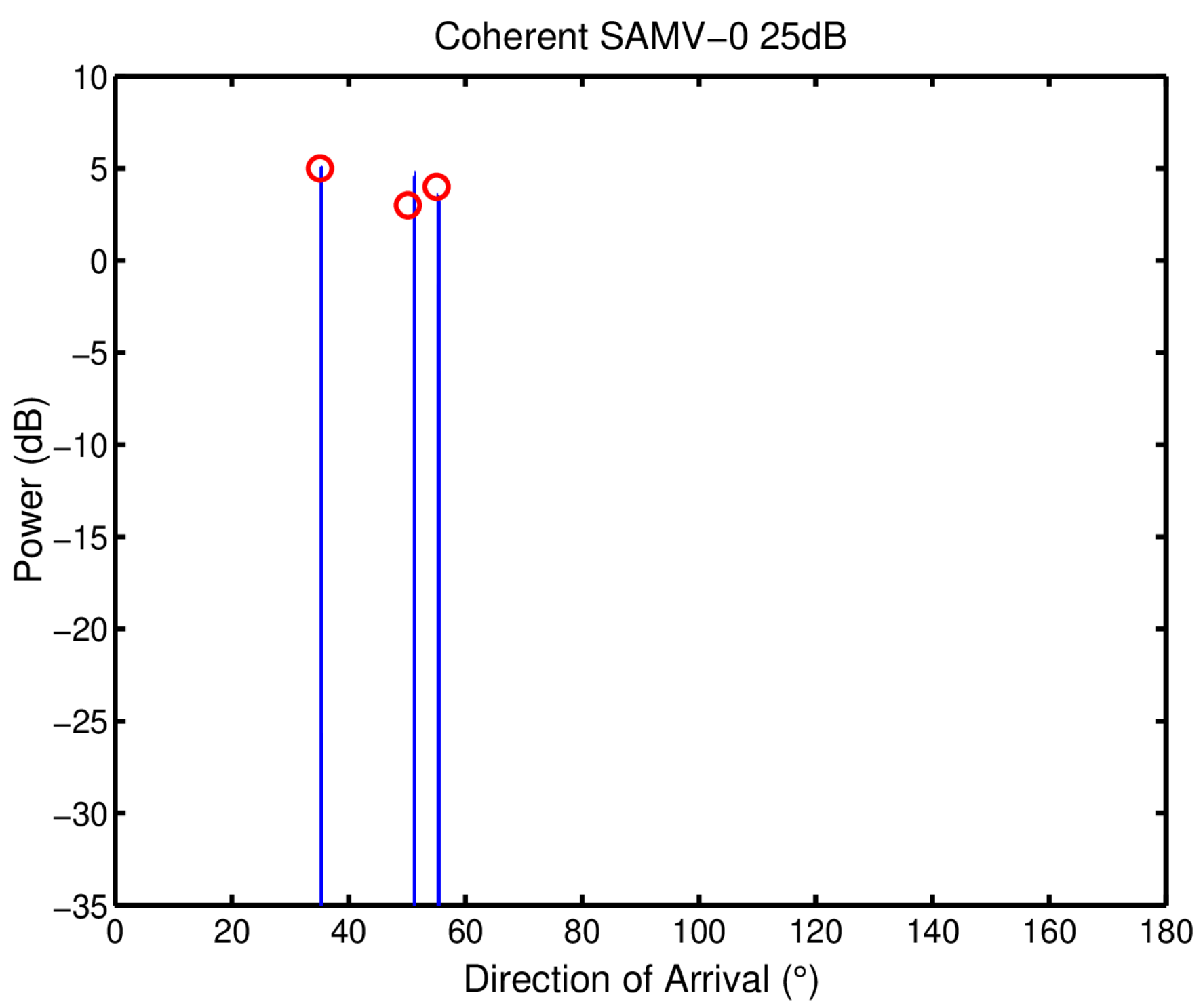}}
&
{\includegraphics[width=2.0in,height=1.5in]{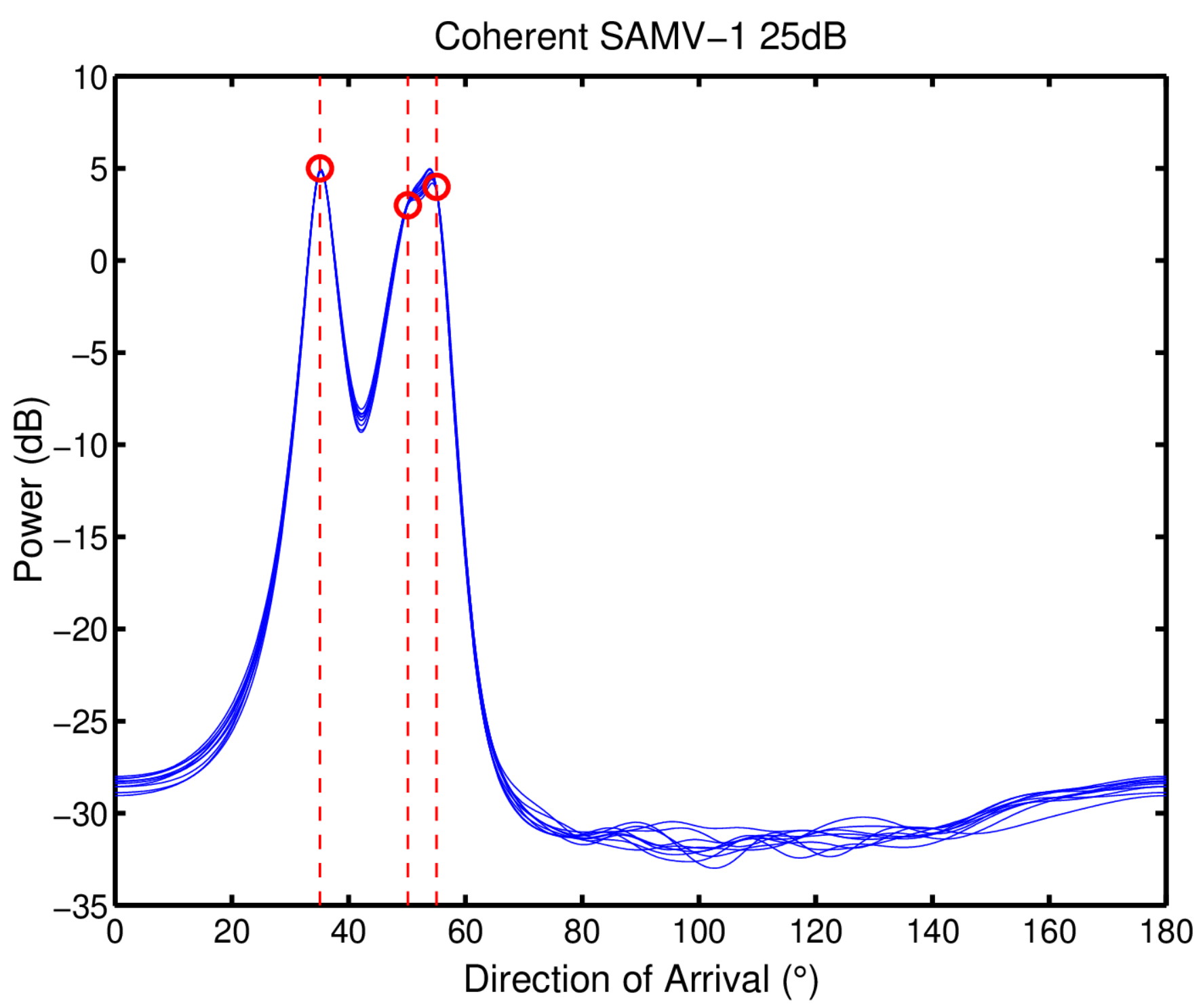}}
&
{\includegraphics[width=2.0in,height=1.5in]{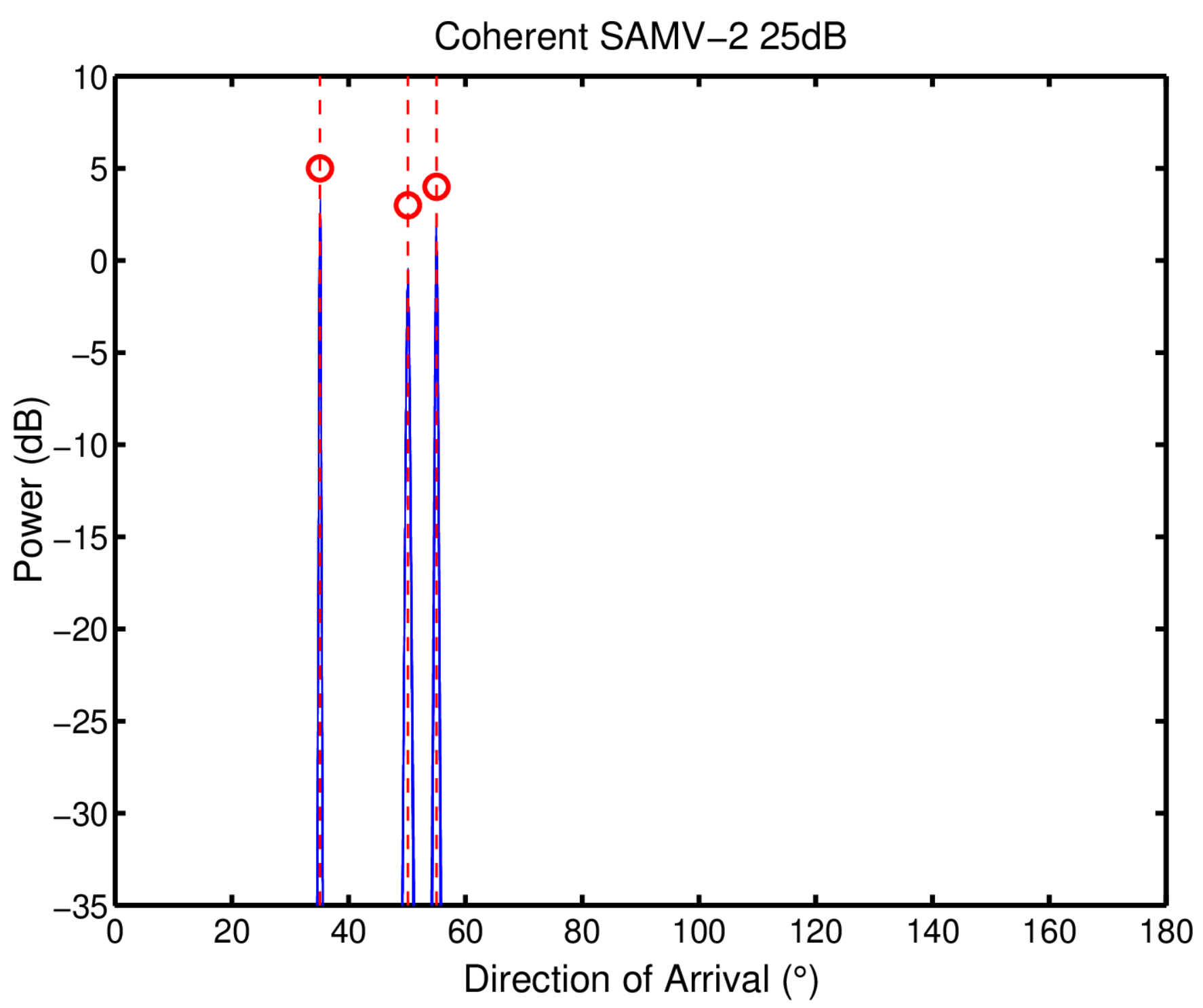}}\\
(e) & (f) & (g) \\
{\includegraphics[width=2.0in,height=1.5in]{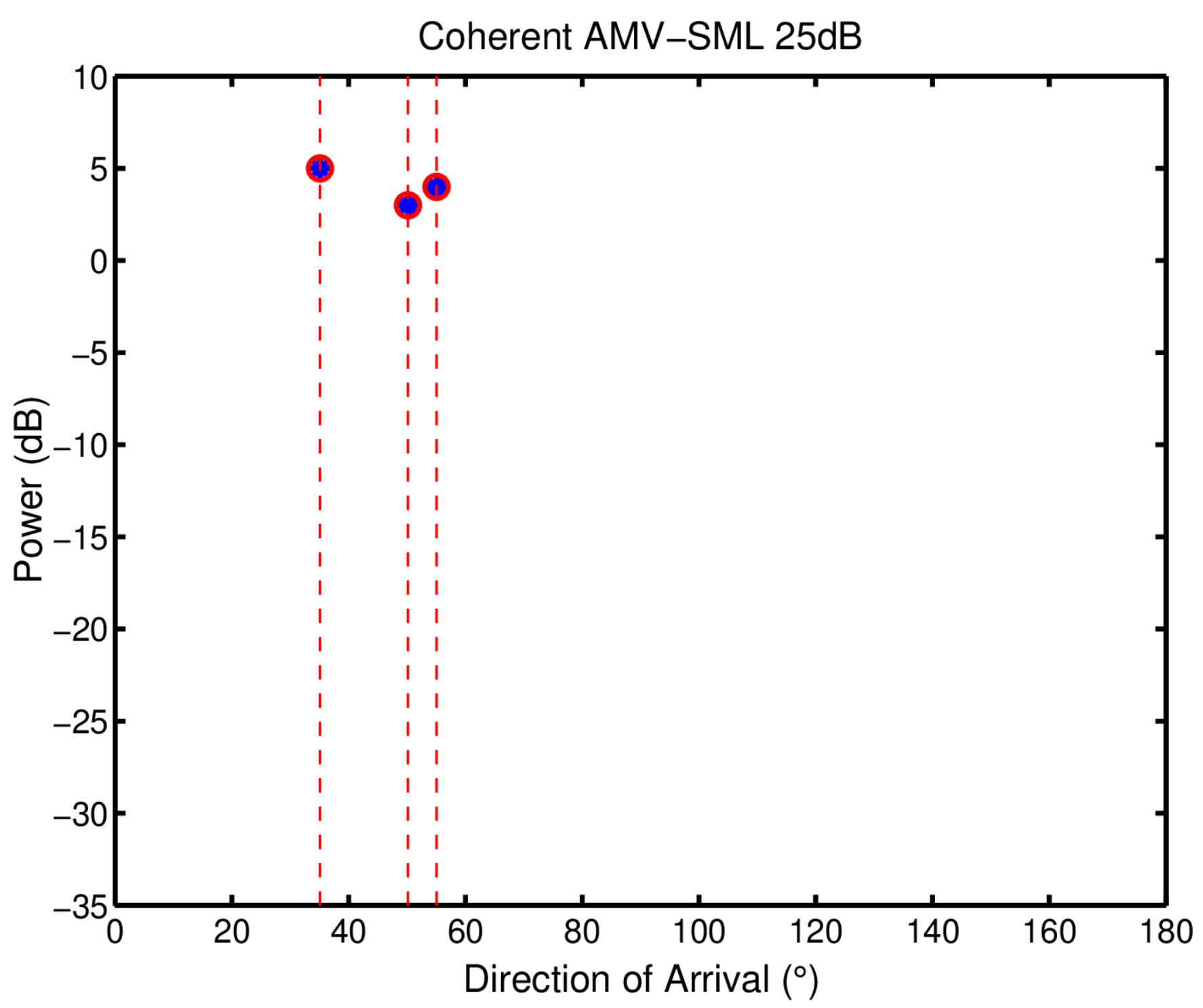}}
&
{\includegraphics[width=2.0in,height=1.5in]{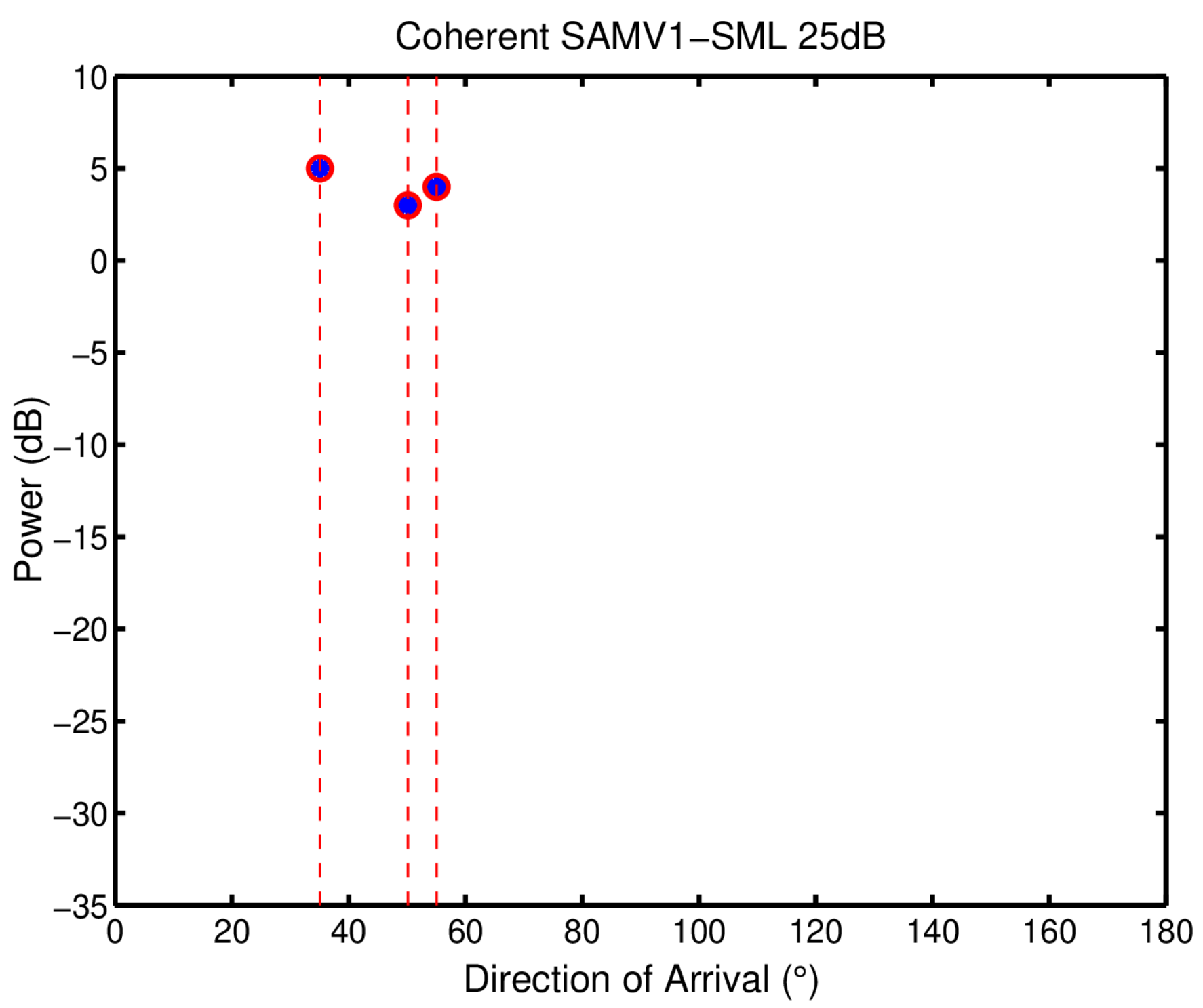}}
&
{\includegraphics[width=2.0in,height=1.5in]{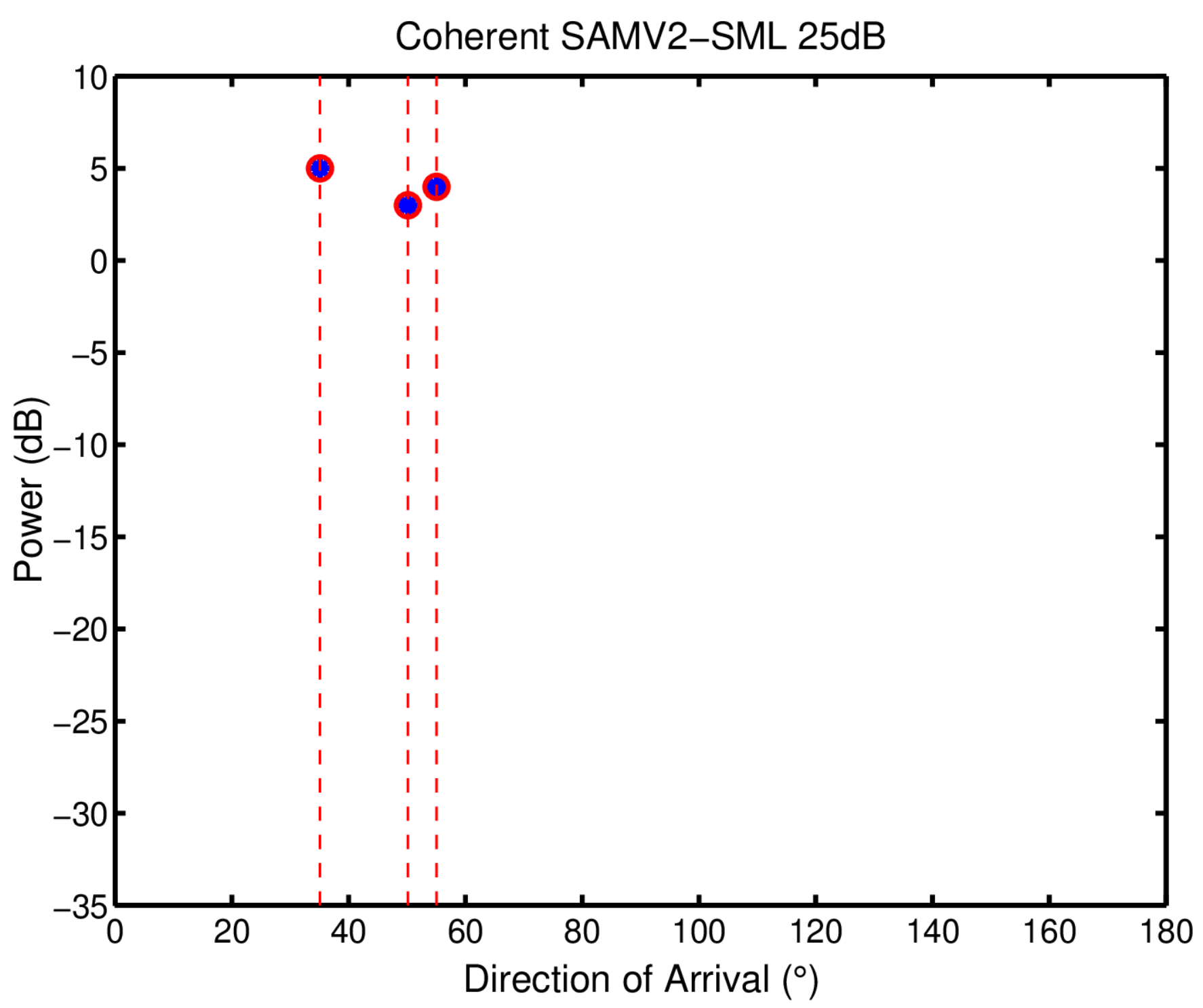}}\\
(h) & (i) & (j) \\
\end{tabular}
\centering \caption{Source localization with a ULA of $M=12$ sensors
and $N = 120$ snapshots, SNR = $25$ dB: Three sources at
$35.11^\circ$, $50.15^\circ$ and $55.05^\circ$, respectively. The
first and the last source are coherent. These sources are
represented by the red circles and vertical dashed lines in each
plot. $10$ Monte Carlo trials are shown in each plot. Spatial
estimates are shown with (a) Periodogram (PER), (b) IAA, (c) SPICE+,
(d) MUSIC, (e) SAMV-0, (f) SAMV-1, (g) SAMV-2, (h) AMV-SML, (i)
SAMV1-SML and (j) SAMV2-SML.} \label{DOAspatial_Cohr}
\end{figure}

First, DOA estimation results using a $12$ element ULA and $N = 120$
snapshots of both independent and coherent sources are given in
Figure \ref{DOAspatial_indp} and Figure \ref{DOAspatial_Cohr},
respectively. Three sources with $5$ dB, $3$ dB and $4$ dB power at
location $\theta_1 = 35.11^\circ$, $\theta_2 = 50.15^\circ$ and
$\theta_3 = 55.05^\circ$ are present in the region-of-interest. For
the coherent sources case in Figure \ref{DOAspatial_Cohr}, the
sources at $\theta_1$ and $\theta_3$ share the same phases but the
source at $\theta_2$ are independent of them. The true source
locations and powers are represented by the circles and vertical
dashed lines that align with these circles. In each plot, the
estimation results of $10$ Monte Carlo trials for each algorithm are
shown together.

Due to the strong smearing effects and limited resolution, the PER
approach fails to correctly separate the close sources at $\theta_2$
and $\theta_3$ (Figure \ref{DOAspatial_indp}(a) and Figure
\ref{DOAspatial_Cohr}(a)). The IAA algorithm has reduced the
smearing effects significantly, resulting lower sidelobe levels in
Figure \ref{DOAspatial_indp}(b) and Figure \ref{DOAspatial_Cohr}(b).
However, the resolution provided by IAA is still not high enough to
separate the two close sources at $\theta_2$ and $\theta_3$.

\begin{figure}[tbp]
\centering
\begin{tabular}{cc}
{\includegraphics[width=3.2in,height=2.4in]{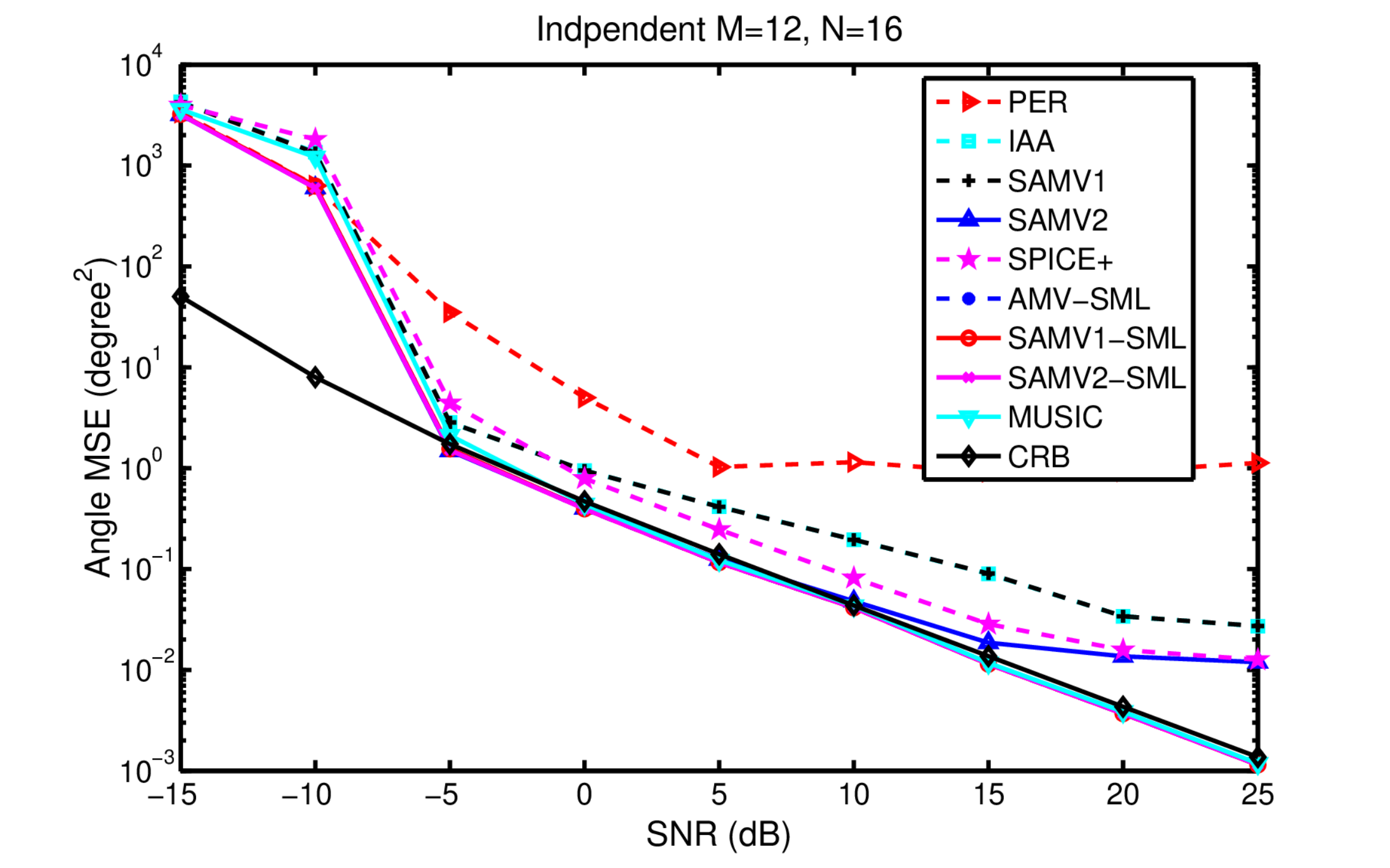}}
&
{\includegraphics[width=3.2in,height=2.4in]{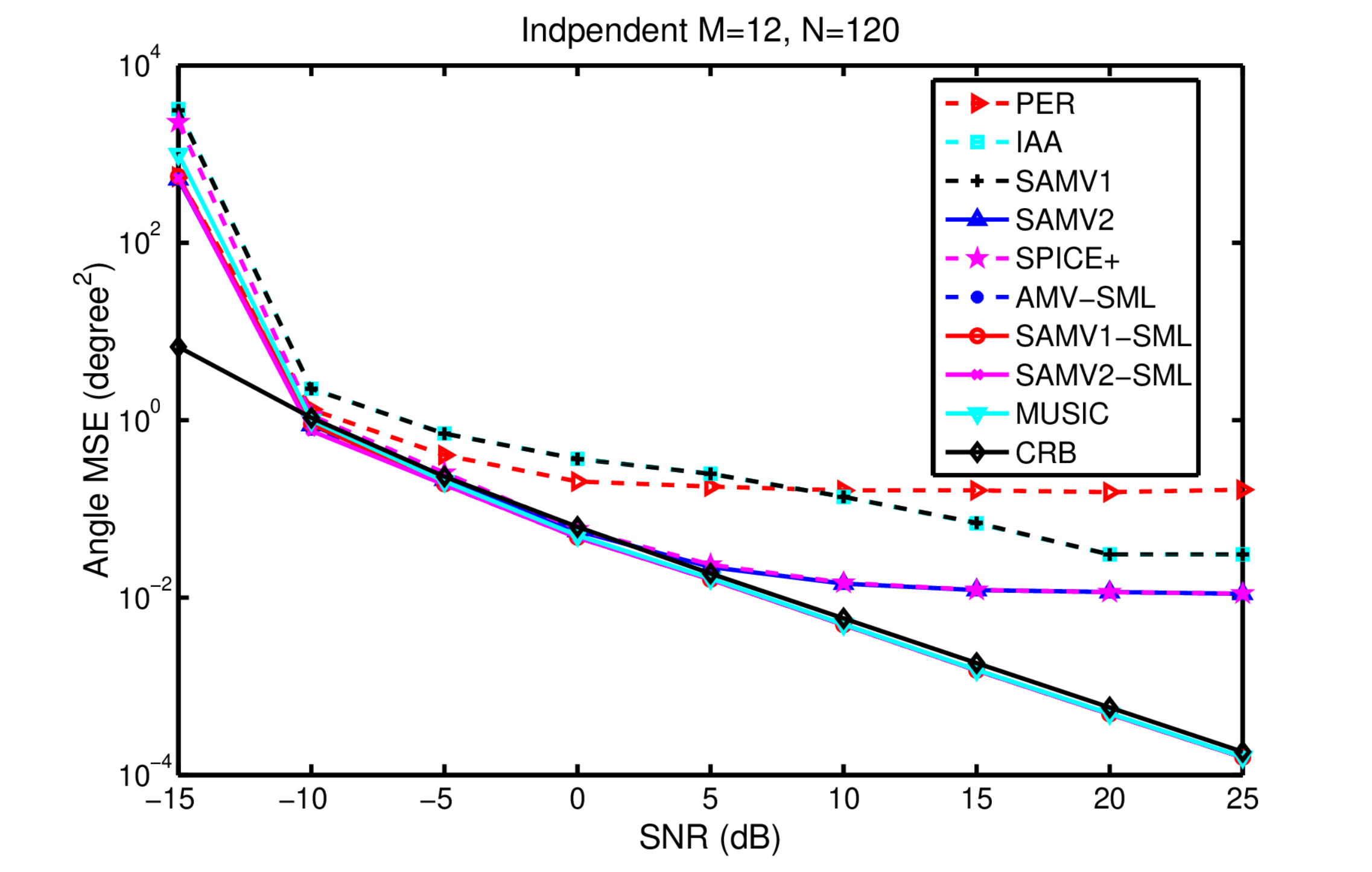}} \\
(a) & (b) \\
\end{tabular}
\centering \caption{Source localization: Two uncorrelated sources at
$35.11^\circ$ and $50.15^\circ$ with a ULA of $M=12$ sensors. (a)
Total angle estimation MSE with $N=16$ snapshots and (b) total angle
estimation MSE with $N=120$ snapshots. } \label{M12IndepFigs}
\end{figure}

\begin{figure}[tbp]
\centering
\begin{tabular}{cc}
{\includegraphics[width=3.2in,height=2.4in]{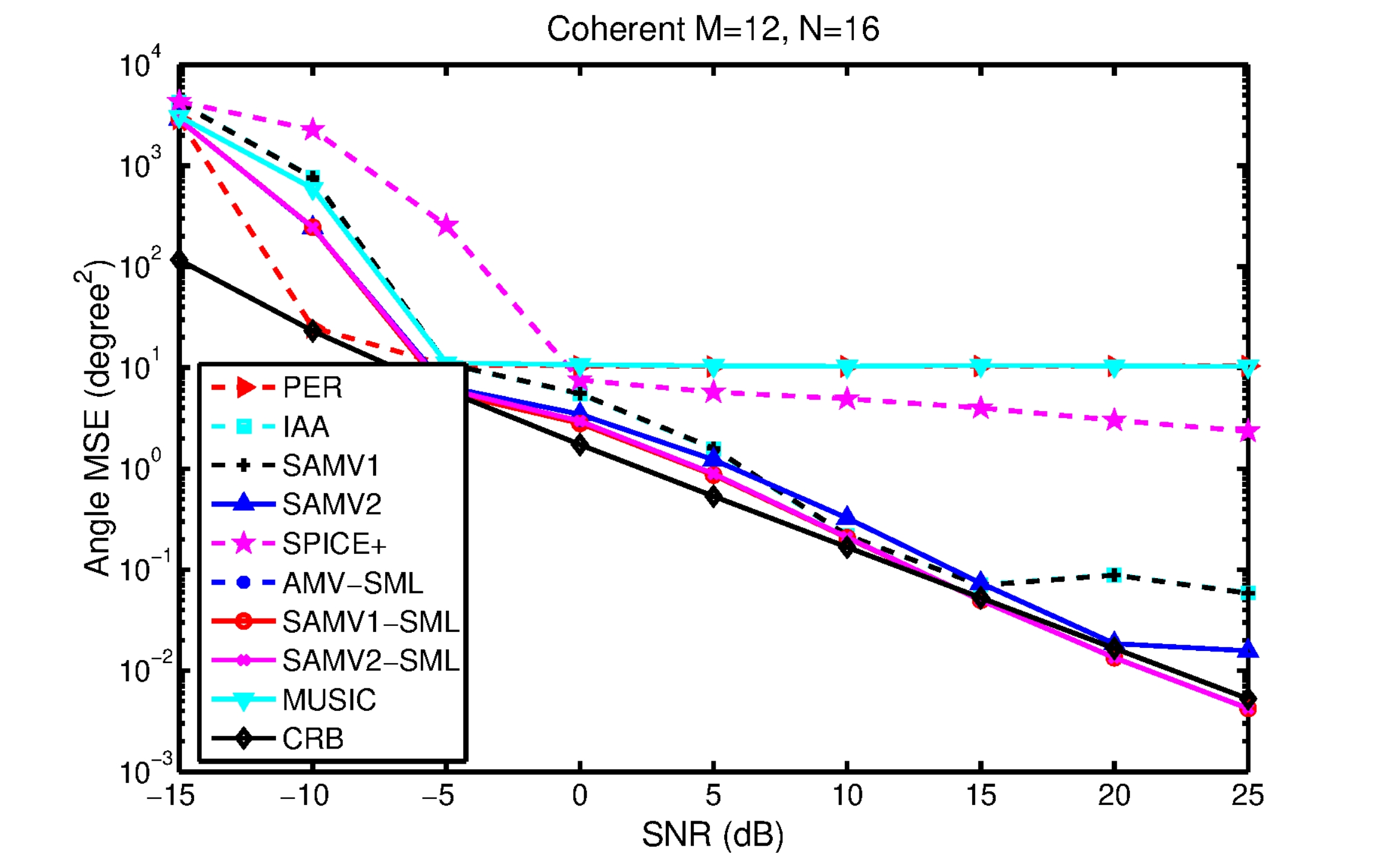}}
&
{\includegraphics[width=3.2in,height=2.4in]{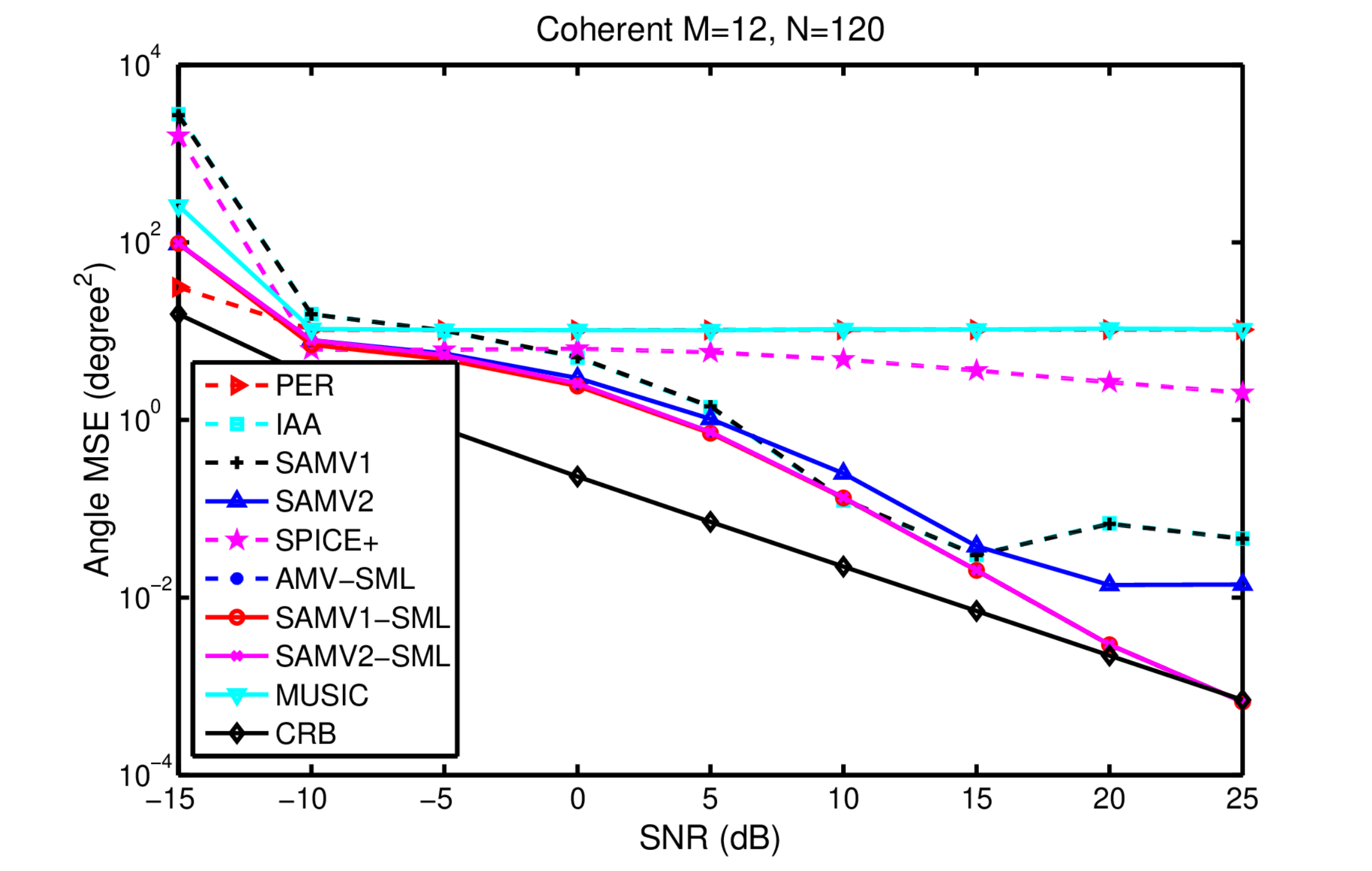}} \\
(a) & (b) \\
\end{tabular}
\centering \caption{Source localization: Two coherent sources at
$35.11^\circ$ and $50.15^\circ$ with a ULA of $M=12$ sensors. (a)
Total angle estimation MSE with $N=16$ snapshots and (b) total angle
estimation MSE with $N=120$ snapshots.} \label{M12CohrFigs}
\end{figure}

In the scenario with independent sources, the eigen-analysis based
MUSIC algorithm and existing sparse methods such as the SPICE+
algorithm, are capable of resolving all three sources in Figure
\ref{DOAspatial_indp}(c)--(d), thanks to their superior resolution.
However, the source coherence degrades their performances
dramatically in Figure \ref{DOAspatial_Cohr}(c)--(d). On contrary,
the proposed SAMV algorithms depicted in Figure
\ref{DOAspatial_Cohr}(e)--(g), are much more robust against signal
coherence. We observe in Figure
\ref{DOAspatial_indp}--\ref{DOAspatial_Cohr} that the SAMV-1
approach generally provides identical spatial estimates to its IAA
counterpart and this phenomenon is again revealed in Figure
\ref{M12IndepFigs}--\ref{M12CohrFigs}, which verifies the comments
in Section \ref{sec:High and low SNR approx}. In Figure
\ref{DOAspatial_indp}--\ref{DOAspatial_Cohr}, the SAMV-0 and SAMV-2
algorithm generate high resolution sparse spatial estimates for both
the independent and coherent sources. However, we notice in our
simulations that the sparsest SAMV-0 algorithm requires a high SNR
to work properly. Therefore, SAMV-0 is not included when comparing
angle estimation mean-square-error over a wide range of SNR in
Figure \ref{M12IndepFigs}--\ref{M12CohrFigs}. From Figure
\ref{DOAspatial_indp}--\ref{DOAspatial_Cohr}, we comment that the
SAMV-SML algorithms (AMV-SML, SAMV1-SML and SAMV2-SML) provide the
most accurate estimates of the source locations and powers
simultaneously.

Next, Figures \ref{M12IndepFigs}--\ref{M12CohrFigs} compare the
total angle mean-square-error (MSE)\footnote{Defined as the
summation of the angle MSE for each source.} of each algorithm with
respect to varying SNR values for both independent and coherent
sources. These DOA localization results are obtained using a $12$
element ULA and $N = 16 \textrm{ or } 120$ snapshots. Two sources
with $5$ dB and $3$ dB power at location $\theta_1 = 35.11^\circ$
and $\theta_2 = 50.15^\circ$ are present\footnote{These DOA true
values are selected so that neither of them is on the direction
grid.}.

While calculating the MSEs for the power-based grid-dependent
algorithms\footnote{Include the IAA, SAMV-1, SAMV-2, SPICE+
algorithms.}, only the highest two peaks in
$\{\hat{p}_{k}\}_{k=1}^{K}$ are selected as the estimates of the
source locations. The grid-independent SAMV-SML algorithms (AMV-SML,
SAMV1-SML and SAMV2-SML) are all initialized by the SAMV-2
algorithm. Each point in Figure
\ref{M12IndepFigs}--\ref{M12CohrFigs} is the average of $1000$ Monte
Carlo trials.

Due to the severe smearing effects (already shown in Figure
\ref{DOAspatial_indp}--\ref{DOAspatial_Cohr}), the PER approach
gives high total angle MSEs in Figure
\ref{M12IndepFigs}--\ref{M12CohrFigs}. Figure \ref{M12IndepFigs}(b)
shows that the SPICE+ algorithm has favorable angle estimation
variance characteristics for independent sources, especially with
sufficient snapshots. However, the source coherence degrades the
SPICE+ performance dramatically in Figure \ref{M12CohrFigs}. On
contrary, the SAMV-2 approach offers lower total angle estimation
MSE, especially for the coherent sources case, and this is also the
main reason why we initialize the SAMV-SML approaches with the
SAMV-2 result. Note that in Figure \ref{M12IndepFigs}, the IAA,
SAMV-1 and SAMV-2 provide similar MSEs at very low SNR, which has
already been investigated in Section \ref{sec:High and low SNR
approx}. The zero-order low SNR approximation shows that the SAMV-1
and SAMV-2 estimates are equivalent to the PER result or a scaled
version of it.

We also observe that there exist the plateau effects for the
power-based grid-dependent algorithms (IAA, SAMV-1, SAMV-2, SPICE+)
in Figure \ref{M12IndepFigs}--\ref{M12CohrFigs} when the SNR are
sufficiently high. These phenomena reflect the resolution limitation
imposed by the direction grid detailed in Section \ref{sec:SAMV-ML
approach}. Since the power-based grid-dependent algorithms estimate
each source location $\theta_{\textrm{source}}$ by selecting one
element from a fixed set of discrete values (i.e., the direction
grid values, $\{\theta_k \}_{k=1}^{K}$), there always exists an
estimation bias provided that the sources are not located precisely
on the direction grid. Theoretically, this bias can be reduced if
the adjacent distance between the grid is reduced. However, a
uniformly fine direction grid with large $K$ values incurs
prohibitive computational costs and is not applicable for practical
applications. In lieu of increasing the value of $K$, some adaptive
grid refinement postprocessing techniques have been developed (e.g.,
\cite{Malioutov}) by refining this grid locally\footnote{This
refinement postprocessing also introduces extra user parameters in
\cite{Malioutov}. }. To combat the resolution limitation without
relying on additional grid refinement postprocessing, the SAMV-SML
approaches\footnote{Include the AMV-SML, SAMV1-SML and SAMV2-SML
approaches.} employ a grid-independent one-dimensional minimization
scheme, and the resulted angle estimation MSEs are significantly
reduced at high SNR compared to the SAMV approaches in Figure
\ref{M12IndepFigs}--\ref{M12CohrFigs}. We also note that the MSE
performances of the SAMV1-SML and SAMV2-SML approaches are identical
to their AMV-SML counterpart, which verifies that the SAMV signal
powers and noise variance updating formulas (Eq.
\ref{eq:SAMV-1}--\ref{eq:SAMV-3}) are good approximations to the ML
estimates (Eq. \ref{eq:pestimate}--\ref{eq:noisepower}). In the
independent sources scenario in Figure \ref{M12IndepFigs}, the MSE
curves of the SAMV-SML approaches agree well with the stochastic
Cram\'{e}r-Rao lower bound (CRB, see, e.g., \cite{Stoica90}) over
most of the indicated range of SNR. Even with coherent sources,
these SAMV-SML approaches are still asymptotically efficient and
they provide lower angle estimation MSEs than competing algorithms
over a wide range of SNR.

\subsection{Active Sensing: Range-Doppler Imaging Examples}
\begin{figure}[tbp]
\centering
\begin{tabular}{cc}
{\includegraphics[width=3.2in,height=2.4in]{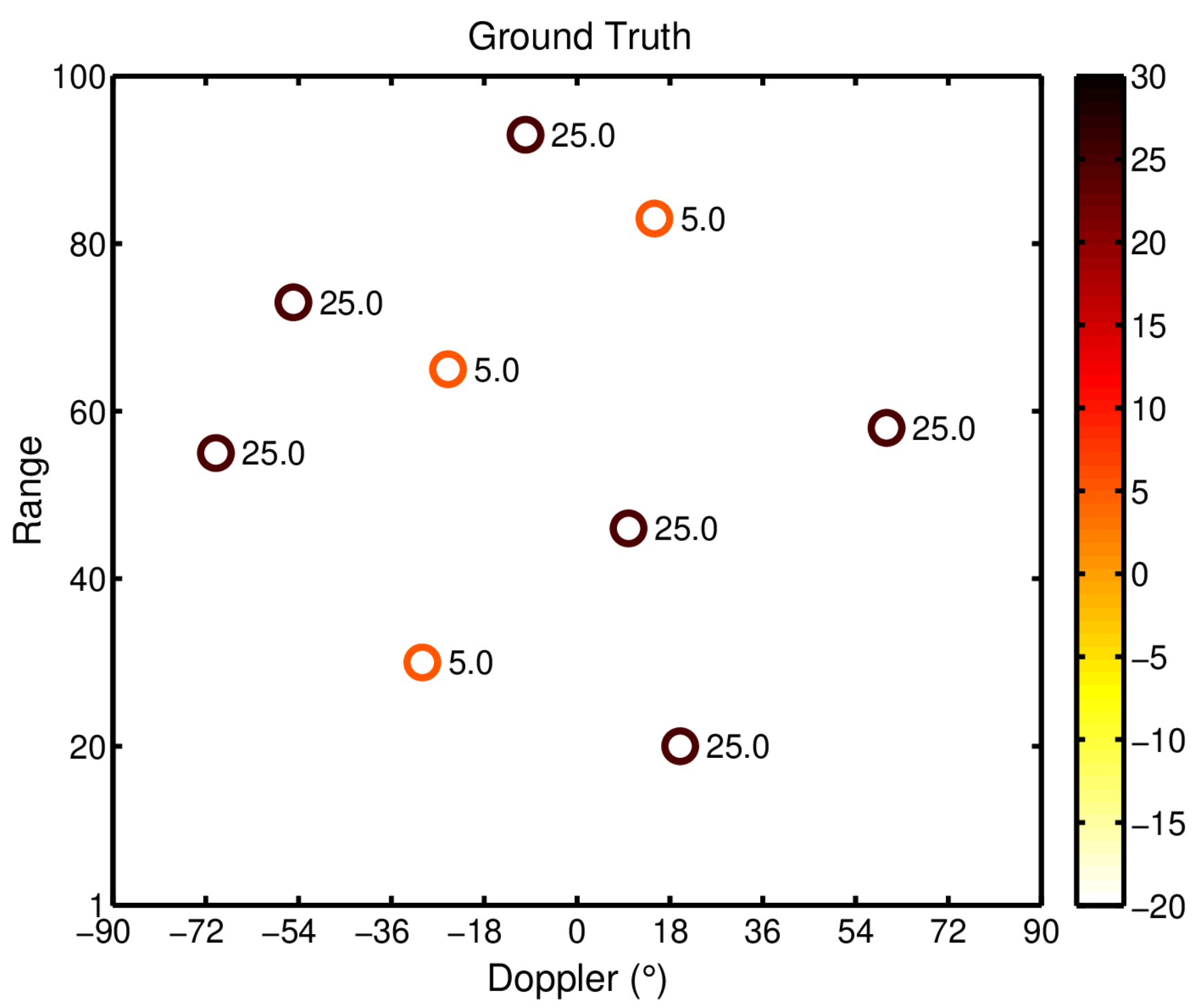}}
&
{\includegraphics[width=3.2in,height=2.4in]{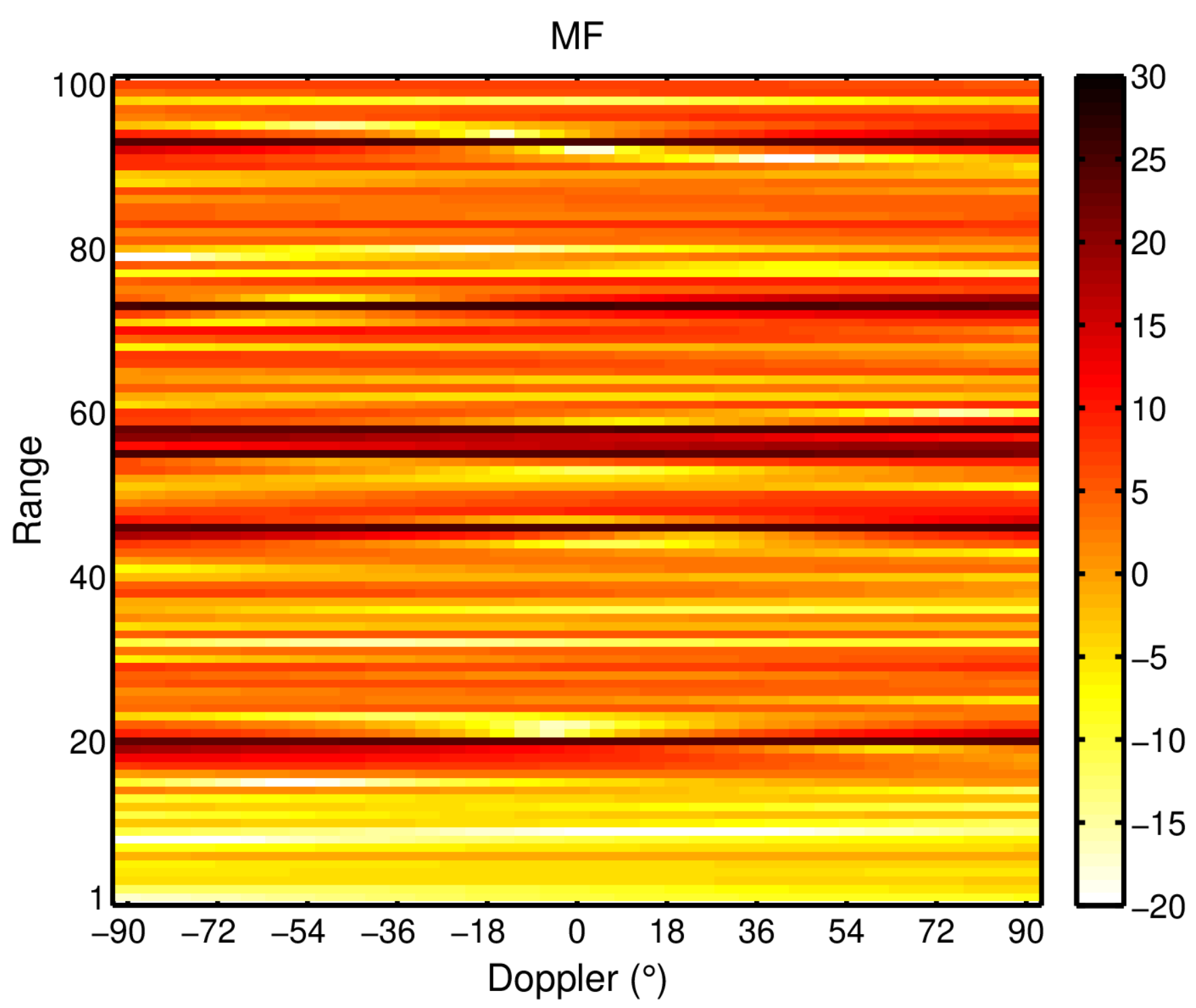}} \\
(a) & (b) \\
{\includegraphics[width=3.2in,height=2.4in]{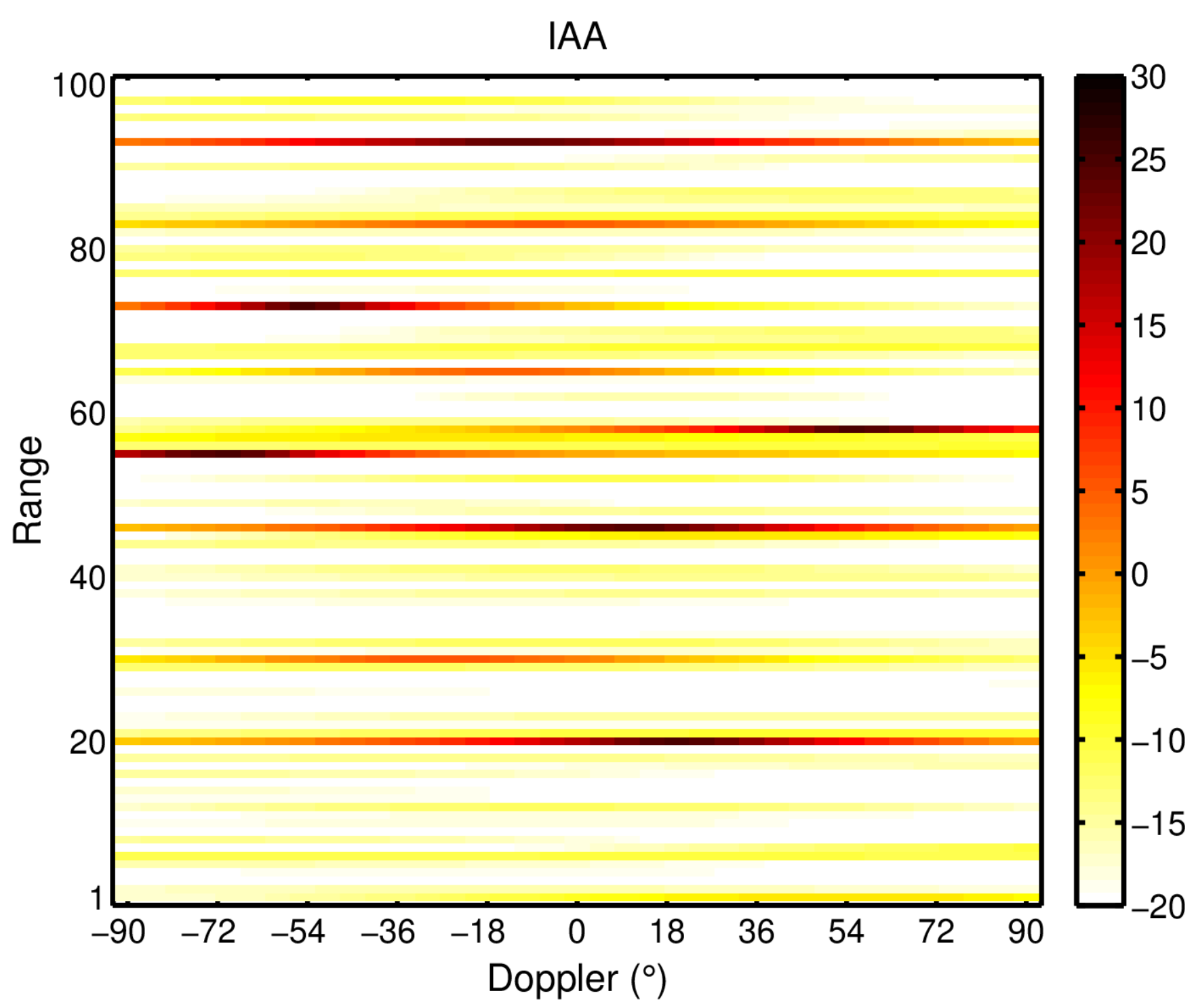}}
&
{\includegraphics[width=3.2in,height=2.4in]{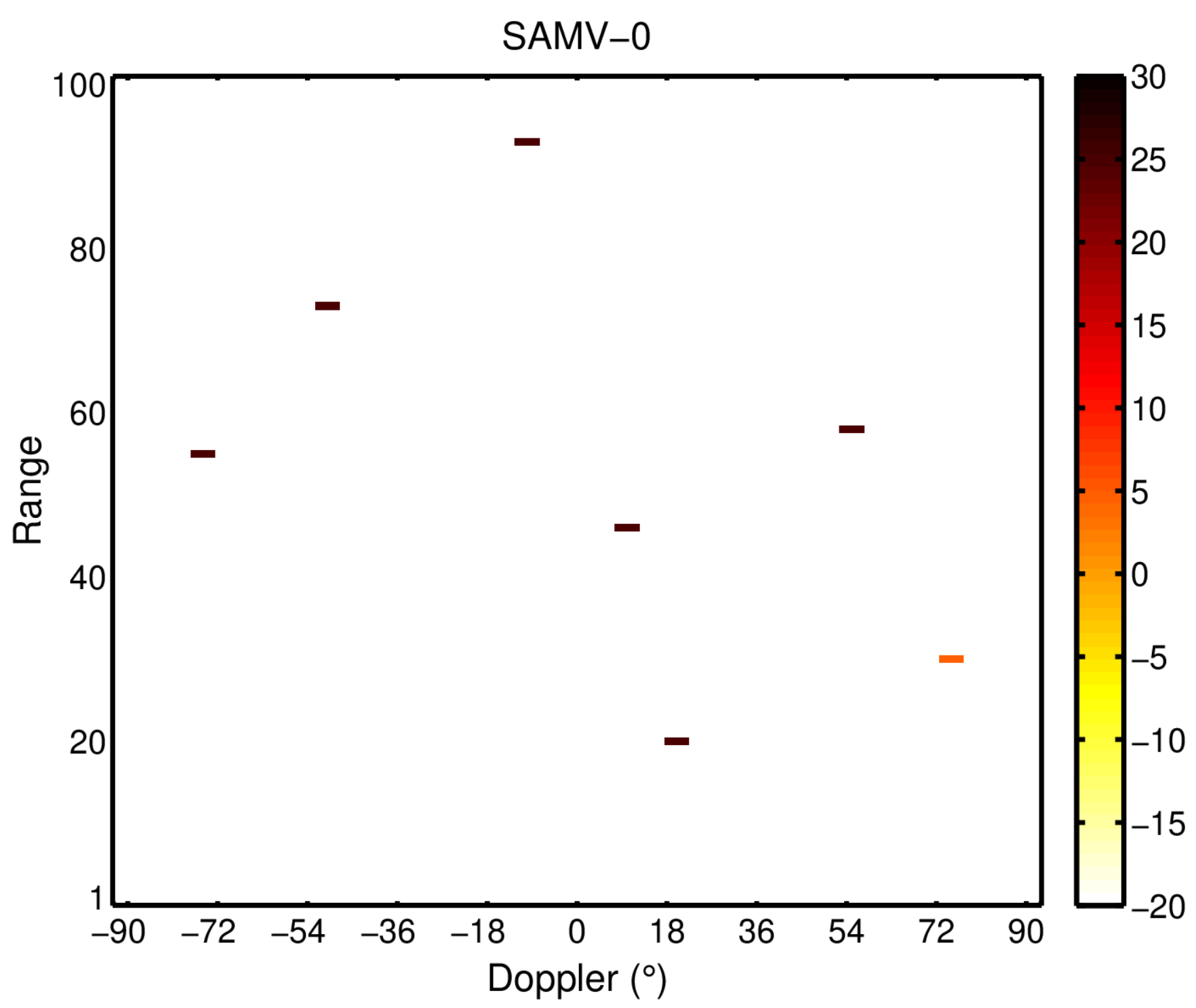}} \\
(c) & (d) \\
{\includegraphics[width=3.2in,height=2.4in]{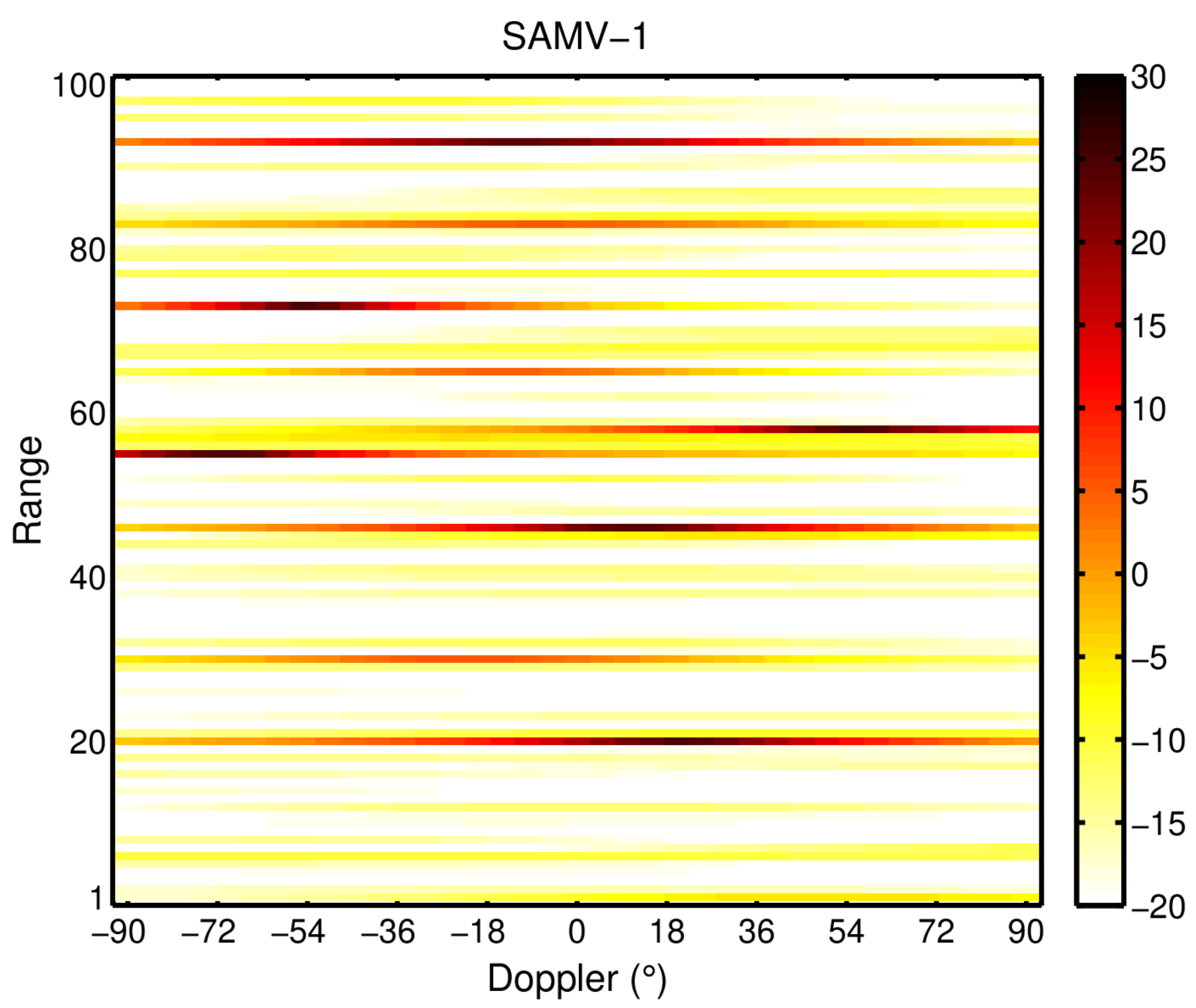}}
&
{\includegraphics[width=3.2in,height=2.4in]{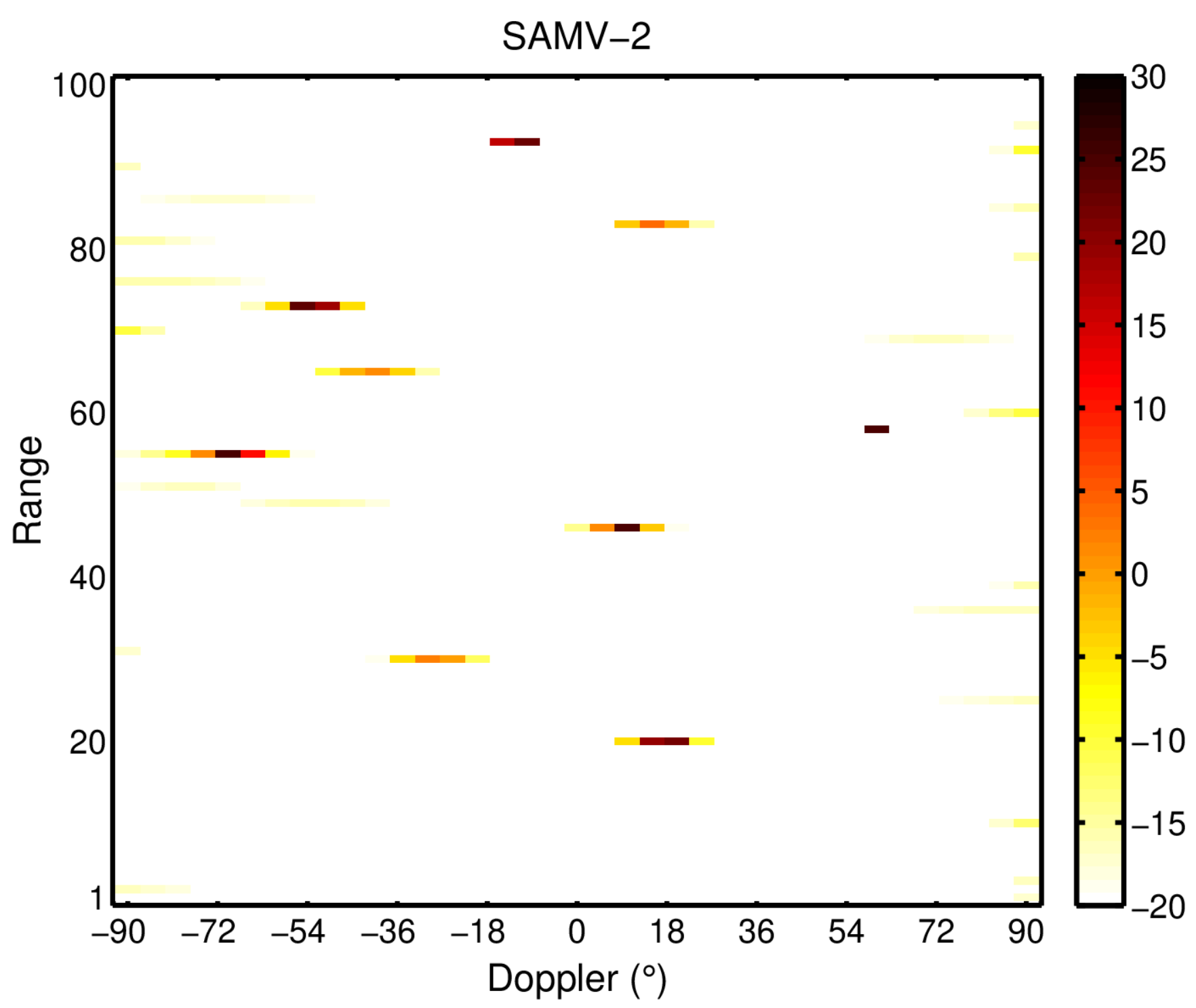}} \\
(e) & (f) \\
\end{tabular}
\centering \caption{SISO range-Doppler imaging with three $5$ dB and
six $25$ dB targets. (a) Ground Truth with power levels, (b) Matched
Filter (MF), (c) IAA, (d) SAMV-0, (e) SAMV-1 and (f) SAMV-2. Power
levels are all in dB. } \label{SISO_RDI}
\end{figure}

This subsection focuses on numerical examples for the SISO
radar/sonar Range-Doppler imaging problem. Since this imaging
problem is essentially a single-snapshot application, only
algorithms that work with single snapshot are included in this
comparison, namely, Matched Filter (MF, another alias of the
periodogram approach), IAA, SAMV-0, SAMV-1 and SAMV-2. First, we
follow the same simulation conditions as in \cite{Yardibi}. A
$30$-element P3 code is employed as the transmitted pulse, and a
total of nine moving targets are simulated. Of all the moving
targets, three are of $5$ dB power and the rest six are of $25$ dB
power, as depicted in Figure \ref{SISO_RDI}(a). The received signals
are assumed to be contaminated with uniform white Gaussian noise of
$0$ dB power. Figure \ref{SISO_RDI} shows the comparison of the
imaging results produced by the aforementioned algorithms.

The Matched Filter (MF) result in Figure \ref{SISO_RDI}(b) suffers
from severe smearing and leakage effects both in the Doppler and
range domain, hence it is impossible to distinguish the $5$ dB
targets. On contrary, the IAA algorithm in Figure \ref{SISO_RDI}(c)
and SAMV-1 in Figure \ref{SISO_RDI}(e) offer similar and greatly
enhanced imaging results with observable target range estimates and
Doppler frequencie. The SAMV-0 approach provides highly sparse
result and eliminates the smearing effects completely, but it misses
the weak $5$ dB targets in Figure \ref{SISO_RDI}(d), which agree
well with our previous comment on its sensitivity to SNR. In Figure
\ref{SISO_RDI}(f), the smearing effects (especially in the Doppler
domain) are further attenuated by SAMV-2, compared with the
IAA/SAMV-1 results. We comment that among all the competing
algorithms, the SAMV-2 approach provides the best balanced result,
providing sufficiently sparse images without missing weak targets.

\begin{figure}[tbp]
\centering
\begin{tabular}{cc}
{\includegraphics[width=3.2in,height=2.4in]{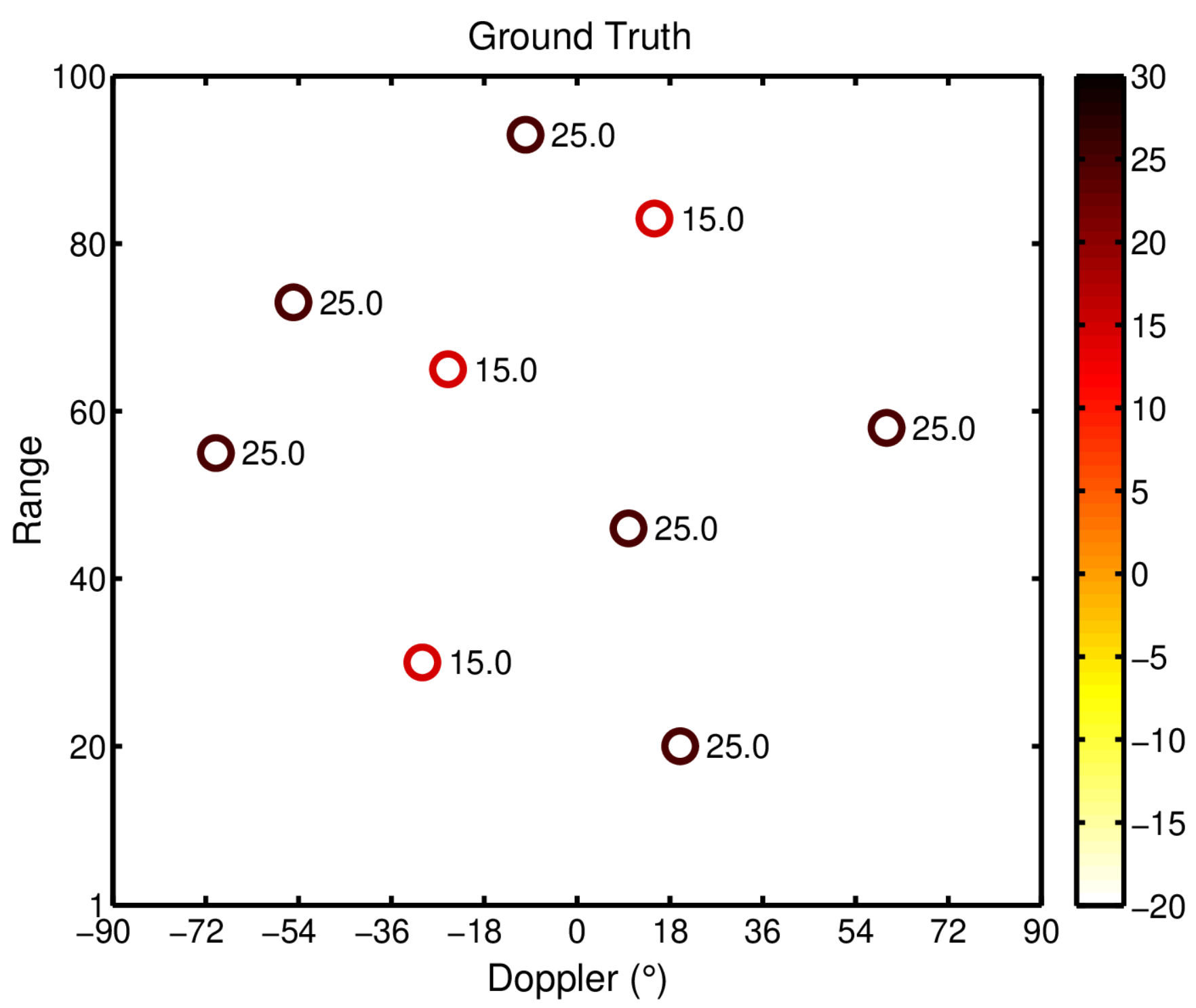}}
&
{\includegraphics[width=3.2in,height=2.4in]{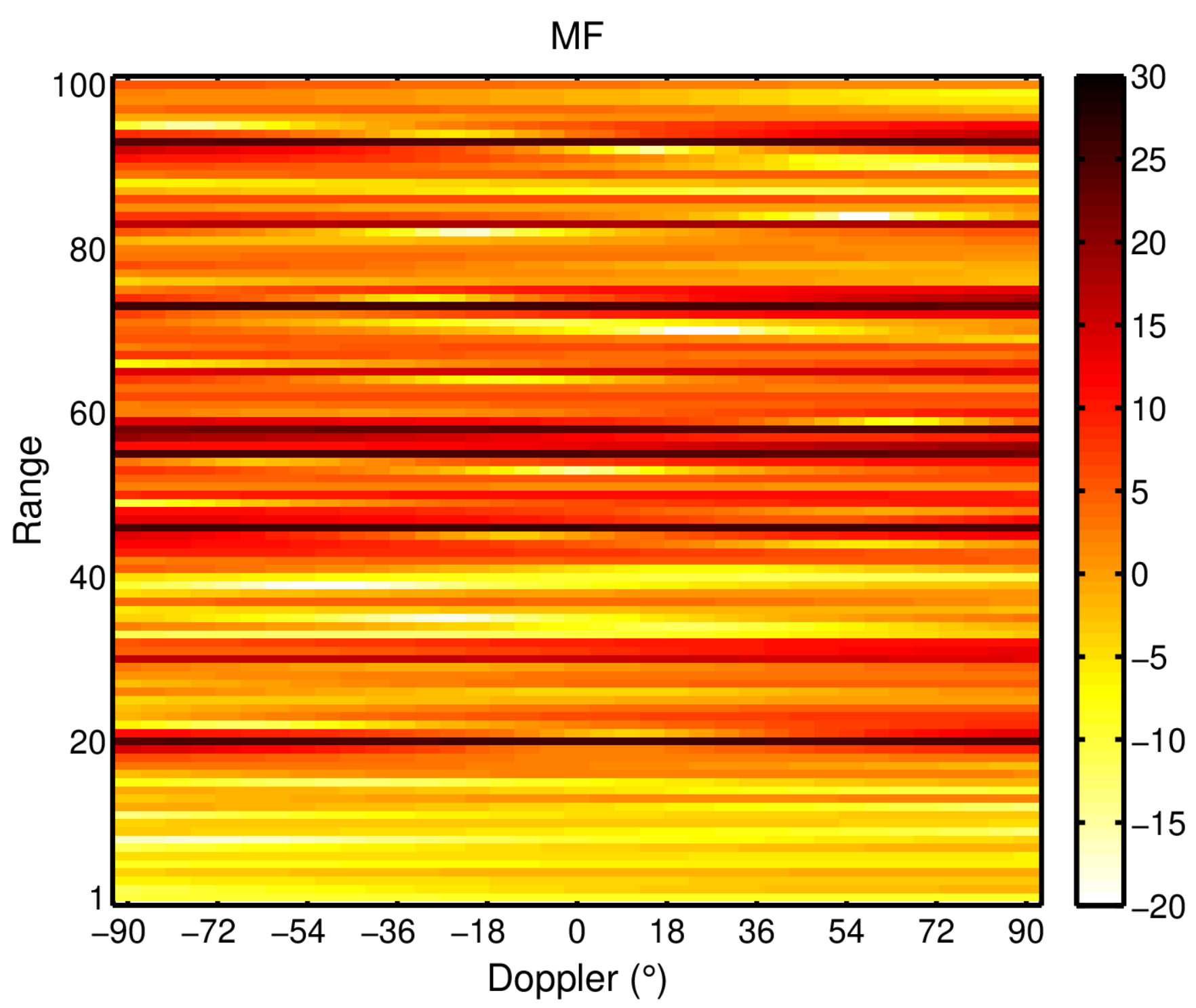}} \\
(a) & (b) \\
{\includegraphics[width=3.2in,height=2.4in]{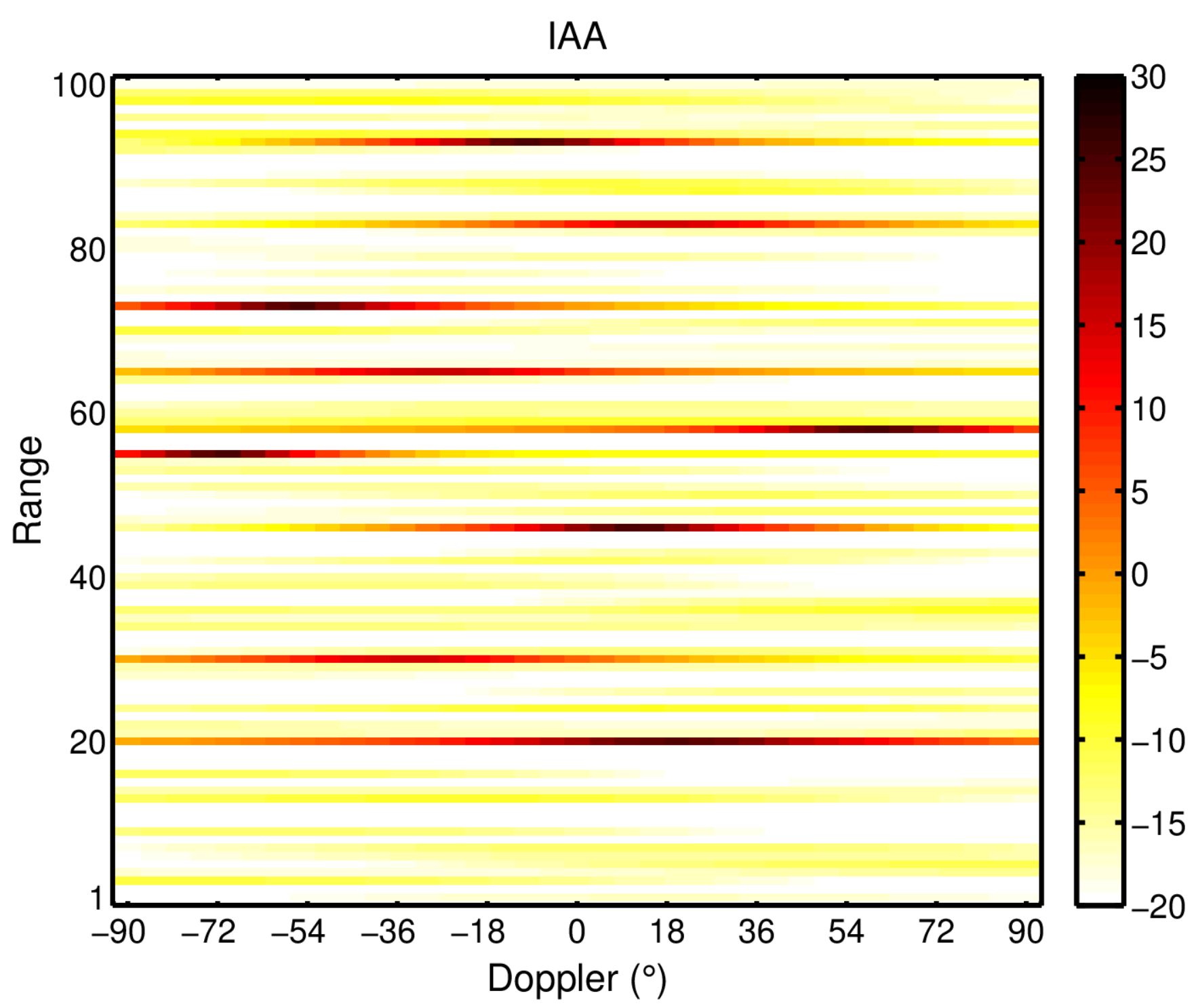}}
&
{\includegraphics[width=3.2in,height=2.4in]{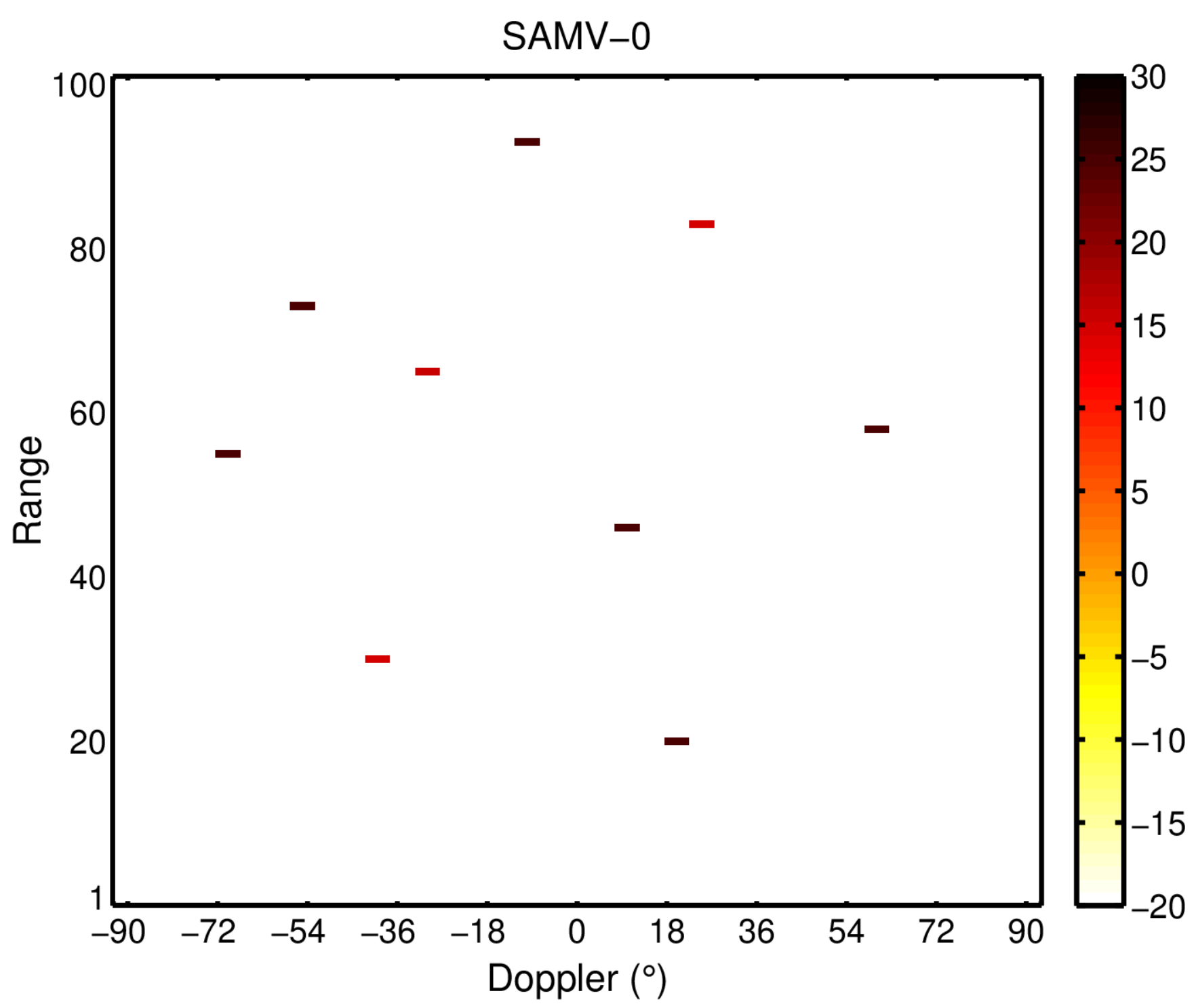}} \\
(c) & (d) \\
{\includegraphics[width=3.2in,height=2.4in]{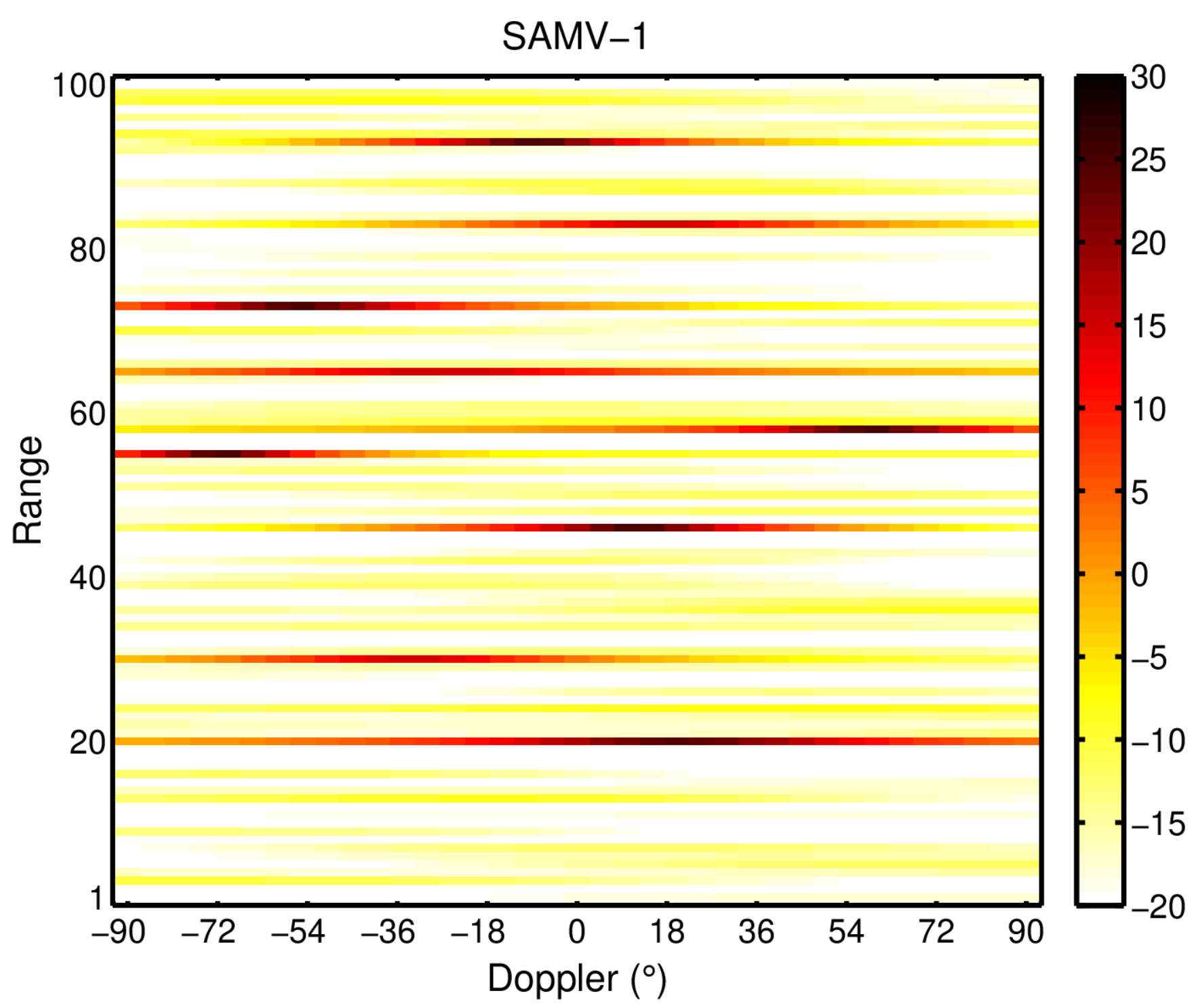}}
&
{\includegraphics[width=3.2in,height=2.4in]{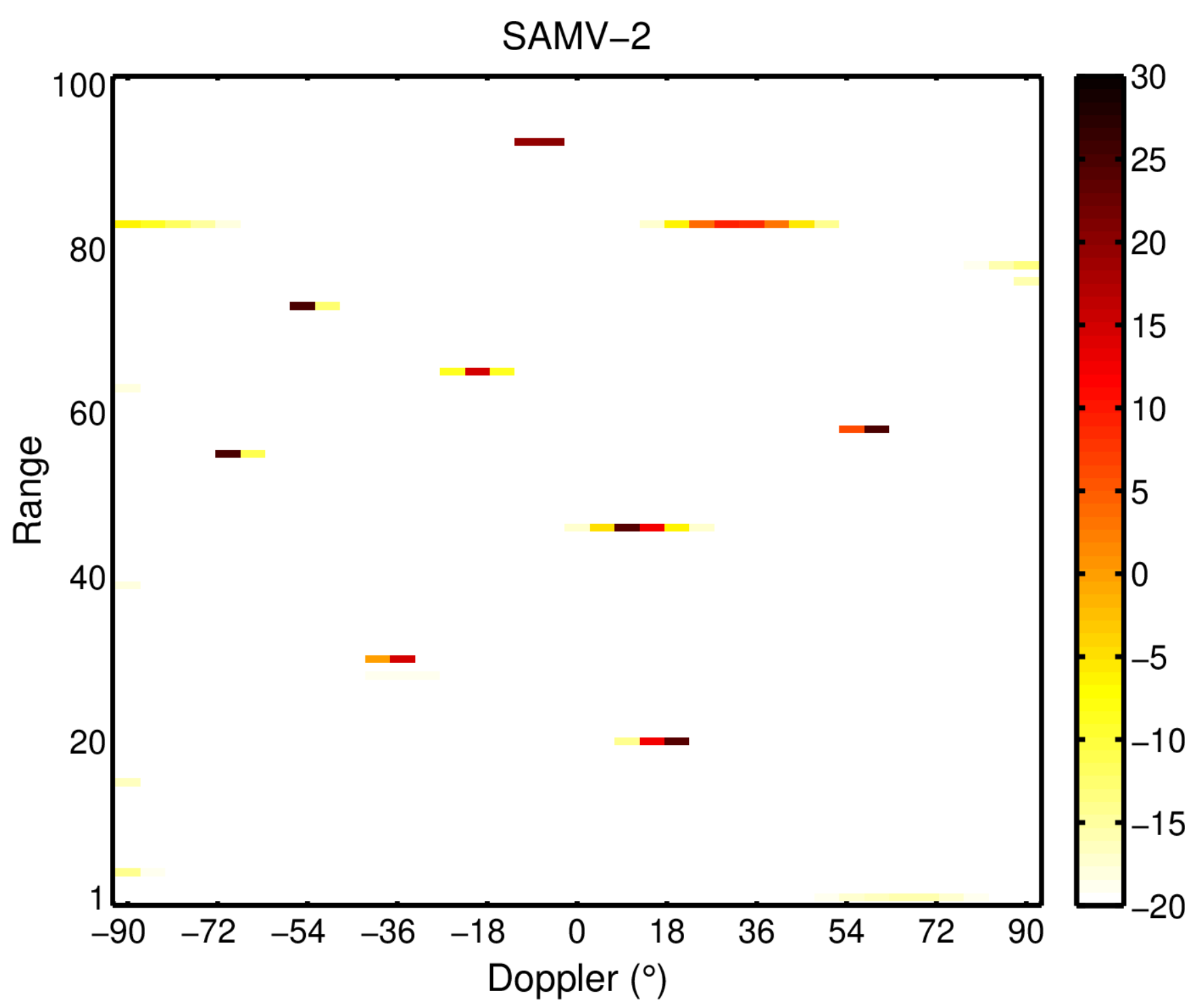}} \\
(e) & (f) \\
\end{tabular}
\centering \caption{SISO range-Doppler imaging with three $15$ dB
and six $25$ dB targets. (a) Ground Truth with power levels, (b)
Matched Filter (MF), (c) IAA, (d) SAMV-0, (e) SAMV-1 and (f) SAMV-2.
Power levels are all in dB. } \label{SISO_RDI_25_15}
\end{figure}

In Figure \ref{SISO_RDI}(d), the three $5$ dB sources are not
resolved by the SAMV-0 approach due to the excessive low SNR. After
increasing the power levels of these sources to $15$ dB (the rest
conditions are kept the same as in Figure \ref{SISO_RDI}), all the
sources can be accurately resolved by the SAMV-0 approach in Figure
\ref{SISO_RDI_25_15}(d). We comment that the SAMV-0 approach
provides the most accurate imaging result provided that all sources
have adequately high SNR.

%
%
%
%
%
%

\section{Conclusions} \label{sec:Conclusion}

We have presented a series of user parameter-free array processing
algorithms, the iterative SAMV algorithms, based on the AMV
criterion. It has been shown that these algorithms have superior
resolution and sidelobe suppression ability and are robust to
practical difficulties such as insufficient snapshots, coherent
source signals, without the need of any decorrelation preprocessing.
Moreover, a series of grid-independent SAMV-SML approaches are
proposed to combat the limitation of the direction grid. It is shown
that these approaches provide grid-independent asymptotically
efficient estimates without any additional grid refinement
postprocessing.

\appendices
%
\section{Proof of Result \ref{res: result1}}
\label{sec:Appendix A}
Given $\{{\hat{p}_k}^{(i)}\}_{k=1}^K$ and $\hat{\sigma}^{(i)}$,
which are the estimates of the first $K$ components and the last
element of ${\bf p}$ at $i$th iteration, the matrix ${\bf
R}^{(i)}={\bf A}{\bf P}^{(i)}{\bf A}^H+\sigma^{(i)}{\bf I}$ is
known, thus the matrix ${\bf C}^{(i)}_r={\bf R}^{*(i)}\otimes{\bf
R}^{(i)}$ is also known. For the notational simplicity, we omit the
iteration index and use ${\bf C}_r$ instead in this section.

Define the vectorized covariance matrix of the interference and
noise as
$${\bf r}'_{k}\pardef{\bf r}-p_k\bar{\bf
a}_k, \; k=1, \ldots, K.$$
Assume that ${\bf r}'_{k}$ is known and substitute ${\bf
r}'_{k}+p_k\bar{\bf a}_k$ for ${\bf r}$ in \eqref{eq:MinVa},
minimizing \eqref{eq:MinVa} is equivalent to minimizing the
following cost function
\begin{equation}
\label{eq:simplifiedminfunc} f(p_k)=[{\bf r}_{N}-p_k\bar{\bf
a}_k]^{H} {\bf C}_{r}^{-1} [{\bf r}_{N}-p_k\bar{\bf a}_k]-[{\bf
r}_{N}-p_k\bar{\bf a}_k]^{H} {\bf C}_{r}^{-1} {\bf r}'_{k} -{\bf
r}^{'H}_{k}{\bf C}_{r}^{-1}[{\bf r}_{N}-p_k\bar{\bf a}_k]+{\bf
r}^{'H}_{k}{\bf C}_{r}^{-1}{\bf r}^{'}_{k}
\end{equation}
Note that ${\bf r}^{'}_{k}$ does not depend on $p_k$.
Differentiating \eqref{eq:MinVa} with respect to ${p}_k$ and setting
the results to zero, we get
\begin{equation}
\label{eq:soultion_p} \hat{p}_k=\frac{1}{\bar{\bf a}^H_k {\bf
C}_{r}^{-1}\bar{\bf a}_k}\left( \bar{\bf a}^H_k {\bf C}_{r}^{-1}{\bf
r}_N -\bar{\bf a}^H_k {\bf C}_{r}^{-1}{\bf r}'_k\right),
\;k=1,\ldots,K+1.
\end{equation}
Replacing ${\bf r}'_{k}$ with its definition in
\eqref{eq:soultion_p} yields
\begin{equation}
\label{eq:soultion_p1} \hat{p}_k=\frac{1}{\bar{\bf a}^H_k {\bf
C}_{r}^{-1}\bar{\bf a}_k}\left( \bar{\bf a}^H_k {\bf
C}_{r}^{-1}{\bf r}_N +p_k\bar{\bf a}^H_k {\bf
C}_{r}^{-1}\bar{\bf a}_k-\bar{\bf a}^H_k {\bf C}_{r}^{-1}{\bf
r}\right).
\end{equation}
Using the following identities (see, e.g., \cite[Th. 7.7,
7.16]{Schott80}),
\begin{eqnarray}
\label{eq:identiesvec1} \ve({\bf A}{\bf B}{\bf C})&=&({\bf
C}^{T}\otimes{\bf A})\ve({\bf B}),  \\\label{eq:identiesvec2}
 ({\bf A}\otimes{\bf
B})\otimes({\bf C} \otimes{\bf D})&=&{\bf A}{\bf C}\otimes {\bf
B}{\bf D},
\end{eqnarray}
Eq. \eqref{eq:soultion_p1} can be simplified as
\begin{eqnarray}
\label{eq:estimatesp}
\hat{p}_k&=&\frac{{\bf a}^{H}_k{\bf R}^{-1{}}{\bf R}_N {\bf
R}^{-1{}}{\bf a}_k}{ ({\bf a}^{H}_k{\bf R}^{-1{}}{\bf
a}_k)^2}+{p}_k-\frac{1}{{\bf a}^{H}_k{\bf
R}^{-1{}}{\bf a}_k}, \; k=1,\ldots,K, \\
\label{eq:estimatessigma}
\hat{\sigma}=\hat{p}_{K+1}&=&\frac{1}{\tra({\bf R}^{-2})}\left(
\tra({\bf R}^{-2}{\bf R}_N)+\sigma\tra({\bf R}^{-2})-\tra({\bf
R}^{-1}) \right).
\end{eqnarray}
%
%
%
%
Computing $\hat{p}_k$ and  $\hat{\sigma}$ requires  the knowledge of
$p_k$, $\sigma$, and ${\bf R}$. Therefore, this algorithm must be
implemented iteratively as is detailed in Table 1.
%
%
\section{Proof of Result \ref{res: result2}}
\label{sec:Appendix B}
Define the covariance matrix of the interference and noise as
 \begin{equation}
 \label{eq:Qkmatrix}
 {\bf Q}_k\pardef{\bf R}-p_k{\bf a}_k{\bf a}^H_k, \; k=1, \ldots, K.
\end{equation}
Applying the matrix inversion lemma to \eqref{eq:Qkmatrix} yields
\begin{equation}
 \label{eq:QkRlemma}
 {\bf R}^{-1}={\bf Q}^{-1}_k-p_k\beta_k{\bf b}_k{\bf b}^H_k, \; k=1,
 \ldots, K,
 \end{equation}
 where
${\bf b}_k\pardef {\bf Q}^{-1}_k{\bf a}_k$ and
$\beta_k\pardef(1+p_k{\bf a}^H_k{\bf Q}^{-1}_k{\bf a}_k)^{-1}$.
Since
\begin{eqnarray}
\label{eq:rel1} \tra({\bf R}^{-1}{\bf R}_N)&=& \tra({\bf
Q}^{-1}_k{\bf R}_N)- p_k\beta_k {\bf b}^H_k{\bf R}_N{\bf b}_k,
\end{eqnarray}
and using the algebraic identity $\det({\bf I}+{\bf A}{\bf
B})=\det({\bf I}+{\bf B}{\bf A})$, we obtain
\begin{eqnarray}
\label{eq:rel2} \ln(\det({\bf R}))&=&\ln(\det({\bf Q}_k+p_k{\bf
a}_k{\bf a}^H_k))=\ln \left[ (1+p_k{\bf a}^H_k{\bf Q}^{-1}_k{\bf
a}_k) \det({\bf Q}_k) \right]\\\nonumber &=&\ln(\det({\bf
Q}_k))-\ln(\beta_k).
\end{eqnarray}

Substituting \eqref{eq:rel1} and \eqref{eq:rel2} into the ML
function \eqref{eq:MLfunc} yields
\begin{eqnarray}
\nonumber
\label{eq:MLfun_se}\mathcal{L}({\bf p})&=&\ln(\det({\bf Q}_k))+\tra({\bf Q}^{-1}_k{\bf
R}_N)-\left(\ln(\beta_k)+p_k\beta_k ({\bf b}^H_k{\bf R}_N{\bf
b}_k)\right)\\&=&\mathcal{L}({\bf p}_{-k})-l(p_k),
\end{eqnarray}
with
\begin{equation}
\label{eq:MLfuncti3} l(p_k)\pardef\ln\left( \frac{1}{1+p_k\alpha_{1,k}}\right)+p_k \frac{\alpha^{N}_{2,k}}{1+p_k\alpha_{1,k}},
\end{equation}
where $\alpha_{1,k} \pardef ({\bf a}^H_k{\bf Q}^{-1}_k{\bf
a}_k)^{-1}$ and $\alpha^N_{2,k} \pardef ({\bf a}^H_k{\bf
Q}^{-1}_k{\bf R}_N{\bf Q}^{-1}_k{\bf a}_k)^{-1}$. The objective
function has now been decomposed into $\mathcal{L}({\bf p}_{-k})$,
the marginal likelihood with $p_k$ excluded, and $l(p_k)$, where
terms concerning $p_k$ are conveniently isolated. Consequently,
minimizing \eqref{eq:MLfunc} with respect to $p_k$ is equivalent to
minimizing the function \eqref{eq:MLfuncti3} with respect to the
parameter $p_k$.

It has been proved in  \cite[Appendix, Eqs. (27) and (28)]{Yardibi}
that the unique minimizer of  the cost function \eqref{eq:MLfuncti3}
is
\begin{equation}
\label{eq:solML} \hat{p}_k=\frac{{\bf a}^H_k{\bf Q}^{-1}_k({\bf
R}_N-{\bf Q}_k){\bf Q}^{-1}_k{\bf a}_k}{ ({\bf a}^H_k{\bf Q}^{-1}_k
{\bf a}_k)^2}, \; k = 1, \ldots, K.
\end{equation}
We note that $\hat{\bf p}$ is strictly positive if ${\bf a}^H_k{\bf
Q}^{-1}_k {\bf R}_N {\bf Q}^{-1}_k{\bf a}_k   >   {\bf a}^H_k{\bf
Q}^{-1}_k {\bf a}_k $. Using \eqref{eq:QkRlemma}, we have
\begin{eqnarray}
\label{eq:relation3} {\bf a}^H_k{\bf Q}^{-1}_k {\bf
a}_k&=&\gamma_k({\bf a}^H_k{\bf
R}^{-1} {\bf a}_k), \\
\label{eq:relation4} {\bf a}^H_k{\bf Q}^{-1}_k {\bf R}_N{\bf
Q}^{-1}_k{\bf a}_k&=&\gamma^2_k({\bf a}^H_k{\bf R}^{-1} {\bf
R}_N{\bf R}^{-1}{\bf a}_k),
\end{eqnarray}
where
$\gamma_k\pardef 1+p_k{\bf a}^H_k{\bf Q}^{-1}_k {\bf a}_k$.
Substituting \eqref{eq:relation3} and \eqref{eq:relation4} into
\eqref{eq:solML}, we obtain the desired expression
\begin{equation}
\label{eq:solML2} \hat{p}_k=\frac{{\bf a}^H_k{\bf R}^{-1}({\bf
R}_N-{\bf R}){\bf R}^{-1}{\bf a}_k}{ ({\bf a}^H_k{\bf R}^{-1} {\bf
a}_k)^2}+p_k=\frac{{\bf a}^H_k{\bf R}^{-1}{\bf R}_N{\bf R}^{-1}{\bf
a}_k}{ ({\bf a}^H_k{\bf R}^{-1} {\bf a}_k)^2}+p_k-\frac{1}{{\bf
a}^H_k{\bf R}^{-1} {\bf a}_k}.
 \end{equation}
%
%
%
Differentiating \eqref{eq:MLfunc} with respect to $\sigma$ and
setting the result to zero, we obtain
\begin{equation}
\hat{\sigma} =  \frac{ \tra{( {\bf R}^{-1}  ( {\bf R}_N - \bar{\bf
R} ) {\bf R}^{-1} })}{\tra{( {\bf R}^{-2} ) } },
\end{equation}
and after substituting ${\bf R} - \sigma {\bf I}$ for $\bar{\bf R}$
in the above equation,
\begin{equation}
\label{eq:solML2} \hat{\sigma} = \frac{ \tra({\bf R}^{-1}({\bf
R}_N-{\bf R}){\bf R}^{-1})}{ \tra({\bf R}^{-2})}+\sigma=\tra({\bf
R}^{-2}{\bf R}_N)/{\tra{({\bf R}^{-2})}}+{\sigma}
 -\tra({\bf R}^{-1})/{\tra{({\bf
R}^{-2})}}.
 \end{equation}
Computing $\hat{p}_k$ and  $\hat{\sigma}$ requires the knowledge of
$p_k$, $\sigma$, and ${\bf R}$. Therefore, the algorithm must
be implemented iteratively as is detailed in Result \ref{res: result1}. %
%
\section{Proof of Result \ref{res: result3}}
\label{sec:Appendix C}
%

Differentiating \eqref{eq:MinVb} with respect to ${p}_k$ and setting
the result to zero, we get
\begin{equation}
\label{eq:estimatspamv1} {p}^{(i+1)}_k=\frac{\bar{\bf a}^H_k {\bf
C'}^{-1}_k{\bf r}_N}{\bar{\bf a}^H_k{\bf C'}^{-1}_k \bar{\bf a}_k}.
\end{equation}
Applying the matrix inversion lemma to ${\bf C'}_k$, the numerator
and denominator of Eq. \eqref{eq:estimatspamv1} can be expressed
respectively, as
\begin{eqnarray*}
\bar{\bf a}^H_k {\bf C'}^{-1}_k{\bf r}_N&=&w_k(\bar{\bf a}^H_k {\bf
C}^{-1}_r{\bf r}_N), \\
\bar{\bf a}^H_k {\bf C'}^{-1}_k\bar{\bf a}_k&=&w_k(\bar{\bf a}^H_k
{\bf C}^{-1}_r\bar{\bf a}_k),
\end{eqnarray*}
where $w_k\pardef 1+\frac{\bar{\bf a}^H_k {\bf C}^{-1}_r\bar{\bf
a}_k}{1/p^2_k+\bar{\bf a}^H_k {\bf C}^{-1}_r\bar{\bf a}_k}$.

Thus,
\begin{eqnarray}
\label{eq:estimatspamv3} {p}^{(i+1)}_k&=&\frac{\bar{\bf a}^H_k {\bf
C'}^{-1}_k{\bf r}_N}{\bar{\bf a}^H_k{\bf C'}^{-1}_k \bar{\bf
a}_k}=\frac{\bar{\bf a}^H_k {\bf C}^{-1}_r{\bf r}_N}{\bar{\bf
a}^H_k{\bf C}^{-1}_r \bar{\bf a}_k}, \;k=1,\ldots,K+1.
\end{eqnarray}
Using the Kronecker product properties and the identities
\eqref{eq:identiesvec1} and \eqref{eq:identiesvec2}, with ${\bf
A}={\bf B}={\bf R}$ and ${\bf C}={\bf R}_N$, the numerator and
denominator of Eq. \eqref{eq:estimatspamv3} can be expressed
respectively, as
\begin{eqnarray}
\label{eq:nopower1} \bar{\bf a}^H_k {\bf C}^{-1}_r{\bf r}_N&=& {\bf
a}^H_k{\bf R}^{-1}{\bf R}_N{\bf R}^{-1}{\bf a}_k, \;k=1,\ldots, K,\\
\label{eq:nopower2}
 \bar{\bf a}^H_k{\bf C}^{-1}_r \bar{\bf a}_k&=&({\bf a}^H_k{\bf
R}^{-1}{\bf a}_k)^2, \;k=1,\ldots, K,
 \end{eqnarray}
 and
\begin{eqnarray}
\label{eq:nosi1} \bar{\bf a}^H_{K+1} {\bf C}^{-1}_r{\bf
r}_N&=&\tra({\bf R}^{-2}{\bf R}_N), \\
\label{eq:nosi2} \bar{\bf a}^H_{K+1}{\bf C}^{-1}_r \bar{\bf
a}_{K+1}&=&\tra({\bf R}^{-2}).
\end{eqnarray}
Therefore, dividing \eqref{eq:nopower1} by \eqref{eq:nopower2} gives
\eqref{eq:SAMV-1}, and dividing \eqref{eq:nosi1} by \eqref{eq:nosi2}
yields \eqref{eq:noisesam}.
%
%
%


\begin{thebibliography}{6}
%
\bibitem{Donoho06}
D. L. Donoho, M. Elad, and V. N. Temlyakov, ``Stable recovery of
sparse overcomplete representations in the presence of noise,'' {\em
IEEE Trans. on Infor.Theory}, vol. 52, no. 1, pp. 6--18, Jan. 2006.
%
\bibitem{Donoho06bis}
 E. Candes, J. Romberg, T. Tao, ``Stable signal recovery from incomplete
and inaccurate measurements,'' {\em Communications on Pure and
Applied Mathematics}, vol. 59, pp. 1207--1223, 2006.
%
\bibitem{Tropp}
 J. A. Tropp, ``Just relax: convex programming methods for identifying
sparse signals in noise,'' {\em IEEE Trans. Infor. Theory}, vol. 52,
no. 3, pp. 1030--1051, Mar. 2006.
%
\bibitem{Donoho03}
 D. L. Donoho and M. Elad, ``Optimally sparse
representation in general (nonorthogonal) dictionaries via $\ell_1$
minimization,'' {\em Proc. Nat. Acad. Sci.}, vol. 100, pp.
2197--2202, 2003.
%
 \bibitem{Gorodnitsky}
I. F. Gorodnitsky and B. D. Rao, ``Sparse signal reconstruction from
limited data using FOCUSS: A re-weighted minimum norm algorithm,''
{\em IEEE Trans. Signal Process.}, vol. 45, no. 3, pp. 600--616,
Mar. 1997.
%
  \bibitem{Rao}
B. D. Rao, K. Engan, S. F. Cotter, J. Palmer, K. Kreutz-Delgado,
``Subset selection in noise based on diversity measure
minimization,'' {\em IEEE Trans. on Signal. Process.}, vol. 51, no.
3, pp. 760--770, 2003.

%
 \bibitem{Fevrier}
I. J. Fevrier, S. B. Gelfand, and M. P. Fitz, ``Reduced complexity
decision feedback equalization for multipath channels with large
delay spreads,'' {\em IEEE Trans. Commun.}, vol. 47, pp. 927--937,
June 1999.
 %
\bibitem{Cotter}
 S. F. Cotter and B. D. Rao, ``Sparse channel
estimation via Matching Pursuit with application to equalization,''
{\em IEEE Trans. Commun.}, vol. 50, pp. 374--377, Mar. 2002.
%
\bibitem{Ling}
J. Ling, T. Yardibi, X. Su, H. He, and J. Li, ``Enhanced channel
estimation and symbol detection for high speed Multi-Input
Multi-Output underwater acoustic communications," {\em Journal of
the Acoustical Society of America}, vol. 125, pp. 3067--3078, May
2009.
\bibitem{Berger}
 C. R. Berger, S. Zhou, J. Preisig, and P. Willett,
``Sparse channel estimation for multicarrier underwater acoustic
communication: From subspace methods to compressed sensing,'' {\em
IEEE Trans. on Signal. Process.}, vol. 58, no. 3, pp. 1708--1721,
March 2010.
%
\bibitem{Cetin01}
M. Cetin and  W. C. Karl, ``Feature-enhanced synthetic aperture
radar image formation based on nonquadratic regularization,'' {\em
IEEE Trans. Image Process.}, vol. 10, no. 4, pp. 623--631, April
2001.
%
\bibitem{Chen10}
 Z. Chen, X. Tan, M. Xue, and J. Li, ``Bayesian SAR imaging,''
{\em In Proc. of SPIE on Technologies and Systems for Defense and
Security,} Orlando, FL, April 2010.
%
\bibitem{Roberts}
W. Roberts, P. Stoica, J. Li, T. Yardibi, and F. A. Sadjadi,
``Iterative adaptive approaches to MIMO radar imaging,'' {\em IEEE
Journal on Selected Topics in Signal Proc.}, vol. 4, no. 1, pp.
5--20, 2010.
%
\bibitem{Austin11}
 C. D. Austin, E. Ertin, and R. L. Moses, ``Sparse signal methods for 3-D radar imaging,''
{\em IEEE Trans. Signal Process.}  vol. 5, no. 3, pp. 408--423, June
2011.
%
\bibitem{Malioutov}
 D. M. Malioutov, M. Cetin, and A. S. Willsky, ``A sparse
signal reconstruction perspective for source localization with
sensor arrays,'' {\em IEEE Trans. Signal Processing}, vol. 53, no.
8, pp. 3010--3022, August 2005.
%
\bibitem{Yardibi}
T. Yardibi, J. Li, P. Stoica, M. Xue, and A. B. Baggeroer, ``Source
localization and sensing: A nonparametric iterative adaptive
approach based on weighted least squares,'' {\em IEEE Trans. Aerosp.
Electron. Syst.}, vol. 46, pp. 425--443, 2010.
%
\bibitem{Schmidt} R. O. Schmidt, ``Multiple emitter location and signal
parameter estimation,'' {\em IEEE Trans. on Antennas and Prop.},
vol. 34, no. 3, pp. 276--280, 1986.
%
\bibitem{Roy}
 R. Roy, A. Paulraj, and T. Kailath, ``ESPRIT--A subspace rotation
approach to estimation of parameters of cisoids in noise,'' {\em
IEEE Transactions on Acoustics, Speech and Signal Processing}, vol.
34, no. 5, pp. 1340--1342, 1986.
%
\bibitem{Pillai} S. U. Pillai and B. H. Kwon, ``Forward/backward spatial
smoothing techniques for coherent signal identification,'' {\em IEEE
Trans. Acoustic, Speech, Signal Processing}, vol. 37, pp. 8--15,
1989.

\bibitem{Stoica10}
P. Stoica, P. Babu, and J. Li, ``SPICE: A sparse covariance-based
estimation method for array processing,'' {\em IEEE Trans. Signal
Processing}, vol. 59, no. 2, pp. 629--638, Feb. 2011.
%
\bibitem{Stoica11} P. Stoica, P. Babu, and J. Li, ``New method of sparse
parameter estimation in separable models and its use for spectral
analysis of irregularly sampled data,'' {\em IEEE Transactions on
Signal Processing}, vol. 59, no. 1, pp. 35--47, 2011.
%
\bibitem{Porat85}
B. Porat and B. Friedlander,
``Asymptotic accuracy of ARMA parameter estimation methods based on
sample covariances,''
{\em Proc.7th IFAC/IFORS Symposium on Identification and System
Parameter Estimation, York}, 1985.
%
\bibitem{Stoica85}
P. Stoica, B. Friedlander and T. S\"oderstr\"om, ``An approximate
maximum approach to ARMA spectral estimation,'' {\em in Proc.
Decision and control, Fort Lauderdale}, 1985.
%
\bibitem{Gershman95} A. B. Gershman, A. L. Matveyev, and J. F. Bohme, ``ML estimation of
signal power in the presence of unknown noise field--simple
approximate estimator and explicit Cramer-Rao bound,'' {\em Proc.
IEEE Int. Conf. on Acoust., Speech, and Signal Processing
(ICASSP'95)}, pp. 1824--1827, Detroit, Apr. 1995.
%
\bibitem{Gershman92} A. B. Gershman, V. I. Turchin, and R. A. Ugrinovsky, ``Simple maximum
likelihood estimator for structured covariance parameters,'' {\em
Electron. Lett.}, vol. 28, no. 18, pp. 1677--1678, Aug. 1992.
%
\bibitem{Stoica90}
P. Stoica and A. Nehorai, ``Performance study of conditional and
unconditional direction of arrival estimation,'' {\em IEEE Trans.
Acoust., Speech, Signal Processing}, vol. 38, pp. 1783--1795, Oct.
1990.
%
\bibitem{Abeida07b}
H. Abeida and J. P. Delmas, ``Efficiency of subspace-based DOA
estimators,'' {\em Signal Process.}, vol. 87, pp. 2075--2084, 2007.

%


\bibitem{Delmas04a}
J.~P.~Delmas, ``Asymptotically minimum variance second-order
estimation for non-circular signals with application to DOA
estimation,'' {\em IEEE Trans. Signal Processing},
 vol. 52, no. 5, pp. 1235--1241, May 2004.
\bibitem{Abeida05a}
H. Abeida and J.~P.~Delmas, ``MUSIC-like estimation of direction of
arrival for non-circular sources,'' {\em IEEE Trans. Signal
Processing}, vol. 54, no. 7, pp. 2678--2690, Jul. 2006.
%
\bibitem{Stoica05b}
P. Stoica and R. Moses, {\em Spectral Analysis of Signals}, Upper
Saddle River, NJ: Prentice-Hall, 2005.
%
\bibitem{Anderson}
 T. W. Anderson, {\em An Introduction to Multivariate Statistical
Analysis}, John Wiley \& Sons, Inc., 1958.

\bibitem{Schott80}
 J. R. Schott, {\em Matrix Analysis for Statistics}, New York:
Wiley, 1980.

\bibitem{Nelder65}
J.~A.~Nelder and R.~Mead, ``A simplex method for function
minimization,'' {\em Computer Journal}, vol. 7, pp. 308--313, 1965. 

\bibitem{zhang2011fast}
Q.~Zhang, H.~Abeida, M.~Xue, W.~Rowe, and J.~Li, ``Fast implementation of
sparse iterative covariance-based estimation for array processing,'' in
\emph{Signals, Systems and Computers (ASILOMAR), 2011 Conference Record of
	the Forty Fifth Asilomar Conference on}.\hskip 1em plus 0.5em minus
0.4em\relax IEEE, 2011, pp. 2031--2035.


\bibitem{zhang2012fast}
Q.~Zhang, H.~Abeida, M.~Xue, W.~Rowe, and J.~Li, ``Fast implementation of sparse iterative covariance-based estimation
for source localization,'' \emph{The Journal of the Acoustical Society of
	America}, vol. 131, no.~2, pp. 1249--1259, 2012.

\end{thebibliography}
\end{document}